\documentclass[11pt]{article}
\usepackage{makeidx}
\usepackage{amssymb, amsmath,amsthm,amsfonts, enumerate}
\usepackage{setspace, graphicx, caption, subcaption, float, relsize, rotating, color, verbatim, multirow}
\usepackage{url, hyperref,epstopdf}
\usepackage[round, comma, longnamesfirst]{natbib}

\makeatletter
\def\maketag@@@#1{\hbox{\m@th\normalfont\normalsize#1}}
\makeatother
\hypersetup{
  colorlinks = true, urlcolor = blue, linkcolor = blue, citecolor = magenta }
\def\maketag@@@#1{\hbox{\m@th\normalfont\normalsize#1}}
\makeatother

\allowdisplaybreaks
\newtheorem {theorem}{Theorem}[section]
\newtheorem {assumption}{Assumption}

\newtheorem{corollary}[theorem]{Corollary}

\newtheorem{lemma}[theorem]{Lemma}

\newtheorem{remark}{Remark}

\newtheorem*{assumption2c'}{Assumption 2.c$^{\prime}$}
\begin{document}

\title{{\huge A Max-Correlation White Noise Test for Weakly Dependent Time
Series\thanks{
We thank three referees and Co-Editor Michael Jansson for helpful comments
and suggestions that led to significant improvements to our manuscript. We
also thank Eric Ghysels, Shigeyuki Hamori, Peter R. Hansen, Yoshihiko
Nishiyama, Kenichiro Tamaki, Kozo Ueda, and Zheng Zhang, seminar
participants at the Kyoto Institute of Economic Research, UNC Chapel Hill,
the University of Essex, and Kobe University, and conference participants at
the 10th Spring Meeting of JSS, 2016 SWET, 2016 AMES, 2016 JJSM, 2016
NBER-NSF Time Series Conference, the 15th International Conference of WEAI,
and SETA 2019 for helpful comments. The second author is grateful for
financial supports from JSPS KAKENHI (Grant Number 16K17104), Kikawada
Foundation, Mitsubishi UFJ Trust Scholarship Foundation, Nomura Foundation,
and Suntory Foundation.}}}
\author{ \ \ Jonathan B. Hill \thanks{%
Corresponding author. Department of Economics, University of North Carolina
at Chapel Hill. E-mail: \texttt{jbhill@email.unc.edu}; web: \texttt{%
https://jbhill.web.unc.edu.}} \qquad\ \ \ \ \ and\qquad\ \ Kaiji Motegi%
\thanks{%
Graduate School of Economics, Kobe University. E-mail: \texttt{%
motegi@econ.kobe-u.ac.jp}} \\
%EndAName
University of North Carolina \qquad\ \qquad Kobe University}
\date{\today \\
}
\maketitle

\setstretch{1}

\begin{center}
\textbf{Abstract}
\end{center}

This paper presents a bootstrapped p-value white noise test based on the
maximum correlation, for a time series that may be weakly dependent under
the null hypothesis. The time series may be prefiltered residuals. The test
statistic is a normalized weighted maximum sample correlation coefficient $%
\max_{1\leq h\leq \mathcal{L}_{n}}\sqrt{n}|\hat{\omega}_{n}(h)\hat{\rho}%
_{n}(h)|$, where $\hat{\omega}_{n}(h)$ are weights and the maximum lag $%
\mathcal{L}_{n}$ increases at a rate slower than the sample size $n$. We
only require uncorrelatedness under the null hypothesis, along with a moment
contraction dependence property that includes mixing and non-mixing
sequences. We show Shao's (\citeyear{Shao2011_JoE}) dependent wild bootstrap
is valid for a much larger class of processes than originally considered. It
is also valid for residuals from a general class of parametric models as
long as the bootstrap is applied to a first order expansion of the sample
correlation.
%The test has non-trivial local power against $\sqrt{n}$-local alternatives, and can detect very weak and distant serial dependence better than a variety of other tests.
%We also prove that our bootstrapped p-value leads to a valid test without exploiting extreme value theoretic arguments (the standard in the literature), or recent Gaussian approximation theory.
%Finally, we extend Escanciano and Lobato's (\citeyear{EscancianoLobato2009}) automatic maximum lag selection to our setting with an unbounded choice set, and find it works very well in controlled experiments.
We prove the bootstrap is asymptotically valid without exploiting extreme
value theory (standard in the literature) or recent Gaussian approximation
theory. Finally, we extend Escanciano and Lobato's (%
\citeyear{EscancianoLobato2009}) automatic maximum lag selection to our
setting with an unbounded lag set that ensures a consistent white noise
test, and find it works extremely well in controlled experiments. \medskip

\setstretch{1.225}

\newpage

\section{Introduction\label{sec:intro}}

We present a bootstrap white noise test based on the maximum (in absolute
value) autocorrelation. The data may be observed, or filtered residuals. A
new asymptotic theory approach is used relative to the literature, one that
sidesteps deriving the asymptotic distribution of a max-correlation
statistic, or working with tools specific to Gaussian approximations and
couplings. We operate solely on the bootstrapped p-value. Convergence in
finite dimensional distributions of the sample correlation is combined with
with new theory for handling convergence of arbitrary arrays. The latter is
applicable for dealing with the maximum of an increasing sequence of
correlations, in particular when residuals based on a plug-in estimator are
used.

The class of time series models considered here is:%
\begin{equation}
y_{t}=f(x_{t-1},\phi _{0})+u_{t}\text{ \ and \ }u_{t}=\epsilon _{t}\sigma
_{t}(\theta _{0})  \label{reg_model}
\end{equation}%
where $\phi $ $\in $ $\mathbb{R}^{k_{\phi }}$, $k_{\phi }$ $\geq $ $0$, and $%
f(x,\phi )$ is a level response function. The error $\epsilon _{t}$
satisfies $E[\epsilon _{t}]$ $=$ $0$, $E[\epsilon _{t}^{2}]$ $<$ $\infty $,
and the regressors are $x_{t}$ $\in $ $\mathbb{R}^{k_{x}}$, $k_{x}$ $\geq $ $%
0$. We assume $\{x_{t},y_{t}\}$ are strictly stationary in order to focus
ideas. Volatility $\sigma _{t}^{2}(\theta _{0})$ is a process measurable
with respect to $\mathcal{F}_{t-1}$ $\equiv $ $\sigma (y_{\tau },x_{\tau }$ $%
:$ $\tau $ $\leq $ $t$ $-$ $1)$, where $\theta _{0}$ is decomposed as $[\phi
_{0}^{\prime },\delta _{0}^{\prime }]$ $\in $ $\mathbb{R}^{k_{\theta }}$, $%
\delta _{0}\in $ $\mathbb{R}^{k_{\delta }}$ are volatility-specific
parameters, and $(k_{\theta },k_{\delta })$ $\geq $ $0$. The dimensions of $%
\phi _{0}$ and $\delta _{0}$ (hence $\theta _{0}$) may be zero, depending on
the model desired and the interpretation of the test variable $\epsilon _{t}$%
. Thus, $k_{\phi }$ $=$ $0$ implies a volatility model $y_{t}$ $=$ $\epsilon
_{t}\sigma _{t}(\theta _{0})$, if $k_{\delta }$ $=$ $0$ then $y_{t}$ $=$ $%
f(x_{t-1},\phi _{0})$ $+$ $\epsilon _{t}$, and $y_{t}$ $=$ $\epsilon _{t}$
when $k_{\theta }$ $=$ $0$ (i.e. a filter is not used). We want to test if $%
\{\epsilon _{t}\}$ is a white noise process:%
\begin{equation*}
H_{0}:E\left[ \epsilon _{t}\epsilon _{t-h}\right] =0\text{ }\forall h\in
\mathbb{N}\text{ against }H_{1}:E\left[ \epsilon _{t}\epsilon _{t-h}\right]
\neq 0\text{ for some }h\in \mathbb{N}.
\end{equation*}

Notice $\epsilon _{t}$ need not have a zero conditional mean: we do not
require, e.g., $E[\epsilon _{t}|x_{t-1}]$ $=$ $0$ $a.s.$ This implies that
we do not require $\sigma _{t}^{2}(\theta _{0})$ to be a conditional
variance. Together, (\ref{reg_model}) allows for model mis-specification.
Nevertheless, (\ref{reg_model}) is assumed correct in some sense, whether $%
H_{0}$ is true or not, in view of $E[\epsilon _{t}]$ $=$ $0$ and possibly
other moment conditions used to identify $\theta _{0}$. Thus, $\theta _{0}$
should be thought of as a pseudo-true value that can be identified, often by
unconditional moment conditions \citep{KullbackLeibler1951,Sawa1978}.
Complete assumptions are given in Section \ref{sec:max_corr}: see especially
Assumption \ref{assum:plug}.

Unless $y_{t}$ $=$ $\epsilon _{t}$ such that $y_{t}$ is known to have a zero
mean, let $\hat{\theta}_{n}$ $=$ $[\hat{\phi}_{n}^{\prime },\hat{\delta}%
_{n}^{\prime }]$ estimates $\theta _{0}$ where $n$ is the sample size, and
define the residual, and its sample serial covariance and correlation at lag
$h$ $\geq $ $1$:
\begin{equation*}
\epsilon _{t}(\hat{\theta}_{n})\equiv \frac{u_{t}(\hat{\phi}_{n})}{\sigma
_{t}(\hat{\theta}_{n})}\equiv \frac{y_{t}-f(x_{t-1},\hat{\phi}_{n})}{\sigma
_{t}(\hat{\theta}_{n})}\text{ \ and \ }\hat{\gamma}_{n}(h)\equiv \frac{1}{n}%
\sum_{t=1+h}^{n}\epsilon _{t}(\hat{\theta}_{n})\epsilon _{t-h}(\hat{\theta}%
_{n})\text{ \ and }\hat{\rho}_{n}(h)\equiv \frac{\hat{\gamma}_{n}(h)}{\hat{%
\gamma}_{n}(0)}.
\end{equation*}%
In the pure volatility model set $f(x_{t-1},\hat{\phi}_{n})$ $=$ $0$, and in
the level model set $\sigma _{t}(\hat{\theta}_{n})$ $=$ $1$.

Our primary test statistic is the normalized weighted sample maximum
correlation,
\begin{equation*}
\mathcal{\hat{T}}_{n}\equiv \sqrt{n}\max_{1\leq h\leq \mathcal{L}%
_{n}}\left\vert \hat{\omega}_{n}(h)\hat{\rho}_{n}(h)\right\vert ,
\end{equation*}%
where $\hat{\omega}_{n}(h)$ $>$ $0$ are possibly stochastic weights with $%
\hat{\omega}_{n}(h)$ $\overset{p}{\rightarrow }$ $\omega (h)$ $>$ $0$, where
$\omega (h)$ are non-stochastic. The weights allow for ($i$) control for
variable dispersion across lags that affect empirical power, or ($ii$) a
decrease in accuracy in probability when $n$ is small and $h$ is large. In
the former case $\hat{\omega}_{n}(h)$ may be an inverted standard deviation
estimator. In the latter case we might use $\hat{\omega}_{n}(h)$ $=$ $(n$ $-$
$2)/(n$ $-$ $h)$ as in \cite{LjungBox1978}. Despite the generality afforded
by weights, we find using $\hat{\omega}_{n}(h)$ $=$ $1$ results in accurate
size and comparably high power in Monte Carlo simulations. Indeed, using an
inverted standard deviation $\hat{\omega}_{n}(h)$ does not improve test
performance in our experiments due to estimation error associated with $\hat{%
\omega}_{n}(h)$.

The number of lags $\mathcal{L}_{n}$\ can converge to a finite positive
integer: the theory follows trivially from the proofs of our main results.
In that case our test would not be a formal test of the white noise
hypothesis. We want $\mathcal{L}_{n}$ $\rightarrow $ $\infty $ as $n$ $%
\rightarrow $ $\infty $ in order to ensure a white noise test, and that $%
\mathcal{L}_{n}$ $=$ $o(n)$ to ensure $\hat{\gamma}_{n}(h)$ $=$ $E[\epsilon
_{t}\epsilon _{t-h}]$ $+$ $O_{p}(1/\sqrt{n})$ for each $h\in \{1,...,%
\mathcal{L}_{n}\}$ and therefore yield a consistent test. The limit theory
in that case requires more than convergence in finite dimensional
distributions based on classic arguments %
\citep[e.g.][]{HoffJorg1984,HoffJorg1991}, which is one of the major
challenges we address in this paper.

Interest in the maximum of an increasing sequence of deviated covariances $%
\sqrt{n}$ $\max_{1\leq h\leq \mathcal{L}_{n}}|$$\hat{\gamma}_{n}$$(h)$ $-$ $%
\gamma $$(h)|$ dates in some form to \cite{Berman1964} and \cite{Hannan1974}%
. See also \cite{XiaoWu2014} and their references. In this literature the
test variable is observed, and the exact asymptotic distribution form of a
suitably normalized $\sqrt{n}\max_{1\leq h\leq \mathcal{L}_{n}}|\hat{\gamma}%
_{n}(h)$ $-$ $\gamma (h)|$ is sought. \cite{XiaoWu2014} impose a moment
contraction property on $y_{t}$, and $\mathcal{L}_{n}$ $=$ $O(n^{\upsilon })$
for some $\upsilon $ $\in $ $(0,1)$ that is smaller with greater allowed
dependence. They show $a_{n}\{\sqrt{n}\max_{1\leq h\leq \mathcal{L}_{n}}|%
\hat{\gamma}_{n}(h)$ $-$ $\gamma (h)|/(\sum_{h=0}^{\infty }\gamma
(h)^{2})^{1/2}$ $-$ $b_{n}\}$ $\overset{d}{\rightarrow }$ $\exp \{-\exp
\{-x\}\}$, a Gumbel distribution, with normalizing sequences $a_{n},b_{n}$ $%
\sim $ $(2\ln (n))^{1/2}$. See, also, \cite{Jirak2011}. \cite{XiaoWu2014} do
not prove their blocks-of-blocks bootstrap is valid under their assumptions,
and only observed data are allowed. The moment contraction property is also
more restrictive than the Near Epoch Dependence [NED] property used here %
\citep[see the supplemental material][Appendix B]{HillMotegi_supp_mat}.

\citet{Chernozhukov_etal2013,Chernozhukov_etal2015,Chernozhukov_etal2016}
significantly improve on results in the literature on Gaussian
approximations and couplings, cf. \cite{Yurinskii1977}, \cite%
{DudleyPhilipp1983}, \cite{Portnoy1986}, and \cite{LeCam1988}. They allow
for arbitrary dependence across the sequence of sample means, and the
sequence length may grow at a rate of order $e^{Kn^{\varsigma }}$ for some $%
K,\varsigma $ $>$ $0$. Sample autocorrelations, however, only exist for lags
$\{0,...,n-1\}$, and are Fisher consistent for the population
autocorrelations for lags $h$ up to order $o(n)$. The independence
assumption, however, is not feasible for a white noise test since $\epsilon
_{t}\epsilon _{t-h}$ is at best a martingale difference, and may be
generally dependent under either hypothesis. Further, a Gaussian
approximation theory cannot handle the maximum distance between $\hat{\rho}%
_{n}(h)$ based on residuals $\epsilon _{t}(\hat{\theta}_{n})\epsilon _{t-h}(%
\hat{\theta}_{n})$, and its version based on $\epsilon _{t}\epsilon _{t-h}$
(and other components due to the plug-in estimator $\hat{\theta}_{n}$)
because $\epsilon _{t}\epsilon _{t-h}$ is typically not Gaussian even if $%
\epsilon _{t}$ is.\footnote{%
When filtered data are used we must \ prove in Lemma \ref{lm:corr_expan}
that $\max_{1\leq h\leq \mathcal{L}_{n}}|1/\sqrt{n}\sum_{t=1}^{n}\epsilon
_{t}(\hat{\theta}_{n})\epsilon _{t-h}(\hat{\theta}_{n})$ $-$ $1/\sqrt{n}%
\sum_{t=1}^{n}z_{t}(h)|$ $\overset{p}{\rightarrow }$ $0$ for some sequence $%
\{\mathcal{L}_{n}\}$, $\mathcal{L}_{n}$ $\rightarrow $ $\infty $, and some
process $\{z_{t}(h)\}$ that is a function of $\epsilon _{t}\epsilon _{t-h}$
and components of $\hat{\theta}_{n}$. We then prove in Lemma \ref{lm:clt_max}
that $\max_{1\leq h\leq \mathcal{L}_{n}}|1/\sqrt{n}\sum_{t=1}^{n}z_{t}(h)|$ $%
\overset{d}{\rightarrow }$ $\max_{1\leq h\leq \infty }|\mathcal{Z}(h)|$ for
some Gaussian process $\{\mathcal{Z}(h)\}$. Under suitable memory and
heterogeneity restrictions, the Gaussian approximation theory of Zhang and
Wu (\citeyear{ZhangWu2017}), cf. \cite{Chernozhukov_etal2013}, can handle $%
\max_{1\leq h\leq \mathcal{L}_{n}}|1/\sqrt{n}\sum_{t=1}^{n}z_{t}(h)|$ $%
\overset{d}{\rightarrow }$ $\max_{1\leq h\leq \infty }|\mathcal{Z}(h)|$
since $\{\mathcal{Z}(h)\}$ is Gaussian. But their theory cannot determine $%
\max_{1\leq h\leq \mathcal{L}_{n}}|1/\sqrt{n}\sum_{t=1}^{n}\epsilon _{t}(%
\hat{\theta}_{n})\epsilon _{t-h}(\hat{\theta}_{n})$ $-$ $1/\sqrt{n}%
\sum_{t=1}^{n}z_{t}(h)|$ $\overset{p}{\rightarrow }$ $0 $ because that would
require $1/\sqrt{n}\sum_{t=1}^{n}z_{t}(h)$ itself to be Gaussian for each $n$%
. The latter generally does not hold because $\epsilon _{t}\epsilon _{t-h}$
is not Gaussian even if $\epsilon _{t}$ is.}
\citet[Appendix
B]{Chernozhukov_etal_manymom_2014}, cf.
\citet[Supplemental
Appendix]{Chernozhukov_etal2018}, allow for \textit{almost surely} bounded $%
\beta $-mixing data, but the above problem involving filtered data is not
resolved, boundedness rules out many time series of practical interest, and
our NED environment eclipses a mixing environment
\citep[see Section \ref{sec:assum_expan}, below, and see, e.g.,][Chapter
17]{Davidson1994}.

\cite{ZhangWu2017} extend results in \cite{Chernozhukov_etal2013} to a large
class of dependent processes
\citep[see also][for an extension to
geometrically dependent data in a bootstrap setting]{ZhangWu2014}. Their
framework is the functional dependence or moment contraction notions
popularized in, e.g., \cite{Wu2005}. The possibility of filtered data is
ignored, which requires a non-Gaussian approximation theory. Further, it is
not obvious which processes satisfy the conditions of their main Theorem 3.2
(e.g. nonlinear ARMA-GARCH, stochastic volatility).

%The maximum lag sequence $\{\mathcal{L}_{n}\}$ in the above methods is
%selected without involvement of a criterion. Evidently the maximum
%correlation literature does not offer data driven or automatic methods for
%selecting $\mathcal{L}_{n}$.

Compared to the above literature, we use a different asymptotic theory
approach. We sidestep extreme value theoretic methods by exploiting
convergence of $\{\sqrt{n}(\hat{\gamma}_{n}(h)$ $-$ $\gamma $$(h))$ $:$ $1$ $%
\leq $ $h$ $\leq $ $\mathcal{L}\}$ to a Gaussian process, for each finite $%
\mathcal{L}$ $\in $ $\mathbb{N}$. Since that is not sufficient for weak
convergence in the classic sense of \citet{HoffJorg1984,HoffJorg1991}, we
develop new theory for double array convergence, which is associated with
arguments dating to \cite{Ramsey1930}. This allows us to prove that under $%
H_{0}$ the maximum distance over $1$ $\leq $ $h$ $\leq $ $\mathcal{L}_{n}$
between $\sqrt{n}\hat{\rho}_{n}(h)$ and its bootstrapped version converges
to zero for some sequence of positive integers $\{\mathcal{L}_{n}\}$, with $%
\mathcal{L}_{n}$ $\rightarrow $ $\infty $ and $\mathcal{L}_{n}$ $=$ $o(n)$,
without using extreme value theoretic arguments or Gaussian approximation
theory. Under additional technical conditions presented in the supplemental
material \citet[Appendix G]{HillMotegi_supp_mat}, we show $\mathcal{L}_{n}$ $%
=$ $O(n^{c}/\ln (n))$ must also hold, for some $c$ $\in $ $(0,1)$ that
depends on the rate of convergence of the weights $\hat{\omega}_{n}(h)$ $%
\overset{p}{\rightarrow }$ $\omega (h)$, and an asymptotic approximation
expansion for the plug-in $\hat{\theta}_{n}$. Under standard regularity
conditions $c$ $=$ $1/2$, hence $\mathcal{L}_{n}$ $=$ $O(\sqrt{n}/\ln (n))$.
These are our primary contributions. As in \citet{Chernozhukov_etal2013}, we
do not require $\sqrt{n}\max_{1\leq h\leq \mathcal{L}_{n}}|\hat{\omega}%
_{n}(h)\hat{\rho}_{n}(h)|$ to converge in law under $H_{0}$ since the
bootstrap is asymptotically valid irrespective of the asymptotic properties
of $\sqrt{n}\max_{1\leq h\leq \mathcal{L}_{n}}|\hat{\omega}_{n}(h)\hat{\rho}%
_{n}(h)|$.

Our asymptotic theory covers a class of continuous transforms of $[\sqrt{n}%
\hat{\omega}_{n}(h)\hat{\rho}_{n}(h)]_{h=1}^{\mathcal{L}_{n}}$, including
the maximum, but also a weighted average $n\sum_{h=1}^{\mathcal{L}_{n}}\hat{%
\omega}_{n}^{2}(h)\hat{\rho}_{n}^{2}(h)$, and therefore portmanteau
statistics \citep[cf.][]{LjungBox1978,Hong1996,Hong2001}. %
\citet{Hong1996,Hong2001} presents spectral density methods for testing for
uncorrelatedness, and the proposed test statistic is simply a normalized
portmanteau. The latter is shown to be asymptotically normal under
regularity conditions that ensure $\sqrt{n}\hat{\rho}_{n}^{2}(h)$ is
asymptotically independent across $h$ under $H_{0}$. The approach taken here
alleviates the necessity for the normalized $n\sum_{h=1}^{\mathcal{L}_{n}}%
\hat{\omega}_{n}^{2}(h)\hat{\rho}_{n}^{2}(h)$ to converge in law under $%
H_{0} $, hence we do not require asymptotic independence. Further, as
opposed to \citet{Hong1996,Hong2001},\ our test statistic achieves the
parametric rate of convergence because we do not use self-normalization. See
Remark \ref{rm:Hong_local} in Section \ref{sec:max_corr}.

We perform a bootstrap p-value test using Shao's (\citeyear{Shao2011_JoE})
dependent wild bootstrap, and prove its validity. In order to control for
the use of filtered sampling errors, the bootstrap is applied to a first
order expansion of the sample covariance. \cite{DelgadoVelasco2011} take a
different approach by using orthogonally transformed jointly standardized
correlations in order to control for residuals and dependence. They assume a
fixed maximum lag $\mathcal{L}$, however, due to joint standardization.

Finally, in order to resolve the choice of $\{\mathcal{L}_{n}\}$ in
practice, we extend Escanciano and Lobato's (\citeyear{EscancianoLobato2009}%
) automatic maximum lag selection method to our setting. They develop a
Q-test with bounded maximum lag that is selected based on the magnitude of
the maximum correlation. We allow for selection from an increasing set of
integers, and provide a new asymptotic theory for the automatic maximum lag.

General dependence under the null is allowed in different ways in \cite%
{Hong1996}, \cite{RomanoThombs1996}, \cite{Shao2011_JoE}, and \cite%
{GuayGuerreLazarova2013}, amongst others. Our NED setting is similar to that
of \cite{Lobato2001} and \citet{NankervisSavin2010,NankervisSavin2012}, but
the former works with observed data and requires a fixed maximum lag, and we
allow for a substantially larger class of filters and parametric estimators
than the latter. NED encompasses mixing and non-mixing processes, hence our
setting is more general than Zhu's (\citeyear{Zhu2015}) for his block-wise
random weighting bootstrap.

\cite{Shao2011_JoE}, \cite{GuayGuerreLazarova2013} and \cite{XiaoWu2014} use
a moment contraction property from \cite{Wu2005} and \cite{WuMin2005} with
(potentially far) greater moment conditions than imposed here %
\citep[e.g][]{Shao2011_JoE,GuayGuerreLazarova2013}. \cite{Shao2011_JoE}
requires a complicated eighth order cumulant condition that is only known to
hold under geometric memory, and residuals are not treated. \cite{XiaoWu2014}
only require slightly more than a $4^{th}$ moment, as we do, but do not
allow for residuals. We show in the supplemental material
\citet[Appendix
B]{HillMotegi_supp_mat} that our NED setting is more general than the moment
contraction properties employed in \cite{Shao2011_JoE} and \cite%
{GuayGuerreLazarova2013}, and allows for slower memory decay than \cite%
{XiaoWu2014}.

Test statistics that combine serial correlations have a vast history dating
to Box and Pierce's (\citeyear{BoxPierce70}) Q-test. Many generalizations
exist, including letting the maximum lag increase \citep{Hong1996,Hong2001};
bootstrapping or re-scaling for size correction under weak dependence %
\citep{RomanoThombs1996,Lobato2001,HorowitzLobatoNankervisSavin2006,KuanLee2006,Zhu2015}%
; using a Lagrange Multiplier type statistic to account for weak dependence %
\citep[e.g.][]{AndrewsPloberger1996,LobatoNankervisSavin2002}; exploiting an
expansion and orthogonal projection to produce pivotal statistics %
\citep{Lobato2001,KuanLee2006,DelgadoVelasco2011};
%and using endogenous maximum lag selection \citep{EscancianoLobato2009,GuayGuerreLazarova2013}.
and using endogenous maximum lag selection (Escanciano and Lobato, %
\citeyear{EscancianoLobato2009}, Guay, Guerre, and Lazarov\'{a}, %
\citeyear{GuayGuerreLazarova2013}).

A related class of estimators exploits the periodogram, an increasing sum of
sample correlations, dating to \cite{GrenanderRossenblatt1952} %
\citep[e.g.][]{Hong1996,Deo2000,DelgadoHidalgoVelasco2005,Shao2011_JoE,ZhuLi2015}%
. \cite{Hong1996} standardizes a periodogram resulting in less-than $\sqrt{n}
$-local power, while Cram\'{e}r-von Mises and Kolmogorov-Smirnov transforms
in \cite{Deo2000}, %\cite{DelgadoHidalgoVelasco2005},
Delgado, Hidalgo, and Velasco \citeyearpar{DelgadoHidalgoVelasco2005}, and
\cite{Shao2011_JoE} result in $\sqrt{n}$-local power. \cite%
{GuayGuerreLazarova2013} show that Hong's (\citeyear{Hong1996}) standardized
portmanteau test (but not a Cram\'{e}r-von Mises test) can detect
local-to-null correlation values at a rate faster than $\sqrt{n}$ provided
an adaptive increasing maximum lag is used. Finally, a weighted sum of
correlations also arises in Andrews and Ploberger's (%
\citeyear{AndrewsPloberger1996}) sup-LM test %
\citep[cf.][]{NankervisSavin2010}.

%The Variance Ratio [VR] test statistic also reduces to a sample correlation
%series \citep[see, e.g.,][]{CampbellMankiw1987}. Classic treatments work
%with a fixed lag. \cite{ChenDeo2006} let the lag diverge to infinity, but
%that does not suffice for a true white noise test. \cite{ChowDenning1993}
%propose a max-VR test over a finite maximum lag.

%A simulation study shows that our test works well for unfiltered or filtered
%data from various time series models. We compare our bootstrap test with
%Hong's (\citeyear{Hong1996}) standardized spectral test; Shao's (%
%\citeyear{Shao2011_JoE}) dependent wild bootstrap spectral Cram\'{e}r-von
%Mises test, which is proposed for observed data; and Zhu and Li's (%
%\citeyear{ZhuLi2015}) block-wise random weighting bootstrap Cram\'{e}r-von
%Mises test, which is proposed for linear regression residuals.\ In %
%\citet[Appendix G]{HillMotegi_supp_mat} we also compare our tests to an
%asymptotic and bootstrapped Ljung-Box test, and a bootstrapped Andrews and
%Ploberger's (\citeyear{AndrewsPloberger1996}) sup-LM white noise test.

A simulation study shows our proposed max-correlation test with Shao's (%
\citeyear{Shao2011_JoE}) dependent wild bootstrap and automatic lag (denoted
$\hat{\mathcal{T}}^{dw}(\mathcal{L}_{n}^{\ast })$) dominates a variety of
other tests. In this paper, we compare $\hat{\mathcal{T}}^{dw}(\mathcal{L}%
_{n}^{\ast })$ and Shao's (\citeyear{Shao2011_JoE}) dependent wild bootstrap
spectral Cram\'{e}r-von Mises test, which is proposed for observed data. In
the supplemental material \citet[][Appendix H]{HillMotegi_supp_mat}, we
consider other tests, including Hong's (\citeyear{Hong1996}) test based on a
standardized periodogram, a CvM test with Zhu and Li's (\citeyear{ZhuLi2015}%
) block-wise random weighting bootstrap, and Andrews and Ploberger's (%
\citeyear{AndrewsPloberger1996}) sup-LM test with the dependent wild
bootstrap. Overall the CvM test is one of the strongest competitors of our
test. First, generally $\hat{\mathcal{T}}^{dw}(\mathcal{L}_{n}^{\ast })$
achieves sharp size. Second, $\hat{\mathcal{T}}^{dw}(\mathcal{L}_{n}^{\ast })
$, the sup-LM, and the CvM tests lead to roughly comparable power when there
exist autocorrelations at small lags. Third, $\hat{\mathcal{T}}^{dw}(%
\mathcal{L}_{n}^{\ast })$ has high power while others have nearly trivial
power when there exist autocorrelations at remote lags. Thus, of the tests
under study, $\hat{\mathcal{T}}^{dw}(\mathcal{L}_{n}^{\ast })$ is the only
white noise test that accomplishes both sharp size in general and high
power. The sharp performance of $\hat{\mathcal{T}}^{dw}(\mathcal{L}%
_{n}^{\ast })$ stems from the fact that the automatic lag selection
mechanism trims redundant lags under $H_{0}$, and hones in on the most
informative lag under $H_{1}$.

The remainder of the paper is as follows. Section \ref{sec:max_corr}
contains the assumptions and main results. Automatic lag selection is
developed in Section \ref{sec:lag_select}, and a Monte Carlo study follows
in Section \ref{sec:sim}. Concluding remarks are left for Section \ref%
{sec:conclude}. Proofs are gathered in Appendix \ref{app:proofs} and the
supplemental material \citet[][Appendix F]{HillMotegi_supp_mat}. \medskip

Throughout $|\cdot |$ is the $l_{1}$-matrix norm; $||\cdot ||$ is the $l_{2}$%
-matrix norm; $||\cdot ||_{p}$ is the $L_{p}$-norm. $I(\cdot )$ is the
indicator function: $I(A)$ $=$ $1$ if $A$ is true, else $I(A)$ $=$ $0$. $%
\mathcal{F}_{t}$ $\equiv $ $\sigma (y_{\tau },x_{\tau }$ $:$ $\tau $ $\leq $
$t)$. All random variables lie in a complete probability measure space $%
(\Omega ,\mathcal{P},\mathcal{F})$, hence $\sigma (\cup _{t\in \mathbb{Z}}%
\mathcal{F}_{t})$ $\subseteq $ $\mathcal{F}$. We drop the (pseudo) true
value $\theta _{0}$ from function arguments when there is no confusion.

\section{Max-Correlation Test\label{sec:max_corr}{}}

We first lay out the assumptions and derive some fundamental properties of
the correlation maximum. We then derive the main results.

\subsection{Assumptions and Asymptotic Expansion\label{sec:assum_expan}}

An expansion of $\epsilon _{t}(\hat{\theta}_{n})$\ around $\theta _{0}$ is
required in order to ensure the bootstrapped statistic captures the
influence of the estimator $\hat{\theta}_{n}$ on $\sqrt{n}\hat{\rho}_{n}(h)$%
. This is accomplished under various regularity assumptions. Let $\{\upsilon
_{t}\}$ be a stationary $\alpha $-mixing process with $\sigma $-fields $%
\mathfrak{V}_{s}^{t}$ $\equiv $ $\sigma (\upsilon _{\tau }$ $:$ $s$ $\leq $ $%
\tau $ $\leq $ $t)$ and $\mathfrak{V}_{t}\equiv \mathfrak{V}_{-\infty }^{t}$%
, and coefficients $\alpha _{m}^{(\upsilon )}$ $=$ $\sup_{\mathcal{A}\subset
\mathfrak{V}_{t}^{\infty },\mathcal{B}\subset \mathfrak{V}_{-\infty
}^{t-m}}|P\left( \mathcal{A}\cap \mathcal{B}\right) $ $-$ $P\left( \mathcal{A%
}\right) P\left( \mathcal{B}\right) |$ $\rightarrow $ $0$ as $m$ $%
\rightarrow $ $\infty $. We say $L_{q}$-bounded $\{\epsilon _{t}\}$ is
stationary $L_{q}$-NED with size $\lambda $ $>$ $0$ on a mixing base $%
\{\upsilon _{t}\}$ when $||\epsilon _{t}$ $-$ $E[\epsilon _{t}|\mathfrak{V}%
_{t-m}^{t+m}]||_{q}$ $=$ $O(m^{-\lambda -\iota })$ for tiny $\iota $ $>$ $0$.%
\footnote{%
This definition of size is slightly different from the conventional one,
e.g. \citet[p. 262]{Davidson1994}. We use de Jong's (\citeyear{deJong1997}:
Definition 1) definition because we use his central limit theorem for NED
arrays.} If $\epsilon _{t}$ $=$ $\upsilon _{t}$ then $||\epsilon _{t}$ $-$ $%
E[\epsilon _{t}|\mathfrak{V}_{t-m}^{t+m}]||_{q}$ $=$ $0$, hence NED includes
mixing sequences, but it also includes non-mixing sequences since it covers
infinite lag functions of mixing sequences that need not be mixing. NED is
related to McLeish's (\citeyear{McLeish1975}) mixingale property. See %
\citet[Chapter 17]{Davidson1994} for historical references and deep results.

\begin{assumption}[data generating process]
\label{assum:dgp}$\medskip $\newline
$a$. $\{x_{t},y_{t}\}$ are stationary, ergodic, and $L_{2+\delta }$-bounded
for tiny $\delta $ $>$ $0.\medskip $\newline
$b$. $\epsilon _{t}$ is stationary, ergodic, $E[\epsilon _{t}]$ $=$ $0$, $%
L_{r}$-bounded, $r$ $>$ $4$, and $L_{4}$-NED with size $1/2$ on stationary $%
\alpha $-mixing $\{\upsilon _{t}\}$ with coefficients $\alpha
_{h}^{(\upsilon )}$ $=$ $O(h^{-r/(r-4)-\iota })$ for tiny $\iota $ $>$ $%
0.\medskip $\newline
$c$. The weights satisfy $\hat{\omega}_{n}(h)$ $>$ $0$ $a.s.$ and $\hat{%
\omega}_{n}(h)$ $=$ $\omega (h)+O_{p}(1/n^{\kappa })$ for some $\kappa $ $>$
$0$ and non-random $\omega (h)$ $\in $ $(0,\infty )$, for each $h$.
\end{assumption}

\begin{remark}
\normalfont The assumption $E[\epsilon _{t}]$ $=$ $0$ is typically imposed
in practice with inclusion of a regression model constant term. It is
important that the necessary steps for ensuring $E[\epsilon _{t}]$ $=$ $0$
are taken, since otherwise a white noise test may reject due merely to $%
E[\epsilon _{t}]$ $\neq $ $0$.
\end{remark}

\begin{remark}
\normalfont Ergodicity is not required in principle, but imposed to allow
easily for laws of large numbers on functions of $f(x_{t},\phi )$ and $%
\sigma _{t}^{2}(\theta )$ and their derivatives. Indeed, NED does not
necessarily carry over to arbitrary measurable transforms of an NED process.
$\alpha $-mixing, for example, implies ergodicity, it extends to measurable
transforms, and is a sub-class of NED. \cite{LobatoNankervisSavin2002}
impose a similar NED property. \cite{NankervisSavin2010}, who generalize the
white noise test of \cite{AndrewsPloberger1996}, allow for NED observed $%
y_{t}$, but mistakenly assume $y_{t}$ is only $L_{2}$-NED.\footnote{%
A Gaussian central limit theorem requires the \emph{product}, in our case $%
\epsilon _{t}\epsilon _{t-h}$, to be $L_{2}$-NED, which holds when $\epsilon
_{t}$ is $L_{p}$-bounded, $p$ $>$ $4$, and $L_{4}$-NED
\citep[Theorem
17.9]{Davidson1994}.}
\end{remark}

\begin{remark}
\normalfont The requirement $\hat{\omega}_{n}(h)$ $=$ $\omega
(h)+O_{p}(1/n^{\kappa })$ will hold under suitable moment conditions,
depending on how $\hat{\omega}_{n}(h)$ is constructed. If $\hat{\omega}%
_{n}(h)$ is a standard deviation for the sample correlation, for example, $%
\kappa $ $=$ $1/2$ can hold under the existence of higher moments and a
broad memory property like $\alpha $-mixing, even for some kernel estimators %
\citep[e.g.][]{AndrewsHAC91}.
\end{remark}

If $y_{t}$ $=$ $\epsilon _{t}$\ is known then a filter is not required and
Assumption \ref{assum:dgp} suffices for our main results. In this case, if $%
y_{t}$ is iid under $H_{0}$, then it only needs to be $L_{2}$-bounded.

The next assumption is required if a filter is used. Let $\boldsymbol{0}_{l}$%
\ be an $l$-dimensional zero vector. Define%
\begin{eqnarray}
&&G_{t}(\phi )\equiv \left[ \frac{\partial }{\partial \phi ^{\prime }}%
f(x_{t-1},\phi ),\boldsymbol{0}_{k_{\delta }}^{\prime }\right] ^{\prime }\in
\mathbb{R}^{k_{\theta }}\text{ \ and \ }s_{t}(\theta )\equiv \frac{1}{2}%
\frac{\partial }{\partial \theta }\ln \sigma _{t}^{2}(\theta )  \label{GsD}
\\
&&  \notag \\
&&\mathcal{D}(h)\equiv E\left[ \left( \epsilon _{t}s_{t}+\frac{G_{t}}{\sigma
_{t}}\right) \epsilon _{t-h}\right] +E\left[ \epsilon _{t}\left( \epsilon
_{t-h}s_{t-h}+\frac{G_{t-h}}{\sigma _{t-h}}\right) \right] \in \mathbb{R}%
^{k_{\theta }}.  \notag
\end{eqnarray}%
We do not require a filter for the above entities to make sense. If $%
y_{t}=\epsilon _{t}$, for example, then $G_{t}(\phi )$, $s_{t}(\theta )$ and
therefore $\mathcal{D}(h)$ are each just zero.

We require notation that makes use of estimating equations $m_{t}$ $\in $ $%
\mathbb{R}^{k_{m}}$ and a matrix $\mathcal{A}$ $\in $ $\mathbb{R}^{k_{\theta
}\times k_{m}}$ defined under Assumption \ref{assum:plug}.c. Define
\begin{eqnarray}
&&r_{t}(h)\equiv \frac{\epsilon _{t}\epsilon _{t-h}-E\left[ \epsilon
_{t}\epsilon _{t-h}\right] -\mathcal{D}(h)^{\prime }\mathcal{A}m_{t}}{E\left[
\epsilon _{t}^{2}\right] }\text{ and }\rho (h)\equiv \frac{E[\epsilon
_{t}\epsilon _{t-h}]}{E[\epsilon _{t}^{2}]}  \label{rz} \\
&&z_{t}(h)\equiv r_{t}(h)-\rho (h)r_{t}(0)=\frac{\epsilon _{t}\epsilon
_{t-h}-\rho (h)\epsilon _{t}^{2}-\left( \mathcal{D}(h)-\rho (h)\mathcal{D}%
(0)\right) ^{\prime }\mathcal{A}m_{t}}{E\left[ \epsilon _{t}^{2}\right] }.
\notag
\end{eqnarray}%
The process that arises in the key approximation is:
\begin{equation}
\mathcal{Z}_{n}(h)\equiv \frac{1}{\sqrt{n}}\sum_{t=1+h}^{n}z_{t}(h).
\label{Zn}
\end{equation}

\begin{assumption}[plug-in: response and identification]
\label{assum:plug} $\ \ \ \medskip $\newline
$a$. \emph{Level response.} $f$ $:$ $\mathbb{R}^{k_{x}}\times \Phi $ $%
\rightarrow $ $\mathbb{R}$, where $\Phi $ is a compact subset of $\mathbb{R}%
^{k_{\phi }}$, $k_{\phi }$ $\geq $ $0$; $f(x,\phi )$ is Borel measurable for
each $\phi $, and for each $x$ three times continuously differentiable,
where $(\partial /\partial \phi )^{j}f(x,\phi )$ is Borel measurable for
each $\phi $ and $j$ $=$ $1,2,3$; $E[\sup_{\phi \in \mathcal{N}_{\phi
_{0}}}|(\partial /\partial \phi )^{j}f(x_{t},\phi )|^{6}]$ $<$ $\infty $ for
$j$ $=$ $0,1,2,3$ and some compact set with positive measure $\mathcal{N}%
_{\phi _{0}}\subseteq \Phi $ containing $\phi _{0}$.\medskip \newline
$b$. \emph{Volatility.} $\sigma _{t}^{2}$ $:\Theta $ $\rightarrow $ $%
[0,\infty )$ where $\Theta $ $=$ $\Phi $ $\times $ $\Delta $ $\in $ $\mathbb{%
R}^{k_{\theta }}$, and $\Delta $\ is a compact subset of $\mathbb{R}%
^{k_{\delta }}$, $k_{\delta }$ $\geq $ $0$; $\sigma _{t}^{2}(\theta )$ is $%
\mathcal{F}_{t-1}$-measurable, continuous, and three times continuously
differentiable, where $(\partial /\partial \theta )^{j}\ln \sigma
_{t}^{2}(\theta )$ is Borel measurable for each $\theta $ and $j$ $=$ $1,2,3$%
; $\inf_{\theta \in \Theta }|\sigma _{t}^{2}(\theta )|$ $\geq $ $\iota $ $>$
$0$ $a.s.$ and $E[\sup_{\theta \in \mathcal{N}_{\theta _{0}}}|(\partial
/\partial \theta )^{j}\ln \sigma _{t}^{2}(\theta )|^{4}]$ $<$ $\infty $ for $%
j$ $=$ $0,1,2,3$ and some compact subset $\mathcal{N}_{\theta _{0}}\subseteq
\Theta $ containing $\theta _{0}$.\medskip \newline
$c$. \emph{Estimator}$.$ $\hat{\theta}_{n}$ $\in $ $\Theta $ for each $n$,
and for a unique interior point $\theta _{0}$ $\in $ $\Theta $ we have $%
\sqrt{n}(\hat{\theta}_{n}$ $-$ $\theta _{0})$ $=$ $\mathcal{A}%
n^{-1/2}\sum_{t=1}^{n}m_{t}(\theta _{0})$ $+$ $\mathcal{R}_{m}(n)$, where
the $k_{m}$ $\times $ $1$ stochastic remainder $\mathcal{R}_{m}(n)$ $=$ $%
O_{p}(n^{-\zeta })$\ for some $\zeta $ $>$ $0$, with $\mathcal{F}_{t}$%
-measurable estimating equations $m_{t}$ $=$ $[m_{i,t}]_{i=1}^{k_{m}}$ $:$ $%
\Theta $ $\rightarrow $ $\mathbb{R}^{k_{m}}$ for $k_{m}$ $\geq $ $k_{\theta }
$, and non-stochastic $\mathcal{A}$ $\in $ $\mathbb{R}^{k_{\theta }\times
k_{m}}$. Moreover, zero mean $m_{t}(\theta _{0})$ is stationary, ergodic, $%
L_{r/2}$-bounded and $L_{2}$-NED with size $1/2$ on $\{\upsilon _{t}\}$,
where $r$ $>$ $4$ and $\{\upsilon _{t}\}$\ appear in Assumption \ref%
{assum:dgp}.b.\medskip \newline
$d$. \emph{Finite dimensional variance}$.$ Let $\mathcal{L}$ $\in $ $\mathbb{%
N}$ be arbitrary, and let $\lambda $ $\equiv $ $[\lambda _{h}]_{h=1}^{%
\mathcal{L}}$ $\in $ $\mathbb{R}^{\mathcal{L}}$. Then\linebreak\ $\lim
\inf_{n\rightarrow \infty }\inf_{\lambda ^{\prime }\lambda =1}E[(\sum_{h=1}^{%
\mathcal{L}}\lambda _{h}\mathcal{Z}_{n}(h))^{2}]$ $>$ $0$.
\end{assumption}

\begin{remark}
\normalfont Smoothness (a) and (b) ensure a stochastic equicontinuity
property for uniform laws of large numbers. Non-differentiability can be
allowed provided certain other smoothness conditions involving, e.g.,
bracketing numbers apply \citep[see, e.g.,][]{PakesPollard1989,ArconesYu1994}%
.
\end{remark}

\begin{remark}
\normalfont$E[\sup_{\phi \in \mathcal{N}_{\phi _{0}}}|(\partial /\partial
\phi )^{j}f(x_{t},\phi )|^{4}]$ $<$ $\infty $ and $E[\sup_{\theta \in
\mathcal{N}_{\theta _{0}}}|(\partial /\partial \theta )^{j}\ln \sigma
_{t}^{2}(\theta )|^{4}]$ $<$ $\infty $ are used to prove a uniform law of
large numbers, where the former can imply higher moment bounds than in
Assumption \ref{assum:dgp}, depending on the response $f$. Fourth moments
are required due to a required residual cross-product expansion. $%
E[\sup_{\theta \in \mathcal{N}_{\theta _{0}}}|(\partial /\partial \theta
)^{j}\ln \sigma _{t}^{2}(\theta )|^{4}]$ $<$ $\infty $ holds for many linear
and nonlinear volatility models, e.g. GARCH, Quadratic GARCH, GJR-GARCH %
\citep{FZ04,FZ10}. The $6^{th}$ moment bound $E[\sup_{\phi \in \mathcal{N}%
_{\phi _{0}}}|(\partial /\partial \phi )^{j}f(x_{t},\phi )|^{6}]$ $<$ $%
\infty $ is used to determine the rate of convergence of the correlation
expansion approximation, which itself is used to bound the rate of increase
of $\mathcal{L}_{n}$ in Lemma \ref{lm:corr_expan}.
\end{remark}

\begin{remark}
\normalfont$\hat{\theta}_{n}$ under (c) asymptotically is a linear function
of some zero mean $\mathcal{F}_{t}$-measurable process $m_{t}(\theta _{0})$.
This includes M-estimators, GMM and (Generalized) Empirical Likelihood with
smooth or nonsmooth estimating equations, and estimators with non-smooth
criteria and asymptotic expansions like LAD and quantile regression.
Typically $m_{t}(\theta _{0})$ is a function of $u_{t}$ or $\epsilon _{t}$
and the gradients $(\partial /\partial \phi )f(x_{t},\phi _{0})$ and/or $%
(\partial /\partial \theta )\sigma _{t}^{2}(\theta _{0})$, in which case $%
E[m_{t}]$ $=$ $0$ represents an orthogonality condition that identifies $%
\theta _{0}$, even if $\epsilon _{t}$ is not white noise. The assumption
that $m_{t}$\ is NED in (c), in conjunction with Assumption \ref{assum:dgp},
implies linear combinations of $\epsilon _{t}\epsilon _{t-h}$ and $m_{t}$
are NED \citep[Theorem 17.8]{Davidson1994}, which promotes Gaussian finite
dimensional asymptotics for the residuals cross-product.
\end{remark}

\begin{remark}
\normalfont The approximation error in (c) $\sqrt{n}(\hat{\theta}_{n}$ $-$ $%
\theta _{0})$ $=$ $\mathcal{A}n^{-1/2}\sum_{t=1}^{n}m_{t}(\theta _{0})$ $+$ $%
O_{p}(n^{-\zeta })$ is of order $n^{-\zeta }$ for some $\zeta $ $>$ $0$. In
many cases $\zeta $ $=$ $1/2$ under suitable regularity conditions. This
allows us to describe the order of convergence for the remainder term in an
asymptotic expansion of the sample correlation, which we require when
deriving an upper bound on $\mathcal{L}_{n}$. All other technical arguments
only require $\sqrt{n}(\hat{\theta}_{n}$ $-$ $\theta _{0})$ $=$ $\mathcal{A}%
n^{-1/2}\sum_{t=1}^{n}m_{t}(\theta _{0})$ $+$ $o_{p}(1)$.
\end{remark}

\begin{remark}
\normalfont(d) is a standard nondegeneracy assumption for finite dimensional
asymptotics.
\end{remark}

The theory developed in this paper extends to a class of measurable
functions of $[\sqrt{n}\hat{\rho}_{n}(h)]_{h=1}^{\mathcal{L}_{n}}$.
Specifically:
\begin{equation}
\vartheta :\mathbb{R}^{\mathcal{L}}\rightarrow \lbrack 0,\infty )\text{ for
arbitrary }\mathcal{L}\in \mathbb{N},  \label{phi_map}
\end{equation}%
which satisfies the following: $\vartheta (x)$ is continuous; lower bound $%
\vartheta (a)$ $=$ $0$ \emph{if and only if} $a$ $=$ $0$; upper bound $%
\vartheta (a)$ $\leq $ $K\mathcal{LM}$ for some $K$ $>$ $0$ and any $a$ $=$ $%
[a_{h}]_{h=1}^{\mathcal{L}}$ such that $|a_{h}|$ $\leq $ $\mathcal{M}$ for
each $h$; divergence $\vartheta (a)$ $\rightarrow $ $\infty $ as $||a||$ $%
\rightarrow $ $\infty $; monotonicity $\vartheta (a_{\mathcal{L}_{1}})$ $%
\leq $ $\vartheta ([a_{\mathcal{L}_{1}}^{\prime },c_{\mathcal{L}_{2}-%
\mathcal{L}_{1}}^{\prime }]^{\prime })$ where $(a_{\mathcal{L}},c_{\mathcal{L%
}})$ $\in $ $\mathbb{R}^{\mathcal{L}}$, $\forall \mathcal{L}_{2}$ $\geq $ $%
\mathcal{L}_{1}$ and any $c_{\mathcal{L}_{2}-\mathcal{L}_{1}}$ $\in $ $%
\mathbb{R}^{\mathcal{L}_{2}-\mathcal{L}_{1}}$; and the triangle inequality $%
\vartheta (a$ $+$ $b)$ $\leq $ $\vartheta (a)$ $+$ $\vartheta (b)$ $\forall
a,b$ $\in $ $\mathbb{R}^{\mathcal{L}_{n}}$. Examples include the maximum $%
\vartheta (a)$ $=$ $\max_{1\leq h\leq \mathcal{L}}|a_{h}|$, and sums $%
\vartheta (a)$ $=$ $\sum_{h=1}^{\mathcal{L}}|a_{h}|$ and $\vartheta (a)$ $=$
$\sum_{h=1}^{\mathcal{L}}a_{h}^{2}$, where $a$ $=$ $[a_{h}]_{h=1}^{\mathcal{L%
}}$. The lower bound $\vartheta (a)$ $=$ $0$ \emph{if and only if} $a$ $=$ $%
0 $ ensures we omit cases where test power is not asymptotically one. As one
example, when $\tilde{\vartheta}(a)$ $=$ $\sum_{h=1}^{\mathcal{L}}a_{h}$ the
statistic $\tilde{\vartheta}([\sqrt{n}\hat{\omega}_{n}(h)\hat{\rho}%
_{n}(h)]_{h=1}^{\mathcal{L}_{n}})$ need not diverge under the alternative
because $\tilde{\vartheta}(a)$ $=$ $0$ is possible for $a$ $\neq $ $0$.

We do not show that $\vartheta $ depends on $\mathcal{L}$ to reduce
notation. The general test statistic is therefore:%
\begin{equation*}
\mathcal{\hat{T}}_{n}\equiv \vartheta \left( \left[ \sqrt{n}\hat{\omega}%
_{n}(h)\hat{\rho}_{n}(h)\right] _{h=1}^{\mathcal{L}_{n}}\right) .
\end{equation*}%
Both $\max_{1\leq h\leq \mathcal{L}_{n}}|\sqrt{n}\hat{\omega}_{n}(h)\hat{\rho%
}_{n}(h)|$ and a weighted portmanteau $n\sum_{h=1}^{\mathcal{L}_{n}}\hat{%
\omega}_{n}^{2}(h)\hat{\rho}_{n}^{2}(h)$ are covered. Note that the
normalization $\vartheta ([\sqrt{n}\hat{\omega}_{n}(h)\hat{\rho}%
_{n}(h)]_{h=1}^{\mathcal{L}_{n}})$ $=$ $(2\mathcal{L}_{n})^{-1/2}\sum_{h=1}^{%
\mathcal{L}_{n}}\hat{\omega}_{n}(h)\{n\hat{\rho}_{n}^{2}(h)$ $-$ $1\}$ and
similar normalized spectral density estimators used in
\citet[eq.
(3)]{Hong1996} and \cite{Hong2001}\ are not covered here because it violates
positivity $\vartheta $ $:$ $\mathbb{R}^{\mathcal{L}}$ $\rightarrow $ $%
[0,\infty )$, lower bound $\vartheta (a)$ $=$ $0$ \emph{if and only if} $a$ $%
=$ $0$, and monotonicity. The fix $\vartheta ([\sqrt{n}\hat{\omega}_{n}(h)%
\hat{\rho}_{n}(h)]_{hP=1}^{\mathcal{L}_{n}})$ $=$ $(2\mathcal{L}%
_{n})^{-1/2}|\sum_{h=1}^{\mathcal{L}_{n}}\hat{\omega}_{n}(h)\{n\hat{\rho}%
_{n}^{2}(h)$ $-$ $1\}|$ still violates $\vartheta (a)$ $=$ $0$ \emph{if and
only if} $a$ $=$ $0$, and monotonicity.

The following result establishes a key (non-Gaussian) approximation theory
for an increasing sequence of serial correlations. See Appendix \ref%
{app:proofs} for all proofs. Recall $\kappa $ $>$ $0$ in $\hat{\omega}%
_{n}(h) $ $=$ $\omega (h)+O_{p}(1/n^{\kappa })$ and $\zeta $ $>$ $0$ in $%
\sqrt{n}(\hat{\theta}_{n}$ $-$ $\theta _{0})$ $=$ $\mathcal{A}%
n^{-1/2}\sum_{t=1}^{n}m_{t}(\theta _{0})$ $+$ $O_{p}(n^{-\zeta })$, cf.
Assumptions \ref{assum:dgp}.c and \ref{assum:plug}.c. These will determine
an upper bound on $\mathcal{L}_{n}$ $\rightarrow $ $\infty $.

\begin{lemma}
\label{lm:corr_expan}Let Assumptions \ref{assum:dgp} and \ref{assum:plug}
hold. Then%
\begin{equation}
\mathcal{\tilde{X}}_{n}(h)\equiv \left\vert \sqrt{n}\hat{\omega}%
_{n}(h)\left\{ \hat{\rho}_{n}(h)-\rho (h)\right\} -\omega (h)\frac{1}{\sqrt{n%
}}\sum_{t=1+h}^{n}\left\{ r_{t}(h)-\rho (h)r_{t}(0)\right\} \right\vert
=O_{p}\left( 1/n^{\min \{\zeta ,\kappa ,1/2\}}\right) .  \label{expans_rate}
\end{equation}%
Moreover, for some non-unique monotonic sequence of positive integers $\{%
\mathcal{L}_{n}\}$, $\mathcal{L}_{n}$ $\rightarrow $ $\infty $ and $\mathcal{%
L}_{n}$ $=$ $o(n)$, we have: $|\vartheta (\sqrt{n}[\hat{\omega}_{n}(h)\{\hat{%
\rho}_{n}(h)$ $-$ $\rho (h)\}]_{h=1}^{\mathcal{L}_{n}})$ $-$ $\vartheta
([\omega (h)\mathcal{Z}_{n}(h)]_{h=1}^{\mathcal{L}_{n}})|$ $\leq $ $%
\vartheta ([\sqrt{n}\hat{\omega}_{n}(h)\{\hat{\rho}_{n}(h)$ $-$ $\rho (h)\}$
$-$ $\omega (h)\mathcal{Z}_{n}(h)]_{h=1}^{\mathcal{L}_{n}})$ $\overset{p}{%
\rightarrow }0$. Therefore, under the null hypothesis:
\begin{equation}
\left\vert \vartheta \left( \left[ \sqrt{n}\hat{\omega}_{n}(h)\hat{\rho}%
_{n}(h)\right] _{h=1}^{\mathcal{L}_{n}}\right) -\vartheta \left( \left[
\omega (h)\frac{1}{\sqrt{n}}\sum_{t=1+h}^{n}\left\{ \frac{\epsilon
_{t}\epsilon _{t-h}-\mathcal{D}(h)^{\prime }\mathcal{A}m_{t}}{E\left[
\epsilon _{t}^{2}\right] }\right\} \right] _{h=1}^{\mathcal{L}_{n}}\right)
\right\vert \overset{p}{\rightarrow }0.  \label{maxmax}
\end{equation}%
Finally, if $\vartheta (\cdot )$ is the maximum transform, and $(n^{\min
\{\zeta ,\kappa ,1/2\}}/\ln (n))\mathcal{\tilde{X}}_{n}(h)$ for all $h$\ is
uniformly integrable, then $\mathcal{L}_{n}$ $=$ $O(n^{\min \{\zeta ,\kappa
,1/2\}}/\ln (n))$ must be satisfied.
\end{lemma}

\begin{remark}
\label{rm: bounded_Ln}\normalfont The sequence $\{\mathcal{L}_{n}\}$ is not
unique because for any other $\{\mathcal{\mathring{L}}_{n}\}$, $\mathcal{%
\mathring{L}}_{n}$ $\rightarrow $ $\infty $ and $\lim \sup_{n\rightarrow
\infty }\{\mathcal{\mathring{L}}_{n}/\mathcal{L}_{n}\}$ $<$ $1$,
monotonicity $\vartheta (a_{k})$ $\leq $ $\vartheta ([a_{k}^{\prime
},c_{l-k}^{\prime }]^{\prime })$ $\forall a_{k}$ $\in $ $\mathbb{R}^{k}$ and
$\forall c_{l-k}$ $\in $ $\mathbb{R}^{l-k}$\ implies as $n$ $\rightarrow $ $%
\infty $:%
\begin{eqnarray}
&&\vartheta \left( \left[ \sqrt{n}\hat{\omega}_{n}(h)\{\hat{\rho}%
_{n}(h)-\rho (h)\}-\omega (h)\mathcal{Z}_{n}(h)\right] _{h=1}^{\mathcal{%
\mathring{L}}_{n}}\right)   \label{phiphi} \\
&&\text{ \ \ \ \ \ \ \ \ \ \ \ \ \ \ \ \ \ \ \ }\leq \vartheta \left( \left[
\sqrt{n}\hat{\omega}_{n}(h)\{\hat{\rho}_{n}(h)-\rho (h)\}-\omega (h)\mathcal{%
Z}_{n}(h)\right] _{h=1}^{\mathcal{L}_{n}}\right) \overset{p}{\rightarrow }0,
\notag
\end{eqnarray}%
hence $|\vartheta (\sqrt{n}[\hat{\omega}_{n}(h)\{\hat{\rho}_{n}(h)$ $-$ $%
\rho (h)\}]_{h=1}^{\mathcal{\mathring{L}}_{n}})$ $-$ $\vartheta (\left[
\omega (h)\mathcal{Z}_{n}(h)\right] _{h=1}^{\mathcal{\mathring{L}}_{n}})|$ $%
\overset{p}{\rightarrow }$ $0$. Indeed, by an identical argument trivially (%
\ref{phiphi}) applies for \emph{any} positive integer sequence $\{\mathcal{%
\mathring{L}}_{n}\}$ that satisfies $\lim \sup_{n\rightarrow \infty }\{%
\mathcal{\mathring{L}}_{n}/\mathcal{L}_{n}\}$ $<$ $1$, covering the case $%
\mathcal{\mathring{L}}_{n}$ $\rightarrow $ $(0,\infty )$. All subsequent
results therefore extend to this general case, but we do not highlight it
because it does not promote a consistent test.
\end{remark}

\begin{remark}
\normalfont An upper bound on $\mathcal{L}_{n}$ requires the mapping $%
\vartheta $ to be specified, so we work with the maximum. By Lemma \ref%
{lm:max_p} in Appendix \ref{app:proofs},\ if $(n^{\min \{\zeta ,\kappa
,1/2\}}/\ln (n))\mathcal{\tilde{X}}_{n}(h)$ is uniformly integrable, then
from standard arguments $\mathcal{L}_{n}$ $=$ $O(n^{\min \{\zeta ,\kappa
,1/2\}}/\ln (n))$ must hold. We do not tackle the case where uniform
integrability fails to hold. Additional technical conditions laid out in
\citet[Appendix
G]{HillMotegi_supp_mat} yield uniform integrability. In particular, we
require $||1/\hat{\gamma}_{n}(0)||_{p}$ $=$ $O(1)$ for some $p$ $>$ $1$,
which generally cannot be easily verified. Further, $E[\epsilon _{t}^{6}]$ $<
$ $\infty $, the plug-in $\sqrt{n}||\hat{\theta}_{n}$ $-$ $\theta _{0}||_{4}$
$=$ $O(1)$ and plug-in remainder\ $||n^{\lambda }\mathcal{R}_{m}(n)||_{q}$ $=
$ $O(1)$ for some $\lambda $ $>$ $0$ and $q$ $>$ $2$, and test statistic
weight $n^{\min \{\kappa ,\zeta ,1/2\}}||\hat{\omega}_{n}(h)-\omega (h)||_{r}
$ $=$ $O(1)$\ for some $r$ $>$ $2$. Conditions like $\sqrt{n}||\hat{\theta}%
_{n}$ $-$ $\theta _{0}||_{4}$ $=$ $O(1)$ generally require moment conditions
higher than $E[\epsilon _{t}^{6}]$ $<$ $\infty $: see
\citet[Appendix
G: Example 1]{HillMotegi_supp_mat}. These are relatively mild conditions and
hold for most of the data generating processes under the simulation study in
Section \ref{sec:sim}.\footnote{%
Some processes in the simulation study evidently fail to have higher
moments, but are used to demonstrate the sensitivity of the proposed test to
moment condition failure. See Section \ref{sec:sim}.} In those cases $\zeta $
$=$ $1/2$, and $\hat{\omega}_{n}(h)$ $=$ $\omega (h)$ $=$ $1$\ so that $%
\kappa $ $=$ $\infty $, hence $\mathcal{L}_{n}$ $=$ $O(\sqrt{n}/\ln (n))$.
\end{remark}

The proof of Lemma \ref{lm:corr_expan}\ relies on a new two-fold argument.
It is new because it cannot rely on Gaussian approximation theory for high
dimensional processes. First we prove $\mathcal{A}_{\mathcal{L},n}$ $\equiv $
$\vartheta ([\sqrt{n}\hat{\omega}_{n}(h)\{\hat{\rho}_{n}(h)$ $-$ $\rho (h)\}$
$-$ $\omega (h)\mathcal{Z}_{n}(h)]_{h=1}^{\mathcal{L}})$ $\overset{p}{%
\rightarrow }$ $0$ for each $\mathcal{L}$\ $\in $ $\mathbb{N}$. Using
standard weak convergence theory, this does not suffice to show $\mathcal{A}%
_{\mathcal{L}_{n},n}$ $\overset{p}{\rightarrow }$ $0$ for some $\mathcal{L}%
_{n}$ $\rightarrow $ $\infty $. This follows because weak convergence, in
the broad sense of \citet{HoffJorg1984,HoffJorg1991}, to a Gaussian limit
with a version that has uniformly bounded and uniformly continuous sample
paths, is equivalent to convergence in finite dimensional distributions, the
existence of a pseudo metric $d$ on $N$\ such that $(N,d)$ is a totally
bounded pseudo metric space, and a stochastic equicontinuity property based
on $d$ holds. If $d$ is the Euclidean distance, for example, then $(N,d)$ is
not totally bounded because $\mathbb{N}$\ is not compact. See %
\citet{Dudley1978,Dudley1984} and \citet[Chapters
9-10]{Pollard1990}. We take an approach different from Hoffman-Jorgensen's %
\citeyearpar{HoffJorg1984} notion of weak dependence. We prove that $%
\mathcal{A}_{\mathcal{L},n}$ $\overset{p}{\rightarrow }$ $0$ for each $%
\mathcal{L}$\ $\in $ $\mathbb{N}$ directly implies $\mathcal{A}_{\mathcal{L}%
_{n},n}$ $\overset{p}{\rightarrow }$ $0$ for some sequence of positive
integers $\{\mathcal{L}_{n}\}$ that satisfies $\mathcal{L}_{n}$ $\rightarrow
$ $\infty $ and $\mathcal{L}_{n}$ $=$ $o(n)$. See Lemmas \ref{lm:array_conv}-%
\ref{lm:max_dist} in Appendix \ref{app:proofs}. Thus, by sidestepping the %
\citet{HoffJorg1984,HoffJorg1991} view of weak dependence, which requires
more than convergence in finite dimensional distributions, we are able to
show that such convergence suffices. Our approach has deep roots in \cite%
{Ramsey1930} theory, based on its implications for monotone subsequences %
\citep[e.g.][]{BoehmeRosenfeld1974,Thomason1988,Myers2002} as applied to
Frech\'{e}t spaces \citep{BoehmeRosenfeld1974}.

The same array argument, coupled with extant central limit theory for NED
arrays, yields the following fundamental Gaussian approximation result for
the Lemma \ref{lm:corr_expan} approximation process $\{\mathcal{Z}_{n}(h)$ $%
: $ $1$ $\leq $ $h$ $\leq $ $\mathcal{L}_{n}\}$. Recall $\mathcal{Z}_{n}(h)$
$\equiv $ $1/\sqrt{n}\sum_{t=1+h}^{n}z_{t}(h)$ where $z_{t}(h)$ $\equiv $ $%
r_{t}(h)$ $-$ $\rho (h)r_{t}(0)$ and $r_{t}(h)$ $\equiv $ $\{\epsilon
_{t}\epsilon _{t-h}$ $-$ $E[\epsilon _{t}\epsilon _{t-h}]$ $-$ $\mathcal{D}%
(h)^{\prime }\mathcal{A}m_{t}\}/E[\epsilon _{t}^{2}]$.

\begin{lemma}
\label{lm:clt_max}Let Assumptions \ref{assum:dgp} and \ref{assum:plug} hold.
Let $\{\mathcal{Z}(h)$ $:$ $h$ $\in $ $\mathbb{N}\}$ be a zero mean Gaussian
process with variance $\lim_{n\rightarrow \infty
}1/n\sum_{s,t=1}^{n}E[z_{s}(h)z_{t}(h)]$ $<$ $\infty $, and covariance
function \linebreak $E[\mathcal{Z}(h)\mathcal{Z}(\tilde{h})]$ $=$ $%
\lim_{n\rightarrow \infty }1/n\sum_{s,t=1}^{n}E[z_{s}(h)z_{t}(\tilde{h})]$.
Then for some $\{\mathcal{Z}(h)$ $:$ $h$ $\in $ $\mathbb{N}\}$ and some
non-unique monotonic sequence of positive integers $\{\mathcal{L}_{n}\}$, $%
\mathcal{L}_{n}$ $\rightarrow $ $\infty $ and $\mathcal{L}_{n}$ $=$ $o(n)$: $%
\vartheta ([\omega (h)\mathcal{Z}_{n}(h)]_{h=1}^{\mathcal{L}_{n}})$ $\overset%
{d}{\rightarrow }$ $\vartheta ([\omega (h)\mathcal{Z}(h)]_{h=1}^{\infty }).$
\end{lemma}

\begin{remark}
\normalfont If an estimator $\hat{\theta}_{n}$ is not required then $%
\mathcal{D}(h)$ $=$ $0$ and the covariance function $E[\mathcal{Z}(h)%
\mathcal{Z}(\tilde{h})]$ reduces accordingly. If additionally $\epsilon _{t}$
is iid under the null then $E[\mathcal{Z}(h)\mathcal{Z}(\tilde{h})]$ $=$ $%
E[\epsilon _{t}^{2}\epsilon _{t-h}^{2}]/(E[\epsilon _{t}^{2}])^{2}$, which
equals $1$ if $h$ $\neq $ $0$, and otherwise $E[\epsilon
_{t}^{4}]/(E[\epsilon _{t}^{2}])^{2}$. If $\hat{\theta}_{n}$ is not required
then we can bypass our array convergence argument and use the Gaussian
approximation argument in \cite{ZhangWu2017}, under their moment contraction
assumptions.
\end{remark}

\begin{remark}
\label{rm:Ln:gaussian}\normalfont An upper bound on the rate $\mathcal{L}_{n}
$ $\rightarrow $ $\infty $\ can be provided in the maximum case under
various dependence settings. For example,
\citet[Appendix
B]{Chernozhukov_etal_manymom_2014} impose boundedness and a $\beta $-mixing
property, and \cite{ZhangWu2014}, and \cite{ZhangWu2017} work with a
functional dependence property. Under their conditions a limit theory that
supports our Lemma \ref{lm:corr_expan} expansion is evidently possible,
while the bound on $\mathcal{L}_{n}$ $\rightarrow $ $\infty $\ follows from
our Lemma \ref{lm:max_p} and results in
\citet[Appendix
G]{HillMotegi_supp_mat}. In that case, their Theorem 3.2 will apply, hence $%
\mathcal{L}_{n}\left( \ln \left( \mathcal{L}_{n}\right) \right) ^{3q/2}$ $=$
$o(n^{q/2-1+\iota })$ for some $\iota $ $>$ $0$, provided $E|z_{t}(h)|^{q}$ $%
<$ $\infty $ for all $h$ and some $q$ $\geq $ $4$.\footnote{%
See \citet[Theorem 3.2 and p. 1900]{ZhangWu2017}. They yield the optimal
bound $\mathcal{L}_{n}\left( \ln \left( \mathcal{L}_{n}\right) \right)
^{3q/2}$ $=$ $o(n^{q/2-1})$, up to a multiplicative logarithmic term that is
trumped by $n^{\iota }$ for any tiny $\iota $ $>$ $0$.} The latter moment
bound generally requires $\epsilon _{t}$ to be $L_{8}$-bounded. Put $q$ $=$ $%
8$ to yield $\mathcal{L}_{n}\left( \ln \left( \mathcal{L}_{n}\right) \right)
^{12}$ $=$ $o\left( n^{3+\iota }\right) $. Hence $\mathcal{L}_{n}$ $%
\rightarrow $ $\infty $ as fast as $Kn^{3-\iota }$ for tiny $\iota $ $>$ $0$%
\ is allowed. Since we require $\mathcal{L}_{n}$ $=$ $o(n)$ for sample
covariance consistency, the binding upper bound on $\mathcal{L}_{n}$ $%
\rightarrow $ $\infty $ comes from Lemma \ref{lm:corr_expan}, e.g. $\mathcal{%
L}_{n}$ $=$ $O(\sqrt{n}/\ln (n))$ under standard regularity conditions and $%
\hat{\omega}_{n}(h)$ $=$ $\omega (h)$ $=$ $1$. We leave for future study a
Gaussian approximation theory for high dimensional, heterogeneous and
possibly non-stationary NED processes.
\end{remark}

Combine Lemmas \ref{lm:corr_expan} and \ref{lm:clt_max} and invoke the
triangle inequality to yield the following main result.

\begin{theorem}
\label{th:max_corr_expan}Under Assumptions \ref{assum:dgp} and \ref%
{assum:plug}, $\vartheta ([\sqrt{n}\hat{\omega}_{n}(h)\{\hat{\rho}_{n}(h)$ $%
- $ $\rho (h)\}]_{h=1}^{\mathcal{L}_{n}})$ $\overset{d}{\rightarrow }$ $%
\vartheta ([\omega (h)\mathcal{Z}(h)]_{h=1}^{\infty })$ for some monotonic
sequence of positive integers $\{\mathcal{L}_{n}\}$ that is not unique, $%
\mathcal{L}_{n}$ $\rightarrow $ $\infty $ and $\mathcal{L}_{n}$ $=$ $o(n)$,
where $\{\mathcal{Z}(h)$ $:$ $h$ $\in $ $\mathbb{N}\}$ is a zero mean
Gaussian process with variance $\lim_{n\rightarrow \infty
}n^{-1}\sum_{s,t=1}^{n}E[z_{s}(h)z_{t}(h)]$ $<$ $\infty $, and covariance
function $\lim_{n\rightarrow \infty }n^{-1}\sum_{s,t=1}^{n}E[z_{s}(h)z_{t}(%
\tilde{h})]$. Therefore under the null hypothesis $\vartheta ([\sqrt{n}\hat{%
\omega}_{n}(h)\hat{\rho}_{n}(h)]_{h=1}^{\mathcal{L}_{n}})$ $\overset{d}{%
\rightarrow }$ $\vartheta ([\omega (h)\mathcal{Z}(h)]_{h=1}^{\infty })$,
where $\{\mathcal{Z}(h)$ $:$ $h$ $\in $ $\mathbb{N}\}$ is a zero mean
Gaussian process with variance $\lim_{n\rightarrow \infty
}n^{-1}\sum_{s,t=1}^{n}E[r_{s}(h)r_{t}(h)]$ $<$ $\infty $ and $r_{t}(h)$ $%
\equiv $ $\{\epsilon _{t}\epsilon _{t-h}-$ $\mathcal{D}(h)^{\prime }\mathcal{%
A}m_{t}\}/E[\epsilon _{t}^{2}]$. Moreover, if $\vartheta (\cdot )$ is the
maximum transform, and $(n^{\min \{\zeta ,\kappa ,1/2\}}/\ln (n))\mathcal{%
\tilde{X}}_{n}(h)$ for all $h$\ is uniformly integrable, where $\mathcal{%
\tilde{X}}_{n}(h)$ is defined in (\ref{expans_rate}), then $\mathcal{L}_{n}$
$=$ $O(n^{\min \{\zeta ,\kappa ,1/2\}}/\ln (n))$ must be satisfied.
\end{theorem}

We now have a fundamental result for the maximum weighted autocorrelation
under white noise.

\begin{corollary}
\label{cor:expand_null}Under Assumptions \ref{assum:dgp} and \ref{assum:plug}%
, $\max_{1\leq h\leq \mathcal{L}_{n}}|\sqrt{n}\hat{\omega}_{n}(h)\{\hat{\rho}%
_{n}(h)$ $-$ $\rho (h)\}|$ $\overset{d}{\rightarrow }$ $\max_{1\leq h\leq
\infty }|\omega (h)\mathcal{Z}(h)|$ for some monotonic sequence of positive
integers $\{\mathcal{L}_{n}\}$ that is not unique, $\mathcal{L}_{n}$ $%
\rightarrow $ $\infty $ and $\mathcal{L}_{n}$ $=$ $o(n)$, where $\{\mathcal{Z%
}(h)$ $:$ $h$ $\in $ $\mathbb{N}\}$ is defined in Theorem \ref%
{th:max_corr_expan}. Therefore, under the white noise null hypothesis $%
\max_{1\leq h\leq \mathcal{L}_{n}}|\sqrt{n}\hat{\omega}_{n}(h)\hat{\rho}%
_{n}(h)|$ $\overset{d}{\rightarrow }$ $\max_{1\leq h\leq \infty }|\omega (h)%
\mathcal{Z}(h)|$. Further, if $(n^{\min \{\zeta ,\kappa ,1/2\}}/\ln (n))%
\mathcal{\tilde{X}}_{n}(h)$ for all $h$\ is uniformly integrable, where $%
\mathcal{\tilde{X}}_{n}(h)$ is defined in (\ref{expans_rate}), then $%
\mathcal{L}_{n}$ $=$ $O(n^{\min \{\zeta ,\kappa ,1/2\}}/\ln (n))$ must be
satisfied.
\end{corollary}

\begin{remark}
\normalfont The conclusions of Theorem \ref{th:max_corr_expan} and Corollary %
\ref{cor:expand_null} do not require $\vartheta ([\sqrt{n}\hat{\omega}_{n}(h)%
\hat{\rho}_{n}(h)]_{h=1}^{\mathcal{L}_{n}})$ to have a well-defined limit
law under the null. This is decidedly different from the max-correlation
literature in which $\lim_{n\rightarrow \infty }\max_{1\leq h\leq \mathcal{L}%
_{n}}|\omega (h)\mathcal{Z}(h)|$ is characterized under suitable conditions
that ensure asymptotic independence $E[\mathcal{Z}(i)\mathcal{Z}(j)]$ $%
\rightarrow $ $0$ as $|i$ $-$ $j|$ $\rightarrow $ $0$. See, e.g., %
\citet[Chapter 6]{Leadbetter_etal1983} and \citet{Husler1986}. We do not
require asymptotic independence, nor therefore convergence in law.
\end{remark}

\subsection{Bootstrapped P-Value Test\label{sec:p_val}}

We work with Shao's (\citeyear{Shao2011_JoE}) dependent wild bootstrap.
Recall $m_{t}(\theta )$ are the estimating equations for $\hat{\theta}_{n}$,
let $\widehat{\mathcal{A}}_{n}$ be a consistent estimator of $\mathcal{A}$\
in Assumption \ref{assum:plug}.c, and define
\begin{equation}
\mathcal{\hat{D}}_{n}(h)\equiv \frac{1}{n}\sum_{t=h+1}^{n}\left\{ \left(
\epsilon _{t}(\hat{\theta}_{n})s_{t}(\hat{\theta}_{n})+\frac{G_{t}(\hat{%
\theta}_{n})}{\sigma _{t}(\hat{\theta}_{n})}\right) \epsilon _{t-h}(\hat{%
\theta}_{n})+\epsilon _{t}(\hat{\theta}_{n})\left( \epsilon _{t-h}(\hat{%
\theta}_{n})s_{t-h}(\hat{\theta}_{n})+\frac{G_{t-h}(\hat{\theta}_{n})}{%
\sigma _{t-h}(\hat{\theta}_{n})}\right) \right\} .  \label{D_hat}
\end{equation}%
We now operate on an approximation of $\epsilon _{t}(\hat{\theta}%
_{n})\epsilon _{t-h}(\hat{\theta}_{n})$ expanded around $\theta _{0}$ under $%
H_{0}$, cf. Lemma \ref{lm:corr_expan}:
\begin{equation*}
\widehat{\mathcal{E}}_{n,t,h}(\hat{\theta}_{n})\equiv \epsilon _{t}(\hat{%
\theta}_{n})\epsilon _{t-h}(\hat{\theta}_{n})-\mathcal{\hat{D}}%
_{n}(h)^{\prime }\widehat{\mathcal{A}}_{n}m_{t}(\hat{\theta}_{n}).
\end{equation*}

In practice $G_{t}(\theta )$ and $\sigma _{t}(\theta )$ are typically
unobserved and must be iteratively approximated based on initial conditions.
Examples include linear and nonlinear AR-GARCH models. In such cases $%
\mathcal{\hat{D}}_{n}(h)$ is infeasible. \cite{Meitz_Saikkonen_2011},
amongst others, lay out sufficient conditions for the QML estimator for a
large class of AR-GARCH models to be consistent and asymptotically normal,
including smoothness conditions similar to Assumption \ref{assum:plug} that
include Lipschitz properties imposed on $f(x_{t},\phi )$ and $\sigma
_{t}(\theta )$. In their setting, initial conditions vanish geometrically
fast and therefore do not play a role in asymptotics both for the QML
estimator, and for sample statistics like a feasible version of $\mathcal{%
\hat{D}}_{n}(h)$. See their Assumptions DGP, E, and C1-C3.

\subsection{Dependent Wild Bootstrap\label{sec:dwb}}

The wild bootstrap is proposed for iid and mds sequences %
\citep{Wu1986,Liu1988,Hansen1996}. \citet{Shao2010,Shao2011_JoE} generalizes
the idea to allow for dependent sequences. \cite{Shao2010} allows for
general dependence by using block-wise iid random draws as weights, with a
covariance function that equals a kernel function. His requirements rule out
a truncated kernel, but allow a Bartlett kernel amongst others. We follow
\cite{Shao2011_JoE} whose draws effectively have a truncated kernel
covariance function.

The algorithm is as follows. Set a block size $b_{n}$ such that $1\leq
b_{n}<n$, $b_{n}$ $\rightarrow $ $\infty $ and $b_{n}/n$ $\rightarrow $ $0$.
Denote the blocks by $\mathcal{B}_{s}=\{(s-1)b_{n}+1,\dots ,sb_{n}\}$ with $%
s=1,\dots ,n/b_{n}$. Assume for simplicity that the number of blocks $%
n/b_{n} $ is an integer. Generate iid random numbers $\{\xi _{1},\dots ,\xi
_{n/b_{n}}\}$ with $E[\xi _{i}]$ $=$ $0$, $E[\xi _{i}^{2}]$ $=$ $1$, and $%
E[\xi _{i}^{4}]$ $<$ $\infty $. Define an auxiliary variable $\varphi
_{t}=\xi _{s}$ if $t$ $\in $ $\mathcal{B}_{s}$. Compute $\mathcal{\hat{T}}%
_{n}^{(dw)}$ $\equiv $ $\vartheta ([\sqrt{n}\hat{\rho}_{n}^{(dw)}(h)]_{h=1}^{%
\mathcal{L}_{n}})$ from:%
\begin{equation}
\hat{\rho}_{n}^{(dw)}(h)\equiv \frac{1}{1/n\sum_{t=1}^{n}\epsilon _{t}^{2}(%
\hat{\theta}_{n})}\frac{1}{n}\sum_{t=1+h}^{n}\varphi _{t}\left\{ \widehat{%
\mathcal{E}}_{n,t,h}(\hat{\theta}_{n})-\frac{1}{n}\sum_{s=1+h}^{n}\widehat{%
\mathcal{E}}_{n,s,h}(\hat{\theta}_{n})\right\} .  \label{R_hat_dwb}
\end{equation}%
Repeat $M$ times, resulting in bootstrapped statistics $\{\mathcal{\hat{T}}%
_{n,i}^{(dw)}\}_{i=1}^{M}$, and an approximate p-value $\hat{p}_{n,M}^{(dw)}$
$\equiv $ $1/M\sum_{i=1}^{M}I(\mathcal{\hat{T}}_{n,i}^{(dw)}$ $\geq $ $%
\mathcal{\hat{T}}_{n})$. The test proposed rejects the null at nominal size $%
\alpha $ when $\hat{p}_{n,M}^{(dw)}$ $<$ $\alpha $. The wild bootstrap has
block size $b_{n}$ $=$ $1$ and no re-centering with $1/n\sum_{s=1+h}^{n}%
\widehat{\mathcal{E}}_{n,s,h}(\hat{\theta}_{n})$.

We use a sample version of the first order expansion variable $\epsilon
_{t}\epsilon _{t-h}$ $-$ $\mathcal{D}(h)^{\prime }\mathcal{A}m_{t}$ from
Lemma \ref{lm:corr_expan}. It is \textit{incorrect} to use just $\epsilon
_{t}(\hat{\theta}_{n})\epsilon _{t-h}(\hat{\theta}_{n})$, as with:
\begin{equation}
\hat{\rho}_{n}^{(dw)}(h)\equiv \frac{1}{1/n\sum_{t=1}^{n}\epsilon _{t}^{2}(%
\hat{\theta}_{n})}\frac{1}{n}\sum_{t=1+h}^{n}\varphi _{t}\left\{ \epsilon
_{t}(\hat{\theta}_{n})\epsilon _{t-h}(\hat{\theta}_{n})-\frac{1}{n}%
\sum_{s=1+h}^{n}\epsilon _{t}(\hat{\theta}_{n})\epsilon _{t-h}(\hat{\theta}%
_{n})\right\} .  \label{rho_no_ex}
\end{equation}%
This follows since $\varphi _{t}$ is mean zero and independent of the data,
hence $1/n\sum_{t=1+h}^{n}\varphi _{t}\epsilon _{t}(\hat{\theta}%
_{n})\epsilon _{t-h}(\hat{\theta}_{n})$ $=$ $1/n\sum_{t=1+h}^{n}\varphi
_{t}\epsilon _{t}\epsilon _{t-h}$ $+$ $o_{p}(1/\sqrt{n})$, yet $%
1/n\sum_{s=1+h}^{n}\epsilon _{t}(\hat{\theta}_{n})\epsilon _{t-h}(\hat{\theta%
}_{n})$ $=$ $E[\epsilon _{t}\epsilon _{t-h}]$ $+$ $O_{p}(1/\sqrt{n})$ by
standard first order arguments and $E[m_{t}]$ $=$ $0$. Hence, $\sqrt{n}\hat{%
\rho}_{n}^{(dw)}(h)$ from (\ref{rho_no_ex}) is equivalent to $1/\sqrt{n}%
\sum_{t=1+h}^{n}\varphi _{t}\epsilon _{t}\epsilon _{t-h}/E[\epsilon
_{t}^{2}] $ asymptotically with probability approaching one, which under the
null has the same asymptotic properties as $1/\sqrt{n}\sum_{t=1+h}^{n}%
\epsilon _{t}\epsilon _{t-h}/E[\epsilon _{t}^{2}]$. The latter is not
equivalent to the Lemma \ref{lm:corr_expan} first order expansion process $\{%
\mathcal{Z}_{n}(h)\}$ because asymptotic information from the estimator $%
\hat{\theta}_{n}$ has been scrubbed out by the bootstrap variable $\varphi
_{t}$. The bootstrapped $\hat{\rho}_{n}^{(dw)}(h)$ in (\ref{R_hat_dwb}),
however, contains the required information.

\cite{Shao2011_JoE} imposes Wu's (\citeyear{Wu2005}) moment contraction
property with an eighth moment, which we denote MC$_{8}$
\citep[see Appendix
B in][for details]{HillMotegi_supp_mat}. He then applies a Hilbert space
approach for weak convergence of a spectral density process $\{\hat{S}%
_{n}(\lambda )$ $:$ $\lambda $ $\in $ $[0,\pi ]\}$ to yield convergence for $%
\int_{0}^{\pi }\hat{S}_{n}^{2}(\lambda )d\lambda $.\footnote{%
See, e.g., \cite{PolitisRomano1994} for applications of weak convergence in
a Hilbert space to bootstrapped statistics.} Only observed data are
considered. There are several reasons why a different approach is required
here. First, $\hat{S}_{n}(\lambda )$ is a sum of all $\{\hat{\gamma}_{n}(h)$
$:$ $1$ $\leq $ $h$ $\leq $ $n$ $-$ $1\}$, and
\citet[proof of Theorem
3.1]{Shao2011_JoE} uses a variance of conditional variance bound for
probability convergence based on Chebyshev's inequality. This requires $%
E[\epsilon _{t}^{8}]$ $<$ $\infty $ and a complicated eighth order joint
cumulant series bound which is only known to hold when $\epsilon _{t}$ is
\textit{geometric} MC$_{8}$ \citep[see][]{ShaoWu2007}. Second, we only need
convergence in distribution of $\sqrt{n}\hat{\gamma}_{n}(h)$, coupled with a
new array convergence result, which are easier to handle than weak
convergence of $\{\hat{S}_{n}(\lambda )$ $:$ $\lambda $ $\in $ $[0,\pi ]\}$
on a Hilbert space. Third, in the Hilbert space approach the supremum is not
a continuous mapping from the space of square integrable (with respect to
Lebesgue measure) functions on $[0,\pi ]$. It is therefore not clear how, or
if, Shao's (\citeyear{Shao2011_JoE}: Theorem 3.1) proof applies to our
statistic.

In order to prove that the bootstrapped $\hat{\rho}_{n}^{(dw)}(h)$ has the
same finite dimensional limit distributions as $\hat{\rho}_{n}(h)$ under the
null, it is helpful to have the equations $m_{t}(\theta )$ in the Assumption %
\ref{assum:plug}.c expansion $\sqrt{n}(\hat{\theta}_{n}$ $-$ $\theta _{0})$ $%
=$ $\mathcal{A}n^{-1/2}\sum_{t=1}^{n}m_{t}(\theta _{0})$ $+$ $o_{p}(1)$ to
be a smooth parametric function for a required uniform law of large numbers.
As with response smoothness under Assumption \ref{assum:plug}.a,b, more
general smoothness properties are achievable at the expense of more intense
notation.\footnote{%
Nonsmoothness can be allowed provided certain bracketing or other smoothness
properties are applied like a Lipschitz condition or the
Vapnick-Chervonenkis class, which ensure a required stochastic
equicontinuity condition. See, e.g., \cite{Andrews1987}, \cite{ArconesYu1994}
and \cite{GaenssslerZiegler1994}.}

\begin{assumption2c'}
$\hat{\theta}_{n}$ $\in $ $\Theta $ for each $n$, and for a unique interior
point $\theta _{0}$ $\in $ $\Theta $ we have $\sqrt{n}(\hat{\theta}_{n}$ $-$
$\theta _{0})$ $=$ $\mathcal{A}n^{-1/2}\sum_{t=1}^{n}m_{t}(\theta _{0})$ $+$
$\mathcal{R}_{m}(n)$ where the $k_{m}$ $\times $ $1$ stochastic remainder $%
\mathcal{R}_{m}(n)$ $=$ $O_{p}(n^{-\zeta })$\ for some $\zeta $ $>$ $0$,
with $\mathcal{F}_{t}$-measurable estimating equations $m_{t}$ $=$ $%
[m_{i,t}]_{i=1}^{k_{m}}$ $:$ $\Theta $ $\rightarrow $ $\mathbb{R}^{k_{m}}$
for $k_{m}$ $\geq $ $k_{\theta }$; and non-stochastic $\mathcal{A}$ $\in $ $%
\mathbb{R}^{k_{\theta }\times k_{m}}$. $m_{t}(\theta )$ is twice
continuously differentiable, $(\partial /\partial \theta )^{j}m_{t}(\theta )$
is Borel measurable for each $\theta $ and $j$ $=$ $1,2$, and $%
E[\sup_{\theta \in \Theta }|(\partial /\partial \theta )^{i}m_{j,t}(\theta
)|]$ $<$ $\infty $ for each $i$ $=$ $0,1,2$ and $j$ $=$ $1,...,k_{m}$.
Moreover, zero mean $m_{t}$ is stationary, ergodic, $L_{r/2}$-bounded and $%
L_{2}$-NED with size $1/2$ on $\{\upsilon _{t}\}$, where $r$ $>$ $4$ and $%
\{\upsilon _{t}\}$\ appear in Assumption \ref{assum:dgp}.b.
\end{assumption2c'}

The bootstrapped p-value leads to a valid and consistent test. Note $\kappa $
$>$ $0$ in $\hat{\omega}_{n}(h)$ $=$ $\omega (h)+O_{p}(1/n^{\kappa })$ and $%
\zeta $ $>$ $0$ in $\sqrt{n}(\hat{\theta}_{n}$ $-$ $\theta _{0})$ $=$ $%
\mathcal{A}n^{-1/2}\sum_{t=1}^{n}m_{t}(\theta _{0})$ $+$ $O_{p}(n^{-\zeta })$%
, cf. Assumptions \ref{assum:dgp}.c and \ref{assum:plug}.c$^{\prime }$.

\begin{theorem}
\label{th:p_dep_wild_boot}Let Assumptions \ref{assum:dgp}, \ref{assum:plug}%
.a,b,c$^{\prime }$,d hold, and let the number of bootstrap samples $M$ $=$ $%
M_{n}$ $\rightarrow $ $\infty $. There exists a non-unique monotonic
sequence of maximum lags $\{\mathcal{L}_{n}\}$, $\mathcal{L}_{n}$ $%
\rightarrow $ $\infty $ and $\mathcal{L}_{n}$ $=$ $o(n)$, such that under $%
H_{0}$, $P(\hat{p}_{n,M}^{(dw)}$ $<$ $\alpha )$ $\rightarrow $ $\alpha $,
and if $H_{0}$ is false then $P(\hat{p}_{n,M}^{(dw)}$ $<$ $\alpha )$ $%
\rightarrow $ $1$. Further, $\mathcal{L}_{n}$ $=$ $O(n^{\min \{\zeta ,\kappa
,1/2\}}/\ln (n))$ must be satisfied.
\end{theorem}

\begin{remark}
\normalfont A similar theory applies to an approximate p-value computed by
wild bootstrap where $\varphi _{t}$ is iid $N(0,1)$, provided $\epsilon _{t}$
forms a mds under the null.
\end{remark}

\begin{remark}
\normalfont The test operates on $\sqrt{n}\hat{\rho}_{n}(h)$ and $\sqrt{n}%
\hat{\rho}_{n}^{(dw)}(h)$ and therefore achieves the parametric rate of
local asymptotic power against the sequence of alternatives: $H_{1}^{L}$ $:$
$\rho (h)$ $=$ $r(h)/\sqrt{n}$ for each $h$ where $r(h)$ are fixed
constants, $|r(h)|$ $\leq $ $\sqrt[\text{ }]{n}$. See
\citet[Appendix D,
especially Theorem D.1]{HillMotegi_supp_mat}.
\end{remark}

\begin{remark}
\normalfont The bound $\mathcal{L}_{n}$ $=$ $O(n^{\min \{\zeta ,\kappa
,1/2\}}/\ln (n))$ generally must hold. A uniform integrability condition is
not imposed here, as it is in Lemma \ref{lm:corr_expan}, cf. Lemma \ref%
{lm:max_p}.a, since the proof operates on conditional probabilities. The
latter are bounded and therefore uniformly integrable. Further, the
conditional probabilities imbed any given transform $\vartheta $ over lags $%
1,...,\mathcal{L}$. The maximum transform requirement from Lemma \ref%
{lm:max_p} for bounding $\mathcal{L}_{n}$ is merely applied to the
conditional probabilities themselves over lags $\mathcal{L}$ $\in $ $\{1,...,%
\mathcal{L}_{n}\}$. See the proof of Theorem \ref{th:p_dep_wild_boot}, cf.
Lemma \ref{lm:max_cor*}.b.
\end{remark}

\begin{remark}
\label{rm:Hong_local}\normalfont Hong's (\citeyear{Hong1996}:\ Section 2)
encompassing class of statistics, which includes \linebreak $(2\mathcal{L}%
_{n})^{-1/2}\sum_{h=1}^{\mathcal{L}_{n}}\hat{\omega}_{n}(h)\{n\hat{\rho}%
_{n}^{2}(h)$ $-$ $1\}$, does not achieve the parametric rate of convergence
due to the normalizing term $\mathcal{L}_{n}^{-1/2}$. The rate logically is $%
n^{1/2}/\mathcal{L}_{n}^{1/4}$ hence Hong's (\citeyear{Hong1996}) class of
statistics have non-trivial power against $n^{1/2}/\mathcal{L}_{n}^{1/4}$%
-local alternatives. We by-pass self-normalization by working solely in a
bootstrap framework based on finite dimensional asymptotics. As noted above,
our transform class $\vartheta $\ does not allow for self-normalized
statistics. Moreover, we do not need to know the limit distribution of $%
\vartheta ([\sqrt{n}\hat{\omega}_{n}(h)\hat{\rho}_{n}(h)]_{h=1}^{\mathcal{L}%
_{n}})$, nor even be guaranteed that it has one. Our approach eases the
burden of self-normalization with an increasing maximum lag: we retain $%
\sqrt{n}$-asymptotics and therefore $\sqrt{n}$-local power, even for Hong's (%
\citeyear{Hong1996}) (non-normalized) $\sum_{h=1}^{\mathcal{L}_{n}}\hat{%
\omega}_{n}(h)n\hat{\rho}_{n}^{2}(h)$.
\end{remark}

\section{Automatic Maximum Lag Selection\label{sec:lag_select}}

We approach lag selection from the perspective of the practitioner by
providing a data-driven, or automatic, lag selection method. Our method
closely follows \cite{EscancianoLobato2009}, whose work is motivated by the
automatic Neyman test proposed in \cite{InglotLedwina2006}. Let $\mathcal{L}%
_{n}^{\ast }$ denote the data-driven lag selected. Under $H_{0}$, Escanciano
and Lobato's (\citeyear{EscancianoLobato2009}) method leads to $P(\mathcal{L}%
_{n}^{\ast }$ $=$ $1)$ $\rightarrow $ $1$ because higher lags do not provide
useful information and incur a high penalty for their use (see below for
details). Contrary to their Q-test method, however, we allow $\mathcal{L}%
_{n} $ $\rightarrow $ $\infty $ and by using a bootstrap we do not need to
standardize the sample autocorrelations.

In order to ease notation, we only work with the max-correlation statistic
and weight $\hat{\omega}_{n}(h)$ $=$ $1$, but all subsequent results carry
over to the general transform $\vartheta $ and general $\hat{\omega}_{n}(h)$
$\overset{p}{\rightarrow }$ $\omega (h)$ $>$ $0$ with few additional proof
steps. Hence $\kappa $ $=$ $\infty $ in Assumption \ref{assum:dgp}.c.

The optimal lag $\mathcal{L}_{n}^{\ast }$ is chosen from a set $\{1,...,%
\mathcal{\bar{L}}_{n}\}$ for some pre-chosen upper-bound $\mathcal{\bar{L}}%
_{n}$ $\rightarrow $ $\infty $. In the case of the maximum and $\hat{\omega}%
_{n}(h)$ $=$ $\omega (h)$ $=$ $1$, we have from expansion Lemma \ref%
{lm:corr_expan} and dependent wild bootstrap Theorem \ref{th:p_dep_wild_boot}
that $\mathcal{\bar{L}}_{n}$ $=$ $O(n^{\min \{\zeta ,1/2\}}/\ln (n))$ must
hold, where $\zeta $ $>$ $0$ appears in the Assumption \ref{assum:plug}.c or %
\ref{assum:plug}.c$^{\prime }$ plug-in expansion $\sqrt{n}(\hat{\theta}_{n}$
$-$ $\theta _{0})$ $=$ $\mathcal{A}n^{-1/2}\sum_{t=1}^{n}m_{t}(\theta _{0})$
$+$ $O_{p}(n^{-\zeta })$. Under standard regularity conditions many plug-in
estimators will satisfy $\zeta $ $=$ $1/2$, hence $\mathcal{\bar{L}}_{n}$ $=$
$O(\sqrt{n}/\ln (n))$. We use the integer part of $\delta \sqrt{n}/(\ln (n))$
for certain $\delta $ $>$ $0$\ in our simulation study below. We only
consider sequences $\{\mathcal{L}_{n}\}$ that satisfy $\mathcal{L}_{n}/%
\mathcal{\bar{L}}_{n}\rightarrow \lbrack 0,K]$ for any finite $K$ $>$ $0$
and we assume the results of Section \ref{sec:max_corr} hold for any such $\{%
\mathcal{L}_{n}\}$. We save notation by fixing $K$ $=$ $1$.

We also need to allow for selection of \textit{any} positive integer
sequence $\{\mathcal{L}_{n}\}$ that satisfies $\mathcal{L}_{n}/\mathcal{\bar{%
L}}_{n}\rightarrow \lbrack 0,1]$, hence $\mathcal{L}_{n}$ $\rightarrow $ $%
(0,\infty ]$ is assumed such that $\mathcal{L}_{n}$ $\rightarrow $ $\mathcal{%
L}$, a finite positive integer, is possible. This is required because
Escanciano and Lobato's (\citeyear{EscancianoLobato2009}) method leads to $P(%
\mathcal{L}_{n}^{\ast }$ $=$ $1)$ $\rightarrow $ $1$ under $H_{0}$. See
Remark \ref{rm: bounded_Ln} for discussion of the validity of our main
results when $\mathcal{L}_{n}$ $\rightarrow $ $(0,\infty )$.

\cite{EscancianoLobato2009} work with a penalized Q-statistic, with a
penalty that is an increasing function of the number of included lags.
Similarly, define the \textit{penalized max-correlation} test statistic%
\begin{equation}
\mathcal{\hat{T}}_{n}^{\mathcal{P}}(\mathcal{L})\equiv \mathcal{\hat{T}}_{n}(%
\mathcal{L})-\mathcal{P}_{n}(\mathcal{L})\text{ where }\mathcal{\hat{T}}_{n}(%
\mathcal{L})\equiv \sqrt{n}\max_{1\leq h\leq \mathcal{L}}\left\vert \hat{\rho%
}_{n}(h)\right\vert  \label{Tp}
\end{equation}%
with penalty function $\mathcal{P}_{n}(\cdot )$:%
\begin{equation}
\mathcal{P}_{n}(\mathcal{L})=\left\{
\begin{array}{ll}
\sqrt{\mathcal{L}\ln n} & \text{if }\mathcal{\hat{T}}_{n}(\mathcal{L})\leq
\sqrt{q\ln n} \\
\sqrt{2\mathcal{L}} & \text{if }\mathcal{\hat{T}}_{n}(\mathcal{L})>\sqrt{%
q\ln n}%
\end{array}%
\right.  \label{Pn}
\end{equation}%
where $q$ is a fixed positive constant. A small value of $q$ leads to the
AIC penalty $\sqrt{2\mathcal{L}}$ being chosen with high probability, while
a large $q$ promotes selection of the BIC penalty. \cite%
{EscancianoLobato2009} use $q$ $=$ $2.4$, a choice motivated by their own
simulation evidence, and evidence from \cite{InglotLedwina2006}. \cite%
{InglotLedwina2006} develop an automatic Neyman test, and the portmanteau
test explored in \cite{EscancianoLobato2009} belongs to a class of smooth
tests proposed in \cite{Neyman1937}. Hence, it is not surprising that their $%
q$ values are similar. We find a slightly larger value $q$ $=$ $3$ leads to
strong results across null and alternative hypotheses for our test: see the
discussion in Section \ref{sec:sim_design}, and see Figure \ref%
{fig:size_power_maxcorr_automatic_lag}.

The chosen maximum lag $\mathcal{L}_{n}^{\ast }$ for each $n$\ is:%
\begin{equation}
\mathcal{L}_{n}^{\ast }=\min \left\{ \mathcal{L}_{n}:1\leq \mathcal{L}%
_{n}\leq \mathcal{\bar{L}}_{n}:\mathcal{\hat{T}}_{n}^{\mathcal{P}}(\mathcal{L%
}_{n})\geq \mathcal{\hat{T}}_{n}^{\mathcal{P}}(l)\text{ for each }l=1,...,%
\mathcal{\bar{L}}_{n}\right\} .  \label{Ln_*}
\end{equation}%
We chose $\{\mathcal{L}_{n}\}$ from those integer sequences satisfying $%
\mathcal{L}_{n}$ $\geq $ $1$ and $\mathcal{L}_{n}$ $\leq $ $\mathcal{\bar{L}}%
_{n}$ to ensure $\mathcal{L}_{n}/\mathcal{\bar{L}}_{n}$ $\rightarrow $ $%
[0,1] $ holds in practice, but in theory we may select \textit{any} $\{%
\mathcal{L}_{n}\}$ such that $\mathcal{L}_{n}$ $\geq $ $1$ and $\mathcal{L}%
_{n}/\mathcal{\bar{L}}_{n}$ $\rightarrow $ $[0,1]$. Notice $l$ may be a
function of $n$, e.g. $l$ $=$ $\mathcal{\bar{L}}_{n}$ $-$ $1$. The penalties
$(\sqrt{\mathcal{L}\ln n},\sqrt{2\mathcal{L}})$ are related to Escanciano
and Lobato's (\citeyear{EscancianoLobato2009}: p. 144) penalties $(\mathcal{L%
}\ln n,2\mathcal{L})$ for a fixed horizon Q-statistic. We need the square
root because the max-correlation operates on $\sqrt{n}\hat{\rho}_{n}(h)$
rather than $n\hat{\rho}_{n}^{2}(h)$. Contrary to \cite{EscancianoLobato2009}%
, however, our test statistic \textit{and} penalty are based on the
max-correlation, we allow for diverging sequences $\{\mathcal{L}_{n}\}$, and
we do not need to standardize the correlations because we use a bootstrap.%
\footnote{\citet[second remark following
Theorem 2]{EscancianoLobato2009} claim that a diverging maximum lag is
possible for their Q-test with an automatic lag, but an asymptotic theory is
not presented. Further, it is not obvious that their current fixed maximum
lag proof can extend to the unbounded maximum lag case. By their eq. (11)
they need $\sum_{l=1}^{\mathcal{\bar{L}}_{n}}P(\mathcal{L}_{n}$\textit{\ }$=$%
\textit{\ }$l)$\textit{\ }$\rightarrow $\textit{\ }$0$\textit{\ as }$n$%
\textit{\ }$\rightarrow $\textit{\ }$\infty $ hence $P(\mathcal{L}_{n}$%
\textit{\ }$=$\textit{\ }$l)$ $\rightarrow $ $0$ fast enough when $\mathcal{%
\bar{L}}_{n}$ $\rightarrow $ $\infty $, which may not hold under their
current assumptions.}

Define $\rho (\infty )$ $\equiv $ $\lim_{h\rightarrow \infty }\rho (h)$, and
$h^{\ast }$ $\equiv $ $\min \{h$ $:$ $h$ $=$ $\arg \max_{1\leq h\leq \infty
}|\rho (h)|\}$, the smallest lag at which the largest correlation in
magnitude occurs.

\begin{theorem}
\label{th:lag_select}Let Assumptions \ref{assum:dgp} and \ref{assum:plug}
hold. $a.$ Under $H_{0}$, if $(n^{\min \{\zeta ,\kappa ,1/2\}}/\ln (n))%
\mathcal{\tilde{X}}_{n}(h)$ for all $h$\ is uniformly integrable, where $%
\mathcal{\tilde{X}}_{n}(h)$ is defined in (\ref{expans_rate}), then $%
\mathcal{L}_{n}$ $=$ $O(n^{\min \{\zeta ,\kappa ,1/2\}}/\ln (n))$ must hold,
and $P(\mathcal{L}_{n}^{\ast }$ $=$ $1)$ $\rightarrow $ $1$. $b$. Under $%
H_{1}$, $\mathcal{L}_{n}^{\ast }$ $\overset{p}{\rightarrow }$ $h^{\ast }$
provided $\mathcal{\bar{L}}_{n}$ $=$ $o(n/\ln (n))$.
\end{theorem}

\begin{remark}
\normalfont We only require the more lenient $\mathcal{\bar{L}}_{n}$ $=$ $%
o(n/\ln (n))$ for the proof of $(b)$. The more stringent restriction $%
\mathcal{\bar{L}}_{n}$ $=$ $O(n^{\min \{\zeta ,1/2\}}/\ln (n))$ arises under
$(a)$ since there we must prove $P(\mathcal{P}_{n}(\mathcal{L}_{n})$ $=$ $%
\sqrt{\mathcal{L}_{n}\ln (n)})$ $\rightarrow $ $1$ by using Lemma \ref%
{lm:corr_expan}.
\end{remark}

\begin{remark}
\normalfont Under $H_{1}$ the optimal lag selected satisfies $\mathcal{L}%
_{n}^{\ast }$ $\overset{p}{\rightarrow }$ $h^{\ast }$. Notice $h^{\ast }$
may be any value in $\mathbb{N}$ because we allow the maximum lag under
consideration for finite samples to diverge $\mathcal{\bar{L}}_{n}$ $%
\rightarrow $ $\infty $. This ensures a consistent white noise test. The
reason $h^{\ast }$ is selected asymptotically is the penalized
max-correlation favors choosing lags that are \emph{at least} as large as
the most informative lag(s), the lag(s) at which the max-correlation takes
place. A nice advantage of the procedure is $\mathcal{L}_{n}^{\ast }$
converges to the \emph{smallest} of such \emph{most informative lags},
ensuring as $n$ $\rightarrow $ $\infty $ that the greatest number of data
points possible are used for computing that correlation magnitude. A
portmanteau statistic, however, sums over \emph{all} squared correlations
over a finite set of lags, hence its version is optimized at the largest
fixed lag $\bar{h}$\ under consideration, hence $P(\mathcal{L}_{n}^{\ast }$ $%
=$ $\bar{h})$ $\rightarrow $ $1$
\citep[see the proof of
Theorem 2 in][]{EscancianoLobato2009}.
\end{remark}

\begin{remark}
\normalfont The proof that $\mathcal{L}_{n}^{\ast }$ converges to $1$ in
probability under $H_{0}$ only exploits the null property $\sqrt{n}\hat{\rho}%
_{n}(h)$ $=$ $O_{p}(1)$. The latter \emph{also} holds under the $\sqrt{n}$%
-local alternative $\rho (h)$ $=$ $r(h)/\sqrt{n}$, since $\sqrt{n}\hat{\rho}%
_{n}(h)$ $=$ $\sqrt{n}(\hat{\rho}_{n}(h)$ $-$ $\rho (h))$ $+$ $r(h)$ $=$ $%
O_{p}(1)$. Thus, $P(\mathcal{L}_{n}^{\ast }$ $=$ $1)$ $\rightarrow $ $1$
under $H_{1}^{L}$ $:$ $\rho (h)$ $=$ $r(h)/\sqrt{n}$ as well. This means the
max-correlation test with our proposed automatic lag selection will have
trivial asymptotic local power against all directions from the null with $%
r(1)$ $=$ $0$.\footnote{%
Simply consider $r(h)$ $=$ $0$ $\forall h$ $\neq $ $2$ and $r(2)$ $\neq $ $0$%
. Since $\mathcal{L}_{n}^{\ast }$ $\overset{p}{\rightarrow }$ $1$, the above
local drift cannot be detected asymptotically (with probability greater than
the size of the test). We thank a referee for pointing this out.}
\end{remark}

\begin{remark}
\normalfont In our proof, e.g., under $H_{0}$, we show $\mathcal{\hat{T}}%
_{n}^{\mathcal{P}}(\mathcal{L}_{n})\geq \mathcal{\hat{T}}_{n}^{\mathcal{P}%
}(l)$ for each $l=1,...,\mathcal{\bar{L}}_{n}$ \emph{if and only if} $%
\mathcal{L}_{n}$ $\rightarrow $ $1$, while by definition $\mathcal{L}%
_{n}^{\ast }$\ is the least of all such sequences. We do this by inspecting
an equivalent expression for $P(\mathcal{\hat{T}}_{n}^{\mathcal{P}}(\mathcal{%
L}_{n})$ $\geq $ $\mathcal{\hat{T}}_{n}^{\mathcal{P}}(l))$ \emph{for each} $%
1 $ $\leq $ $l$ $\leq $ $\mathcal{\bar{L}}_{n}$ (equivalence holds
asymptotically with probability approaching one).
\citet[proof of Theorem
1]{EscancianoLobato2009}, by contrast, look at the joint probability that $%
\mathcal{L}_{n}^{\ast }$ $\neq $ $1$, hence they must show $\sum_{l=2}^{%
\mathcal{\bar{L}}}P(\mathcal{L}_{n}^{\ast }$ $=$ $l)$ $\rightarrow $ $0$
where $\mathcal{\bar{L}}$ is fixed and finite
\citep[see][eq.
(11)]{EscancianoLobato2009}. The joint probability argument does not
obviously transfer to the case where $\mathcal{\bar{L}}$ $\mathcal{=}$ $%
\mathcal{\bar{L}}_{n}$ $\rightarrow $ $\infty $ since $\sum_{l=2}^{\mathcal{%
\bar{L}}_{n}}P(\mathcal{L}_{n}^{\ast }$ $=$ $l)$ $\rightarrow $ $0$ requires
$P(\mathcal{L}_{n}^{\ast }$ $=$ $l)$ $\rightarrow $ $0$ sufficiently fast,
which need not hold under their assumptions.
\end{remark}

\section{Monte Carlo Experiments \label{sec:sim}}

We now perform a Monte Carlo experiment to gauge the merits of the
max-correlation test and automatic lag (labeled $\hat{\mathcal{T}}^{dw}(%
\mathcal{L}_{n}^{\ast })$). A main competitor studied here is a Shao's (%
\citeyear{Shao2011_JoE}) dependent wild bootstrap spectral Cram\'{e}r-von
Mises test (labeled $CvM^{dw}$). See Section \ref{sec:sim_design} for the
simulation design and Section \ref{sec:sim_results} for results. In the
supplemental material \citet[][Appendix H]{HillMotegi_supp_mat} we study
other tests, including the max-correlation with a pre-chosen non-random lag $%
\mathcal{L}_{n}$, the Ljung-Box test, Hong's (\citeyear{Hong1996}) test
based on a standardized periodogram, a CvM test with Zhu and Li's (%
\citeyear{ZhuLi2015}) block-wise random weighting bootstrap, and Andrews and
Ploberger's (\citeyear{AndrewsPloberger1996}) sup-LM test with the dependent
wild bootstrap. $CvM^{dw}$ is one of the strongest competitors in terms of
empirical size and power.

\subsection{Simulation Design \label{sec:sim_design}}

We consider a variety of data generating processes, filters, and estimation
methods. We first construct an error term $e_{t}$ that drives an observed
variable $y_{t}$. Let $\nu _{t}$ be iid $N(0,1)$. We consider iid $e_{t}=\nu
_{t}$; GARCH(1,1) $e_{t}=\nu _{t}w_{t}$ with random volatility process $%
w_{1}^{2}=1$ and $w_{t}^{2}=1+0.2e_{t-1}^{2}+0.5w_{t-1}^{2}$ for $t\geq 2$;
MA(2) $e_{t}=\nu _{t}+0.5\nu _{t-1}+0.25\nu _{t-2}$ for $t$ $\geq $ $3$,
with initial values $e_{1}$ $=$ $0$ and $e_{2}$ $=$ $\nu _{2}+0.5\nu _{1}$;
and AR(1) $e_{t}=0.7e_{t-1}+\nu _{t}$ for $t$ $\geq $ $2$ with initial $%
e_{1} $ $=$ $0$. Each error process is strictly stationary and ergodic.%
\footnote{%
Ergodicity follows since each error process is stationary $\alpha $-mixing.
See, e.g., \cite{KolmogorovRozanov1960} for processes with continuous
bounded spectral densities (e.g. stationary Gaussian AR, Gaussian MA(2));
\cite{Nelson1990} for GARCH process stationarity; and \cite{CarrascoChen2002}
for mixing properties of stationary GARCH processes.} We use each of the
four error terms in each of the following six scenarios.

\renewcommand\labelitemi{}

\begin{itemize}
\item \textbf{Scenario \#1: Simple} $y_{t}=e_{t}$; mean filter $\epsilon
_{t}=y_{t}-E[y_{t}]$; $\hat{\phi}_{n}=1/n\sum_{t=1}^{n}y_{t}$.

\item \textbf{Scenario \#2: Bilinear} $y_{t}=0.5e_{t-1}y_{t-2}+e_{t}$; mean
filter $\epsilon _{t}=y_{t}-E[y_{t}]$; $\hat{\phi}_{n}=1/n%
\sum_{t=1}^{n}y_{t} $.

\item \textbf{Scenario \#3: AR(2)} $y_{t}=0.3y_{t-1}-0.15y_{t-2}+e_{t}$;
AR(2) filter $\epsilon _{t}=y_{t}-\phi _{1}y_{t-1}-\phi _{2}y_{t-2}$; least
squares.

\item \textbf{Scenario \#4: AR(2)} $y_{t}=0.3y_{t-1}-0.15y_{t-2}+e_{t}$;
AR(1) filter $\epsilon _{t}=y_{t}-\phi _{1}y_{t-1}$; least squares.

\item \textbf{Scenario \#5: GARCH(1,1)} $y_{t}=\sigma _{t}e_{t}$, $\sigma
_{t}^{2}=1+0.2y_{t-1}^{2}+0.5\sigma _{t-1}^{2}$; no filter.

\item \textbf{Scenario \#6: GARCH(1,1)} $y_{t}=\sigma _{t}e_{t}$, $\sigma
_{t}^{2}=1+0.2y_{t-1}^{2}+0.5\sigma _{t-1}^{2}$; GARCH(1,1) filter $\epsilon
_{t}=y_{t}/\sigma _{t}$ with $\sigma _{t}^{2}=\omega +\alpha
y_{t-1}^{2}+\beta \sigma _{t-1}^{2}$; quasi-maximum likelihood.\footnote{%
QML is performed using the iterated process $\tilde{\sigma}_{1}^{2}(\theta )$
$=$ $\omega $ and $\tilde{\sigma}_{t}^{2}(\theta )$ $=$ $\omega $ $+$ $%
\alpha y_{t-1}^{2}$ $+$ $\beta \tilde{\sigma}_{t-1}^{2}(\theta )$ for $t$ $=$
$2,\dots ,n$. We impose $(\omega ,\alpha ,\beta )$ $>$ $0$ and $\alpha $ $+$
$\beta $ $\leq $ $1$ during estimation.}
\end{itemize}

In \#5 and \#6, $e_{t}$ is standardized so that $E[e_{t}^{2}]=1$.

The null is true for \#1, \#2, \#3, \#5 and \#6 when the error $e_{t}$ is
iid or GARCH. For \#4 the null is false for any error $e_{t}$ because a
misspecified AR(1) filter is used. This results in an AR(1) test variable $%
\epsilon _{t}$, with geometrically decaying autocorrelations when $e_{t}$ is
iid or GARCH.

In \#1--\#4, $y_{t}$ is stationary for each error. The GARCH(1,1) process in
\#5--\#6 is strong when $e_{t}$ is iid, and semi-strong when $e_{t}$ is
GARCH(1,1) since it is an adapted mds \citep{DrostNijman1993}, hence in
those cases $y_{t}$ is stationary \citep{Nelson1990,LeeHansen1994}. If $%
e_{t} $ is MA(2) or AR(1), then both $\{e_{t},y_{t}\}$ are serially
correlated. In the MA(2) error case, it can be verified that GARCH $y_{t}$
is stationary due to the finite feedback structure. It is unknown whether
GARCH $y_{t}$ with a GARCH or AR(1) error has a stationary solution {%
\citep[see,
e.g.,][]{DrostNijman1993,StraumannMikosch2006}.}

All of our chosen tests require a finite fourth moment on the test variable $%
\epsilon _{t}$, and in all cases $E[e_{t}^{4}]$ $<$ $\infty $. In \#1--\#4, $%
E[\epsilon _{t}^{4}]$ $<$ $\infty $ holds for each error type $e_{t}$. In
Scenario \#6 we test the standardized error $\epsilon _{t}$ $=$ $e_{t}$ $=$ $%
y_{t}/\sigma _{t}$ which has a finite fourth moment in all cases.

In Scenario \#3 we do not include a constant term in the filter in order to
reduce estimator dispersion, and because $E[y_{t}]$ $=$ $0$ is known to be
correct within this experiment. In practice a constant term would be
included to ensure $E[\epsilon _{t}]$ $=$ $0$.

In Scenario \#5, however, we test GARCH $\epsilon _{t}$ $=$ $y_{t}$ itself. $%
E[\epsilon _{t}^{4}]$ $<$ $\infty $ holds when $e_{t}$ is iid or MA(2), but
it is unknown in theory whether a fourth moment exists when $e_{t}$ is
GARCH(1,1) or AR(1).\footnote{%
As an experiment not presented in this paper, we simulated $J$ $=$ $10,000$
sample paths $\{y_{t}\}_{t=1}^{250}$ from GARCH $\epsilon _{t}$ $=$ $y_{t}$
with GARCH(1,1) or AR(1) error $e_{t}$. We inspected the median over all $J$
samples of the $4^{th}$ moment for subsamples $\{y_{t}\}_{t=1}^{T}$ with $T$
$=$ $50,...,250$. Denote this statistic as $k(T)$. $k(T)$ grows
exponentially in $T$, suggesting a $4^{th}$ moment does not exist for either
process.} Test results in the latter case should therefore be interpreted
with some caution.

We also consider three additional scenarios in which remote autocorrelations
are present. Only an iid error $e_{t}$ is used for the following processes
in order to focus on autocorrelation remoteness.

\begin{itemize}
\item \textbf{Scenario \#7: Remote MA(6)} $y_{t}=e_{t}+0.25e_{t-6}$; mean
filter $\epsilon _{t}=y_{t}-E[y_{t}]$; $\hat{\phi}_{n}=1/n%
\sum_{t=1}^{n}y_{t} $.

\item \textbf{Scenario \#8: Remote MA(12)} $y_{t} = e_{t} + 0.25 e_{t-12}$;
mean filter $\epsilon_{t} = y_{t} - E[y_{t}]$; $\hat{\phi}_{n} = 1/n
\sum_{t=1}^{n} y_{t}$.

\item \textbf{Scenario \#9: Remote MA(24)} $y_{t} = e_{t} + 0.25 e_{t-24}$;
mean filter $\epsilon_{t} = y_{t} - E[y_{t}]$; $\hat{\phi}_{n} = 1/n
\sum_{t=1}^{n} y_{t}$.
\end{itemize}

In Remote MA($q$), $\rho (h)\neq 0$ \textit{if and only if} $h=q$. Hence,
any test with a maximum lag less than $q$ should fail to detect serial
dependence.

We draw $J=1000$ Monte Carlo samples of size $n\in \{100,250,500,1000\}$. We
draw $2n$ observations and retain the last $n$ observations for analysis.
The rejection frequency of any test corresponds to its empirical size when
the tested variable $\epsilon _{t}$ is white noise, and empirical power when
$\epsilon _{t}$ is correlated. In Table \ref{table:sim_scenarios_summary} we
summarize the dependence property of $\epsilon _{t}$ under each scenario and
error $e_{t}$.

%\begin{table}[th]
%\caption{Dependence of Test Variable $\protect\epsilon _{t}$ under Each
%Scenario, Error $e_{t}$, and Filter}
%\label{table:sim_scenarios_summary}
%\begin{center}
%{\fontsize{11pt}{17pt}
%\selectfont{\small
%\begin{tabular}{r|c|c|c|c|c|c|c}
%& \multicolumn{7}{|c}{Scenario: Model and Filter} \\ \hline\hline
%\multicolumn{1}{c|}{} & \#1 & \#2 & \#3 & \#4 & \#5 & \#6 & \#7, \#8, \#9 \\
%& Simple & Bilinear & AR(2) & AR(2) & GARCH & GARCH & Remote MA \\
%\multicolumn{1}{r|}{$e_{t}$ \textbackslash \text{ filter}} & - & - & AR(2) filter & AR(1) filter & - &
%GARCH filter & - \\ \hline\hline
%\multicolumn{1}{l|}{iid} & \textbf{iid} & \textbf{wn} & \textbf{iid} & corr & \textbf{mds} & \textbf{iid} & remote corr
%\\
%\multicolumn{1}{l|}{GARCH} & \textbf{mds} & \textbf{wn} & \textbf{mds} & corr & \textbf{mds} & \textbf{mds} & not considered \\
%\multicolumn{1}{l|}{MA(2)} & corr & corr & corr & corr & corr & corr & not
%considered \\
%\multicolumn{1}{l|}{AR(1)} & corr & corr & corr & corr & corr & corr & not
%considered \\ \hline\hline
%\end{tabular}}}
%\end{center}
%\par
%{\fontsize{10pt}{14pt} \selectfont wn $=$ non-mds white noise. corr $=$
%autocorrelated. remote corr $=$ autocorrelation is present at a remote lag.
%\textbf{bold} text is used to highlight when the null is true.}
%\end{table}

\begin{table}[th]
\caption{Dependence of Test Variable $\protect\epsilon _{t}$ under Each
Scenario and Error $e_{t}$}
\label{table:sim_scenarios_summary}
\begin{center}
{\fontsize{10pt}{16pt} \selectfont
\begin{tabular}{r|c|c|c|c|c|c|c}
\hline
Scenario & \#1 & \#2 & \#3 & \#4 & \#5 & \#6 & \#7, \#8, \#9 \\ \hline
DGP & Simple & Bilinear & AR(2) & AR(2) & GARCH & GARCH & Remote MA \\
Filter & Mean & Mean & AR(2) & AR(1) & None & GARCH & Mean \\ \hline
iid $e_{t}$ & \textbf{iid} & \textbf{wn} & \textbf{iid} & corr & \textbf{mds}
& \textbf{iid} & remote corr \\
GARCH $e_{t}$ & \textbf{mds} & \textbf{wn} & \textbf{mds} & corr & \textbf{%
mds} & \textbf{mds} & not considered \\
MA(2) $e_{t}$ & corr & corr & corr & corr & corr & corr & not considered \\
AR(1) $e_{t}$ & corr & corr & corr & corr & corr & corr & not considered \\
\hline
\end{tabular}%
}
\end{center}
\par
{\fontsize{10pt}{14pt} \selectfont wn $=$ non-mds white noise. corr $=$
autocorrelated. remote corr $=$ autocorrelation is present at a remote lag.
\textbf{Bold} text is used to highlight when the null is true.}
\end{table}

Our proposed test is the max-correlation test with the dependent wild
bootstrap and automatic lag, $\hat{\mathcal{T}}^{dw}(\mathcal{L}_{n}^{\ast })
$. The test statistic is $\mathcal{\hat{T}}_{n}(\mathcal{L}_{n}^{\ast })$ $%
\equiv $ $\sqrt{n}\max_{1\leq h\leq \mathcal{L}_{n}^{\ast }}|\hat{\omega}%
_{n}(h)\hat{\rho}_{n}(h)|$ with weight $\hat{\omega}_{n}(h)$ $=$ $1$.%
\footnote{%
Other plausible weights include an inverted standard deviation based on a
HAC estimator, and/or the \cite{LjungBox1978} weights. In the present paper,
we demonstrate that the uniform weight leads to accurate size and high
power. In simulations not reported here we also find that an inverted
standard deviation, either parametric (when known) or nonparametric, is
suboptimal due to the added sampling error.} We compute the bootstrapped
statistic $\mathcal{\hat{T}}_{n,i}^{(dw)}(\mathcal{L}_{n,i}^{\ast })$ $%
\equiv $ $\sqrt{n}\max_{1\leq h\leq \mathcal{L}_{n,i}^{\ast }}|\hat{\rho}%
_{n,i}^{(dw)}(h)|$ for each bootstrap sample $i\in \{1,\dots ,M\}$ with $%
M=500$. $\hat{\rho}_{n,i}^{(dw)}(h)$ is computed via (\ref{R_hat_dwb}) based
on the Lemma \ref{lm:corr_expan} correlation expansion, which correctly
accounts for the first order (asymptotic) impact of the $i^{th}$ sample's
plug-in $\hat{\theta}_{n,i}$. Note that $\mathcal{L}_{n,i}^{\ast }$ is the
automatic lag for the $i^{th}$ bootstrap sample specifically. The dependent
wild bootstrap requires a choice of block size $b_{n}$. \cite{Shao2011_JoE}
uses $b_{n}$ $=$ $b\sqrt{n}$ with $b$ $\in $ $\{.5,1,2\}$, leading to
qualitatively similar results. We therefore use the middle value $b=1$.%
\footnote{%
We compared $b_{n}=b\sqrt{n}$ across $b$ $\in $ $\{.5,1,2\}$ in unreported
simulations and found there is little difference in test performance.} The
approximate p-value is computed as $\hat{p}_{n,M}^{(dw)}=1/M\sum_{i=1}^{M}I(%
\mathcal{\hat{T}}_{n,i}^{(dw)}(\mathcal{L}_{n,i}^{\ast })\geq \hat{\mathcal{T%
}}_{n}(\mathcal{L}_{n}^{\ast }))$.

The automatic lag selection requires a choice set $\{1,..,\mathcal{\bar{L}}%
_{n}\}$ with maximum possible lag length $\mathcal{\bar{L}}_{n}$, and the
tuning parameter $q$ (cf. (\ref{Pn}) and (\ref{Ln_*})). Let $[z]$ denote the
integer part of $z$. We set $\mathcal{\bar{L}}_{n}=[\delta \sqrt{n}/(\ln n)]$
with $\delta =10$ so that $\bar{\mathcal{L}}_{n}\in \{21,28,35,45\}$ for $%
n\in \{100,250,500,1000\}$, respectively. In the present simulation design, $%
\mathcal{\bar{L}}_{n}$ satisfies the Lemma \ref{lm:corr_expan} and Theorem %
\ref{th:lag_select}\ requirement $\mathcal{\bar{L}}_{n}$ $=$ $O(\sqrt{n}/\ln
(n))$ for all processes except possibly when the test variable is GARCH $%
y_{t}$ with GARCH or AR(1) error $e_{t}$. Similar and larger values lead to
qualitatively similar results.\footnote{%
In experiments not reported here we also used $\mathcal{\bar{L}}_{n}$ $=$ $%
[\delta n/(\ln (n))^{c}]$ for various $c$ and $\delta $ and found
essentially the same results. Thus, the value $\mathcal{\bar{L}}_{n}=[\delta
\sqrt{n}/(\ln n)]$ is not essential to test performance, but does satisfy
Lemma \ref{lm:corr_expan} and Theorem \ref{th:lag_select} for most processes
under study.}

In order to choose a plausible value of $q$ for the penalty function in (\ref%
{Pn}), we perform a preliminary simulation study that computes empirical
size and size-adjusted power for the max-correlation test with $\hat{%
\mathcal{T}}_{n}(\mathcal{L}_{n}^{\ast })$\ across $q\in \{1.50,1.75,\dots
,4.50\}$. We consider two cases in order to highlight empirical size and
power properties. In Case 1, size is computed under Scenario \#1 with an iid
error; and size-adjusted power is computed under \#4 with an iid error. In
Case 2, size is computed under \#5 with an iid error; and size-adjusted
power is computed under \#5 with MA(2) error. For each case, sample size is $%
n\in \{100,500\}$; nominal size is $\alpha =0.05$; $J=1000$ Monte Carlo
samples and $M=500$ bootstrap samples are generated. See Figure \ref%
{fig:size_power_maxcorr_automatic_lag} for results. Variation of empirical
size and size-adjusted power for the test based on $\hat{\mathcal{T}}^{dw}(%
\mathcal{L}_{n}^{\ast })$ across the values of $q$ is fairly small in each
experiment, implying that a choice of $q$ should not have a critical impact
on the test performance. For each case and sample size, we obtain relatively
accurate size and high power around $q$ $=$ $3$, hence $q$ $=$ $3$ is used.

We also perform the dependent wild bootstrap Cram\'{e}r-von Mises test in
\cite{Shao2011_JoE}, $CvM^{dw}$. This test is based on the sample spectral
distribution function $F_{n}(\lambda )$ $\equiv $ $\int_{0}^{\lambda
}I_{n}(\omega )d\omega $ with periodogram $I_{n}(\omega )$ $\equiv $ $(2\pi
)^{-1}\sum_{h=1-n}^{n-1}\hat{\gamma}_{n}(h)e^{-h\omega }$. Define:
\begin{equation*}
S_{n}(\lambda )\equiv \sqrt{n}(F_{n}(\lambda )-\hat{\gamma}_{n}(0)\psi
_{0}(\lambda ))=\sum_{h=1}^{n-1}\sqrt{n}\hat{\gamma}_{n}(h)\psi _{h}(\lambda
),
\end{equation*}%
where $\psi _{h}(\lambda )$ $=$ $(h\pi )^{-1}\sin (h\lambda )$ if $h$ $\neq $
$0$,\ else $\psi _{h}(\lambda )$ $=$ $\lambda (2\pi )^{-1}$. The CvM test
statistic is $\mathcal{C}_{n}$ $=$ $\int_{0}^{\pi }S_{n}^{2}(\lambda
)d\lambda $, which has a non-standard limit distribution under the null.%
\footnote{%
In practice we use a numerical integral based on the midpoint approximation
with an increment of $.01$.} We then use Shao's (\citeyear{Shao2011_JoE},
Section 3) dependent wild bootstrap based on the Lemma \ref{lm:corr_expan}
correlation expansion to compute an approximate p-value. Note that all $%
\mathcal{L}_{n}=n-1$ lags are used by construction. \cite{Shao2011_JoE} does
not consider the use of a filter, but we apply the test to all scenarios for
the sake of comparison, and use the correlation expansion to control for a
filter when used.

\subsection{Simulation Results \label{sec:sim_results}}

\subsubsection{Automatic Lag}

We first check the performance of the automatic lag selection itself. Recall
that by Theorem \ref{th:lag_select} $\mathcal{L}_{n}^{\ast }$ $\overset{p}{%
\rightarrow }$ $1$ under $H_{0}$, and under $H_{1}$ $\mathcal{L}_{n}^{\ast}$
$\overset{p}{\rightarrow }$ $h^{\ast }$, the smallest lag at which the
largest correlation occurs. Under Scenarios \#1-\#6 when the error $e_{t}$
is iid or GARCH the null is false only for \#4. In the latter case, the test
variable $\epsilon _{t}$ is AR(1) hence $h^{\ast }$ $=$ $1$.

In Table \ref{table:median_automatic_lag} we report the median of optimal
lags $\{\mathcal{L}_{n}^{\ast (1)},\dots ,\mathcal{L}_{n}^{\ast (J)}\}$ for
each scenario, where $\mathcal{L}_{n}^{\ast (j)}$ is the $j^{th}$ sample's
optimal lag. We also report the smallest lag at which the largest
correlation occurs, $h^{\ast }$. In most cases we compute $h^{\ast }$
analytically. In a few cases an analytic solution is not feasible so we use
a large sample simulation. We generate $50000$ samples of size $n$ $=$ $%
50000 $, and the autocorrelations for $\epsilon _{t}$ for each sample. We
then report the median computed $h^{\ast }$ across all samples.

In \#1--\#6, when $H_{0}$ is true or autocorrelations exist at small lags,
the median of $\mathcal{L}_{n}^{\ast (j)}$ is 1 or 2. This (nearly) matches
the predictions of Theorem \ref{th:lag_select} and the reported $h^{\ast }$
in most cases. In just two cases, (i) bilinear with GARCH error and (ii)
GARCH with GARCH error and without a filter, the reported $h^{\ast }$\ is $4$%
. This is higher than the optimally selected lag (1 or 2). These are the
only cases where the median of $\mathcal{L}_{n}^{\ast (j)}$ deviates by more
than 0 or 1 from $h^{\ast }$. In both of these cases the process is highly
volatile, possibly causing the aberrant deviation of $\mathcal{L}_{n}^{\ast
(j)}$ from $h^{\ast }$, and the low empirical size of the max-correlation
test. As suggested in Section \ref{sec:sim_design}, either of these
processes may fail the required moment conditions for the underlying theory
surrounding $\mathcal{L}_{n}^{\ast }$.

In \#7--\#9, where autocorrelations exist at remote lags, the median of $%
\mathcal{L}_{n}^{\ast (j)}$ pinpoints those lags given a large enough sample
size. Under Remote MA(12), for example, the median is 1 for $n\leq 250$ but
exactly 12 for $n\geq 500$.

\subsubsection{Empirical Size \label{sec:sim_size}}

We now report rejection frequencies associated with nominal size $\alpha \in
\{.01,.05,.10\}$. See Table \ref%
{table:main_paper_rf_automatic_scenario123456} for $\hat{\mathcal{T}}^{dw}(%
\mathcal{L}_{n}^{\ast })$ under \#1--\#6; see Table \ref%
{table:main_paper_rf_CvM_scenario123456} for $CvM^{dw}$ under \#1--\#6; and
see Table \ref{table:main_paper_rf_scenario789} for both tests under
\#7--\#9.

We begin with Scenario \#1 (simple), $n=100$, and iid error. The empirical
size with respect to nominal sizes $\alpha \in \{.010,.050,.100\}$ is $%
\{.017,.068,.128\}$ for $\hat{\mathcal{T}}^{dw}(\mathcal{L}_{n}^{\ast })$
and $\{.023,.081,.138\}$ for $CvM^{dw}$, hence $\hat{\mathcal{T}}^{dw}(%
\mathcal{L}_{n}^{\ast })$ has reasonably sharp size that is sharper than $%
CvM^{dw}$. A similar implication holds for \#2 (bilinear), $n=100$, and iid
error, where the empirical size is $\{.008,.047,.090\}$ for $\hat{\mathcal{T}%
}^{dw}(\mathcal{L}_{n}^{\ast })$ and $\{.018,.076,.149\}$ for $CvM^{dw}$. In
general, the empirical size of the test based on $\hat{\mathcal{T}}^{dw}(%
\mathcal{L}_{n}^{\ast })$ is at least as good as (and often better than)
size associated with $CvM^{dw}$.

The reason why $\hat{\mathcal{T}}^{dw}(\mathcal{L}_{n}^{\ast })$ achieves
fairly sharp size in most cases is that, as confirmed in Table \ref%
{table:median_automatic_lag}, $\mathcal{L}_{n}^{\ast }$ is sufficiently
close to $1$ in most samples under $H_{0}$. That feature cuts redundant lags
and improves the size of the test. In fact, we find in the supplemental
material \citet[][Appendix H]{HillMotegi_supp_mat} that $\hat{\mathcal{T}}%
^{dw}(\mathcal{L}_{n}^{\ast })$ achieves the sharpest size among a variety
of tests.\footnote{%
In Scenario \#2 (bilinear) with a GARCH error, the max-correlation test is
undersized, even in large samples $n=1000$. The primary cause is the
bilinear process combined with a GARCH error results in extreme volatility,
which undermines the efficacy of the bootstrap. The test is even more
undersized under Scenario \#5 (GARCH) with a GARCH error. The CvM test is
also undersized for Scenario \#2 with a GARCH error. It is, however, less
affected than the max-correlation test in Scenario \#5 with a GARCH error.
Weighting the correlations for a max-correlation test might alleviate the
under-rejection, for example using weights equal to the inverted standard
errors. The least volatile correlations in this case are given the greatest
weight. We leave that possibility for a future project.} $CvM^{dw}$ uses all
$\mathcal{L}_{n}=n-1$ lags, but the greatest weight is assigned to small
lags by construction. Hence $CvM^{dw}$ leads to have fairly accurate size in
most cases, although generally the max-correlation test dominates.

\subsubsection{Empirical Power \label{sec:sim_power}}

In \#1--\#6, the relative performance of $\hat{\mathcal{T}}^{dw}(\mathcal{L}%
_{n}^{\ast })$ and $CvM^{dw}$ under $H_{1}$ varies across cases. The former
is more powerful than the latter in some cases, but not in other cases. In
general, there is not a drastic gap between the two tests. See \#2, $n=1000$%
, and AR(1) error, for example. The empirical power with respect to $\alpha
\in \{.010,.050,.100\}$ is $\{.723,.823,.864\}$ for $\hat{\mathcal{T}}^{dw}(%
\mathcal{L}_{n}^{\ast })$ and $\{.474,.697,.810\}$ for $CvM^{dw}$. But in
\#3, with $n=1000$, and an AR(1) error, power is $\{.599,.847,.922\}$ for $%
\hat{\mathcal{T}}^{dw}(\mathcal{L}_{n}^{\ast })$ and $\{.688,.876,.923\}$
for $CvM^{dw}$.

In \#7--\#9, however, $\hat{\mathcal{T}}^{dw}(\mathcal{L}_{n}^{\ast })$
dominates $CvM^{dw}$ completely (see Table \ref%
{table:main_paper_rf_scenario789}). $\hat{\mathcal{T}}^{dw}(\mathcal{L}%
_{n}^{\ast })$ successfully detects remote autocorrelations given a large
enough sample size, while $CvM^{dw}$ fails to detect them for any $n$. The
power of $\hat{\mathcal{T}}^{dw}(\mathcal{L}_{n}^{\ast })$ under \#8 (Remote
MA(12)), for instance, is $\{.013,.067,.117\}$ for $n=100$, $%
\{.024,.134,.244\}$ for $n=250$, $\{.371,.673,.770\}$ for $n=500$, and $%
\{.983,.997,.997\}$ for $n=1000$. Logically power increases as $n$ grows.
The reason that $\hat{\mathcal{T}}^{dw}(\mathcal{L}_{n}^{\ast })$\ detects
remote autocorrelations is confirmed in Table \ref%
{table:median_automatic_lag} (cf. Theorem \ref{th:lag_select}.b): $\mathcal{L%
}_{n}^{\ast }$ converges to $h^{\ast }=12$ when $n\geq 500$ under \#8. The
power of $CvM^{dw}$, by contrast, is $\{.034,.110,.179\}$ for $n=100$, $%
\{.025,.087,.155\}$ for $n=250$, $\{.026,.092,.161\}$ for $n=500$, and $%
\{.017,.083,.166\}$ for $n=1000$. $CvM^{dw}$ has (almost) no power against
the remote autocorrelation even when $n=1000$. In fact, we find in %
\citet[][Appendix H]{HillMotegi_supp_mat} that $\hat{\mathcal{T}}^{dw}(%
\mathcal{L}_{n}^{\ast })$ is the only test that has power against remote
autocorrelations among a variety of tests which have decent size.

The reason why $CvM^{dw}$ fails to capture remote autocorrelations is that
it incorporates \textit{all} available sample correlations, while assigning
the greatest weight to small lags. That feature delivers sharp size and high
power against adjacent correlations like Scenarios \#1--\#6, but critically
low power against remote correlations like Scenarios \#7--\#9.

The (non-weighted) max-correlation, by contrast, operates on the most
informative serial correlation over a range of lags $\{1,...,\mathcal{L}%
_{n}^{\ast }\}$. The optimal maximum lag selected $\mathcal{L}_{n}^{\ast }$
asymptotically hones in on the most informative lag range: the range that
includes the smallest lag at which the greatest correlation in magnitude
occurs. Thus, in large samples in particular, $\hat{\mathcal{T}}^{dw}(%
\mathcal{L}_{n}^{\ast })$ delivers the single most informative serial
correlation for test purposes, as opposed to a weighted sum of all, and
therefore potentially less useful, correlations. That feature itself
generally delivers accurate size (or under-rejections in some cases) and
competitive power for Scenarios \#1-\#6, and dominant power against remote
correlations.

In some cases against adjacent correlations power is not dominant when a
large pre-chosen non-random $\mathcal{L}_{n}$ is used
\citep[see][Appendix
H]{HillMotegi_supp_mat}, but such a shortcoming is alleviated by using our
proposed automatic lag $\mathcal{L}_{n}^{\ast }$. The combined
max-correlation with automatic lag and bootstrapped p-value leads to a
dominant test over all when size and power are considered, in comparison to
a variety of tests.

\section{Conclusion\label{sec:conclude}}

We present a bootstrap max-correlation test of the white noise hypothesis
for regression model residuals. The maximum correlation over an increasing
lag length has a long history in the statistics literature, but only in
terms of characterizing its limit distribution using extreme value theory
and only for observed data. We apply a bootstrap method to a first order
correlation expansion in order to account for the impact of a plug-in $\hat{%
\theta}_{n}$ used to compute model residuals. We prove that Shao's (%
\citeyear{Shao2011_JoE}) dependent wild bootstrap yields a valid test in a
more general environment than \cite{Shao2011_JoE} or \cite{XiaoWu2014}
considered. Our approach does not require showing that the original and
bootstrapped max-correlation test statistics have the same limit properties
under the null, allowing us to bypass the extreme value theory approach
altogether. We also extend Escanciano and Lobato's (%
\citeyear{EscancianoLobato2009}) automatic lag selection to our setting with
an (asymptotically) unbounded lag set. We prove that the automatic lag
converges in probability to one under the null, and the smallest lag at
which the largest correlation in magnitude occurs under the alternative. In
both cases, the procedure hones in on the most informative lag, offering the
greatest number of data points for analysis, for the given hypothesis.

Simulation experiments show that our test with the automatic lag generally
out-performs a variety of other tests. It achieves sharper empirical size in
most cases than other tests since the automatic lag $\mathcal{L}_{n}^{\ast }$
is sufficiently close to 1 under the null hypothesis. When there exist
serial correlations at small lags, the max-correlation test and some strong
competitors such as the Cram\'{e}r-von Mises test with the dependent wild
bootstrap lead to roughly comparable empirical power. When there exist
correlations only at remote lags, the max-correlation test has (potentially)
high power while the Cram\'{e}r-von Mises test has nearly trivial power for
any sample size due to its weighting structure. Other tests also have
comparatively lower power. This striking difference stems from the fact that
the automatic lag $\mathcal{L}_{n}^{\ast }$ pinpoints the relevant remote
lag, while other tests by construction incorporate many lags into a test
statistic (the CvM test gives the greatest weight to low lags, making it
useless against remote lags).

\setcounter{equation}{0} \renewcommand{\theequation}{{\thesection}.%
\arabic{equation}} \setcounter{remark}{0} \renewcommand{\theremark}{%
\Alph{section}.\arabic{remark}}

\appendix

\section{Appendix: Proofs\label{app:proofs}}

We assume all random variables exist on a complete measure space such that
majorants and integrals over uncountable families of measurable functions
are measurable, and probabilities where applicable are outer probability
measures. See Pollard's (\citeyear{Pollard1984}: Appendix C) \textit{%
permissibility} criteria, and see Dudley's (\citeyear{Dudley1984}: p. 101)
\textit{admissible Suslin} property.

We use the following variance bound for NED sequences repeatedly. If $w_{t}$%
\ is zero mean, $L_{p}$-bounded for some $p$ $>$ $2$, and $L_{2}$-NED with
size $1/2$, on an $\alpha $-mixing base with decay $O(h^{-p/(p-2)-\iota })$,
then by Theorem 17.5 in \cite{Davidson1994} and Theorem 1.6 in \cite%
{McLeish1975}:%
\begin{equation}
E\left[ \left( 1/\sqrt{n}\sum\nolimits_{t=1}^{n}w_{t}\right) ^{2}\right]
=O(1).  \label{NED_var}
\end{equation}

The following results are key steps toward sidestepping extreme value theory
and Gaussian approximations when working with the maximum. The first result
expands on a result in \citet[Lemma 1]{BoehmeRosenfeld1974} for first
countable topological spaces. The latter is intimately linked to array
convergence implications of theory developed in \cite{Ramsey1930}, cf. \cite%
{BoehmeRosenfeld1974}, \cite{Thomason1988} and \cite{Myers2002}. Recall that
any metric space is a first countable topological space
\citep[see,
e.g.,][p. 131]{Lipschitz1965}.

\begin{lemma}
\label{lm:array_conv}Assume the array $\{\mathcal{A}_{k,n}$ $:$ $1$ $\leq $ $%
k$ $\leq $ $\mathcal{I}_{n}\}_{n\geq 1}$ lies in a first countable
topological space, where $\{\mathcal{I}_{n}\}_{n\geq 1}$ is a sequence of
positive integers, $\mathcal{I}_{n}$ $\rightarrow $ $\infty $ as $n$ $%
\rightarrow $ $\infty $. Let $\lim_{n\rightarrow \infty }\mathcal{A}_{k,n}$ $%
=$ $a_{k}$ for each fixed $k$, and $\lim_{k\rightarrow \infty }a_{k}$ $=$ $a$%
. Then $\lim_{n\rightarrow \infty }\mathcal{A}_{\mathcal{L}_{n},n}$ $=$ $a$
for some non-unique sequence of positive integers $\{\mathcal{L}_{n}\}$,
where $\mathcal{L}_{n}$ $\leq $ $\mathcal{I}_{n}$, and $\mathcal{L}_{n}$ $%
\rightarrow $ $\infty $. If $\mathcal{I}_{n}$ $=$ $n$ then $\mathcal{L}_{n}$
$\leq $ $n$. Moreover, $\lim_{n\rightarrow \infty }\mathcal{A}_{\mathcal{%
\tilde{L}}_{n},n}$ $=$ $a$ for any other monotonic sequence of positive
integers $\{\mathcal{\tilde{L}}_{n}\}_{n=1}^{\infty }$, $\mathcal{\tilde{L}}%
_{n}$ $\rightarrow $ $\infty $ that satisfies $\mathcal{\tilde{L}}_{n}/%
\mathcal{L}_{n}$ $\rightarrow $ $0$, hence $\mathcal{L}_{n}$ $=$ $o(n)$ can
always be assured.
\end{lemma}

\noindent \textbf{Proof.}\qquad In view of $\lim_{n\rightarrow \infty }%
\mathcal{A}_{k,n}$ $=$ $a_{k}$ for each fixed $k$, and $\lim_{k\rightarrow
\infty }a_{k}$ $=$ $a$, there always exists a monotonic sequence of positive
integers $\{\mathcal{L}_{n}\}_{n=1}^{\infty }$, $\mathcal{L}_{n}$ $%
\rightarrow $ $\infty $ as $n$ $\rightarrow $ $\infty $, satisfying%
\begin{equation}
\left\vert \mathcal{A}_{\mathcal{L}_{n},n}-a_{\mathcal{L}_{n}}\right\vert
\leq \frac{1}{\mathcal{L}_{n}}.  \label{ALaL}
\end{equation}%
Similarly, for any $\{\mathcal{L}_{n}\}_{n=1}^{\infty }$ that satisfies (\ref%
{ALaL}), any other monotonic sequence of positive integers $\{\mathcal{%
\tilde{L}}_{n}\}_{n=1}^{\infty }$, $\mathcal{\tilde{L}}_{n}$ $\rightarrow $ $%
\infty $ and $\mathcal{\tilde{L}}_{n}/\mathcal{L}_{n}$ $\rightarrow $ $0$
also satisfies (\ref{ALaL}). Hence $\{\mathcal{L}_{n}\}_{n=1}^{\infty }$ is
not unique, and $\mathcal{L}_{n}$ $=$ $o(n)$ is always feasible.

Inequality (\ref{ALaL}) yields $a_{\mathcal{L}_{n}}$ $-$ $1/\mathcal{L}_{n}$
$\leq $ $\mathcal{A}_{\mathcal{L}_{n},n}$ $\leq $ $a_{\mathcal{L}_{n}}$ $+$ $%
1/\mathcal{L}_{n}$ for each $n$. Then, by the definition of a limit:%
\begin{equation*}
\lim_{n\rightarrow \infty }a_{\mathcal{L}_{n}}-\frac{1}{\mathcal{L}_{n}}\leq
\lim_{n\rightarrow \infty }\mathcal{A}_{\mathcal{L}_{n},n}\leq
\lim_{n\rightarrow \infty }a_{\mathcal{L}_{n}}+\frac{1}{\mathcal{L}_{n}}.
\end{equation*}%
The Sandwich theorem, $\mathcal{L}_{n}$ $\rightarrow $ $\infty $ and
therefore $\lim_{n\rightarrow \infty }a_{\mathcal{L}_{n}}$ $=$ $a$, now
yields:%
\begin{equation*}
\lim_{n\rightarrow \infty }\mathcal{A}_{\mathcal{L}_{n},n}=\lim_{n%
\rightarrow \infty }a_{\mathcal{L}_{n}}=a.
\end{equation*}%
See also \citet[Lemma 1]{BoehmeRosenfeld1974}. This proves $%
\lim_{n\rightarrow \infty }\mathcal{A}_{\mathcal{L}_{n},n}$ $=$ $a$ for some
non-unique monotonic sequence of positive integers $\{\mathcal{L}%
_{n}\}_{n=1}^{\infty }$, where $\mathcal{L}_{n}$ $=$ $o(n)$ is always
feasible. By construction $\mathcal{L}_{n}$ $\leq $ $\mathcal{I}_{n}$ must
be satisfied, hence if $\mathcal{I}_{n}$ $=$ $n$ then $\mathcal{L}_{n}$ $%
\leq $ $n$. $\mathcal{QED}$.\bigskip

The next result uses Lemma \ref{lm:array_conv} as the basis for deriving
\textit{in probability} convergence of a function of an increasing set of
random variables. This result forms the basis for the proof of the
non-Gaussian correlation expansion Lemma \ref{lm:corr_expan}.

Recall the continuous $\vartheta $ $:$ $\mathbb{R}^{\mathcal{L}_{n}}$ $%
\rightarrow $ $[0,\infty )$ satisfies: lower bound $\vartheta (a)$ $=$ $0$
\emph{if and only if} $a$ $=$ $0$; upper bound $\vartheta (a)$ $\leq $ $K%
\mathcal{LM}$ for some $K$ $>$ $0$ and any $a$ $=$ $[a_{h}]_{h=1}^{\mathcal{L%
}}$ such that $|a_{h}|$ $\leq $ $\mathcal{M}$ for each $h$; divergence $%
\vartheta (a)$ $\rightarrow $ $\infty $ as $||a||$ $\rightarrow $ $\infty $;
monotonicity $\vartheta (a_{\mathcal{L}_{1}})$ $\leq $ $\vartheta ([a_{%
\mathcal{L}_{1}}^{\prime },c_{\mathcal{L}_{2}-\mathcal{L}_{1}}^{\prime
}]^{\prime })$ where $(a_{\mathcal{L}},c_{\mathcal{L}})$ $\in $ $\mathbb{R}^{%
\mathcal{L}}$, $\forall \mathcal{L}_{2}$ $\geq $ $\mathcal{L}_{1}$ and any $%
c_{\mathcal{L}_{2}-\mathcal{L}_{1}}$ $\in $ $\mathbb{R}^{\mathcal{L}_{2}-%
\mathcal{L}_{1}}$; and the triangle inequality $\vartheta (a$ $+$ $b)$ $\leq
$ $\vartheta (a)$ $+$ $\vartheta (b)$ $\forall a,b$ $\in $ $\mathbb{R}^{%
\mathcal{L}_{n}}$.

\begin{lemma}
\label{lm:max_p}Let $\{\mathcal{X}_{n}(i),\mathcal{Y}_{n}(i)$ $:$ $i$ $\in $
$\mathbb{N}\}_{n\geq 1}$ be arrays of random variables.\medskip \newline
$a.$ If $\mathcal{X}_{n}(i)$ $\overset{p}{\rightarrow }$ $0$ as $n$ $%
\rightarrow $ $\infty $\ for each $i$ $\in $ $\mathbb{N}$, then $\vartheta ([%
\mathcal{X}_{n}(i)]_{i=1}^{\mathcal{L}_{n}})$ $\overset{p}{\rightarrow }$ $0$
for some non-unique sequence of positive integers $\{\mathcal{L}_{n}\}$ with
$\mathcal{L}_{n}$ $\rightarrow $ $\infty $. Moreover $\mathcal{L}_{n}$ $=$ $%
o(n)$ can always be assured.

Further, let $\vartheta $ be the maximum transform: $\vartheta ([\mathcal{X}%
_{n}(i)]_{i=1}^{\mathcal{L}_{n}})$ $=$ $\max_{1\leq i\leq \mathcal{L}_{n}}|%
\mathcal{X}_{n}(i)|$. If for some non-stochastic $g$ $:$ $\mathbb{N}$\ $%
\rightarrow $ $[0,\infty )$ with $g(n)$ $\rightarrow $ $\infty $ as $n$ $%
\rightarrow $ $\infty $, $g(n)\mathcal{X}_{n}(i)$ $\overset{p}{\rightarrow }$
$0$ and $g(n)\mathcal{X}_{n}(i)$\ is uniformly integrable for each $i$, then
$\mathcal{L}_{n}$ $=$ $O(g(n))$ must hold.\medskip \newline
$b.$ If $\mathcal{X}_{n}(i)$ $-$ $\mathcal{Y}_{n}(i)$ $\overset{p}{%
\rightarrow }$ $0$ as $n$ $\rightarrow $ $\infty $\ for each $i$ $\in $ $%
\mathbb{N}$, then for some non-unique sequence of positive integers $\{%
\mathcal{L}_{n}\}$ with $\mathcal{L}_{n}$ $\rightarrow $ $\infty $: $%
|\vartheta ([\mathcal{X}_{n}(i)]_{i=1}^{\mathcal{L}_{n}})$ $-$ $\vartheta ([%
\mathcal{Y}_{n}(i)]_{i=1}^{\mathcal{L}_{n}})|$ $\leq $ $|\vartheta ([%
\mathcal{X}_{n}(i)$ $-$ $\mathcal{Y}_{n}(i)]_{i=1}^{\mathcal{L}_{n}})|$ $%
\overset{p}{\rightarrow }$ $0$. Moreover $\mathcal{L}_{n}$ $=$ $o(n)$ can
always be assured.

Further, let $\vartheta $ be the maximum transform: $\vartheta ([\mathcal{X}%
_{n}(i)]_{i=1}^{\mathcal{L}_{n}})$ $=$ $\max_{1\leq i\leq \mathcal{L}_{n}}|%
\mathcal{X}_{n}(i)|$. If for some non-stochastic $g$ $:$ $\mathbb{N}$\ $%
\rightarrow $ $[0,\infty )$ with $g(n)$ $\rightarrow $ $\infty $ as $n$ $%
\rightarrow $ $\infty $, $g(n)(\mathcal{X}_{n}(i)$ $-$ $\mathcal{Y}_{n}(i))$
$\overset{p}{\rightarrow }$ $0$ and $g(n)(\mathcal{X}_{n}(i)$ $-$ $\mathcal{Y%
}_{n}(i))$ is uniformly integrable for each $i$, then $\mathcal{L}_{n}$ $=$ $%
O(g(n))$ must hold.
\end{lemma}

\begin{remark}
\normalfont$\mathcal{L}_{n}$ $=$ $o(n)$ is required for sample correlation
consistency.
\end{remark}

\begin{remark}
\normalfont Due to difficulties associated with having a general plug-in and
therefore non-Gaussian approximation theory without imposing dependence or
heterogeneity restrictions, we only bound $\mathcal{L}_{n}$ for the maximum
case using classic arguments. An improved upper bound on $\mathcal{L}_{n}$ $%
\rightarrow $ $\infty $ is possible with additional assumptions. The
Gaussian approximation or extreme value theoretic approaches yield (possibly
far) sharper bounds, but require information that generally does not hold
when $X_{n}(i)$ is a filtered functional of a plug-in estimator as discussed
in Section \ref{sec:intro}. Indeed, under Gaussianicity the
Sudakov--Fernique inequality is available
\citep[see,
e.g.,][]{Chatterjee2005,Chernozhukov_etal2015}. See also the tools developed
in \citet[Sections 6 and 7]{ZhangWu2017}: in their stationary functional
dependence setting, a sharper bound than (\ref{Pmax}), below, is available
for a Gaussian approximation \citep[see, e.g.,][Theorem 6.2]{ZhangWu2017}.
\end{remark}

\begin{remark}
\normalfont In principle we can let $\{g(n)\}_{n=1}^{\infty }$ be an array $%
\{g_{n}(i)$ $:$ $i$ $=$ $1,2,...\}_{n=1}^{\infty }$ allowing for a different
rate of convergence for each $i$, e.g. $g_{n}(i)\mathcal{X}_{n}(i)$ $\overset%
{p}{\rightarrow }$ $0$. We only treat a single sequence $\{g(n)\}_{n=1}^{%
\infty }$ for brevity, and given the environment in which we apply the
theory.\
\end{remark}

\noindent \textbf{Proof.}\medskip \newline
\textbf{Claim (a).}\qquad By assumption each $\mathcal{X}_{n}(i)$ $\overset{p%
}{\rightarrow }$ $0$, therefore $\vartheta ([\mathcal{X}_{n}(i)]_{i=1}^{k})$
$\overset{p}{\rightarrow }$ $0$ for each $k$. Define $\mathcal{A}_{k,n}$ $%
\equiv $ $1$ $-$ $\exp \{-\vartheta ([\mathcal{X}_{n}(i)]_{i=1}^{k})\}$ $\in
$ $[0,1]$ $a.s$. $\forall (k,n)$, and $\mathcal{P}_{k,n}$ $\equiv $ $%
\int_{0}^{\infty }P(\mathcal{A}_{k,n}$ $>$ $\epsilon )d\epsilon $. Note that
$\mathcal{P}_{k,n}$ $\leq $ $\mathcal{P}_{k+1,n}$ because $\mathcal{A}_{k,n}$
$\leq $ $\mathcal{A}_{k+1,n}$ by monotonicity of $\vartheta $. Lebesgue's
dominated convergence theorem, and $\mathcal{A}_{k,n}$ $\overset{p}{%
\rightarrow }$ $0$, therefore yield for each $k$:
\begin{equation*}
\lim_{n\rightarrow \infty }\mathcal{P}_{k,n}=\lim_{n\rightarrow \infty
}\int_{0}^{\infty }P\left( \mathcal{A}_{k,n}>\epsilon \right) d\epsilon
=\lim_{n\rightarrow \infty }\int_{0}^{1}P\left( \mathcal{A}_{k,n}>\epsilon
\right) d\epsilon =\int_{0}^{1}\lim_{n\rightarrow \infty }P\left( \mathcal{A}%
_{k,n}>\epsilon \right) d\epsilon =0.
\end{equation*}%
Hence $\lim_{n\rightarrow \infty }\mathcal{P}_{k,n}$ $=$ $0$ for each $k$,
and therefore $\lim_{k\rightarrow \infty }\lim_{n\rightarrow \infty }%
\mathcal{P}_{k,n}$ $=$ $0$.

Now apply Lemma \ref{lm:array_conv} to $\mathcal{P}_{k,n}$ to deduce that
there exists a positive integer sequence $\{\mathcal{L}_{n}\}$ that is not
unique, where $\mathcal{L}_{n}$ $\rightarrow $ $\infty $, and $\mathcal{L}%
_{n}$ $=$ $o(n)$ is always feasible, such that $\lim_{n\rightarrow \infty }%
\mathcal{P}_{\mathcal{L}_{n},n}$ $=$ $\lim_{n\rightarrow \infty
}\int_{0}^{1}P(\mathcal{A}_{\mathcal{L}_{n},n}$ $>$ $\epsilon )d\epsilon $ $%
= $ $0$. Therefore, by construction $E[\mathcal{A}_{\mathcal{L}_{n},n}]$ $=$
$\int_{0}^{1}P(\mathcal{A}_{\mathcal{L}_{n},n}$ $>$ $\epsilon )d\epsilon $ $%
\rightarrow $ $0$. Hence $\mathcal{A}_{\mathcal{L}_{n},n}$ $\overset{p}{%
\rightarrow }$ $0$ by Markov's inequality, which yields $\vartheta ([%
\mathcal{X}_{n}(i)]_{i=1}^{\mathcal{L}_{n}})$ $\overset{p}{\rightarrow }$ $0$
as claimed.

Now consider an upper bound on $\mathcal{L}_{n}$ $\rightarrow $ $\infty $
when $\vartheta $ is the maximum transform $\vartheta ([\mathcal{X}%
_{n}(i)]_{i=1}^{\mathcal{L}_{n}})$ $=$ $\max_{1\leq i\leq \mathcal{L}_{n}}|%
\mathcal{X}_{n}(i)|$. By assumption $g(n)|\mathcal{X}_{n}(i)|$ $\overset{p}{%
\rightarrow }$ $0$ $\forall i$ $\in $ $\mathbb{N}$ for some non-random
positive $g(n)$ $\rightarrow $ $\infty $. Bonferroni and Markov inequalities
yield:
\begin{eqnarray}
P\left( \max_{1\leq i\leq \mathcal{L}_{n}}\left\vert \mathcal{X}%
_{n}(i)\right\vert >\eta \right)  &=&P\left( \bigcup_{i=1}^{\mathcal{L}%
_{n}}\left\vert \mathcal{X}_{n}(i)\right\vert >\eta \right) \leq \sum_{i=1}^{%
\mathcal{L}_{n}}P\left( \left\vert \mathcal{X}_{n}(i)\right\vert >\eta
\right)   \label{Pmax} \\
&\leq &\frac{1}{\eta }\sum_{i=1}^{\mathcal{L}_{n}}E\left\vert \mathcal{X}%
_{n}(i)\right\vert =\frac{\mathcal{L}_{n}}{\eta g(n)}\frac{1}{\mathcal{L}_{n}%
}\sum_{i=1}^{\mathcal{L}_{n}}g(n)E\left\vert \mathcal{X}_{n}(i)\right\vert .
\notag
\end{eqnarray}%
Since $g(n)|\mathcal{X}_{n}(i)|$ $\overset{p}{\rightarrow }$ $0$ $\forall i$
$\in $ $\mathbb{N}$, if $g(n)\mathcal{X}_{n}(i)$ is also uniformly
integrable for each $i$ then $g(n)E|\mathcal{X}_{n}(i)|$ $\rightarrow $ $0$ $%
\forall i$ $\in $ $\mathbb{N}$ \citep[e.g.][Theorem
3.5]{Billingsley1999}. Thus $1/\mathcal{L}_{n}\sum_{i=1}^{\mathcal{L}%
_{n}}g(n)E|\mathcal{X}_{n}(i)|$ $\rightarrow $ $0$, hence $P(\max_{1\leq
i\leq \mathcal{L}_{n}}|\mathcal{X}_{n}(i)|$ $>$ $\eta )$ $=$ $o(\mathcal{L}%
_{n}/g(n))$. Therefore any $\mathcal{L}_{n}$ $=$ $O(g(n))$ yields $%
\max_{1\leq i\leq \mathcal{L}_{n}}|\mathcal{X}_{n}(i)|$ $\overset{p}{%
\rightarrow }$ $0$ as required.\medskip \newline
\textbf{Claim (b).}\qquad The mapping $\vartheta $ satisfies the triangle
inequality and $\vartheta (\cdot )$ $\geq $ $0$. Apply the inequality twice
to yield
\begin{equation*}
\vartheta \left( \left[ \mathcal{X}_{n}(i)\right] _{i=1}^{\mathcal{L}%
_{n}}\right) \leq \vartheta \left( \left[ \mathcal{Y}_{n}(i)\right] _{i=1}^{%
\mathcal{L}_{n}}\right) +\vartheta \left( \left[ \mathcal{X}_{n}(i)-\mathcal{%
Y}_{n}(i)\right] _{i=1}^{\mathcal{L}_{n}}\right)
\end{equation*}%
and%
\begin{equation*}
\vartheta \left( \left[ \mathcal{Y}_{n}(i)\right] _{i=1}^{\mathcal{L}%
_{n}}\right) \leq \vartheta \left( \left[ \mathcal{X}_{n}(i)\right] _{i=1}^{%
\mathcal{L}_{n}}\right) +\vartheta \left( \left[ \mathcal{X}_{n}(i)-\mathcal{%
Y}_{n}(i)\right] _{i=1}^{\mathcal{L}_{n}}\right)
\end{equation*}%
hence%
\begin{equation*}
\left\vert \vartheta \left( \left[ \mathcal{X}_{n}(i)\right] _{i=1}^{%
\mathcal{L}_{n}}\right) -\vartheta \left( \left[ \mathcal{Y}_{n}(i)\right]
_{i=1}^{\mathcal{L}_{n}}\right) \right\vert \leq \vartheta \left( \left[
\mathcal{X}_{n}(i)-\mathcal{Y}_{n}(i)\right] _{i=1}^{\mathcal{L}_{n}}\right)
.
\end{equation*}%
Now apply (a) to $\mathcal{X}_{n}(i)$ $-$ $\mathcal{Y}_{n}(i)$ to yield $%
\vartheta ([\mathcal{X}_{n}(i)$ $-$ $\mathcal{Y}_{n}(i)]_{i=1}^{\mathcal{L}%
_{n}})$ $\overset{p}{\rightarrow }$ $0$. The upper bound on $\mathcal{L}_{n}$
follows from the Claim (a) argument. $\mathcal{QED}$.\bigskip

Lemma \ref{lm:max_dist} similarly uses Lemma \ref{lm:array_conv} as the
basis for \textit{in distribution} convergence of a function of an
increasing set of random variables. This result is used to prove the
Gaussian approximation Lemma \ref{lm:clt_max}. The following result,
however, does not use any distributional assumptions other than convergence.

\begin{lemma}
\label{lm:max_dist}Let $\{\mathcal{X}_{n}(i)$ $:$ $i$ $\in $ $\mathbb{N}%
\}_{n\geq 1}$ be an array of random variables, and assume $\{\mathcal{X}%
_{n}(i)$ $:$ $1$ $\leq $ $i$ $\leq $ $\mathcal{L}\}$ $\overset{d}{%
\rightarrow }$ $\{\mathcal{X}(i)$ $:$ $1$ $\leq $ $i$ $\leq $ $\mathcal{L}\}$
for each $\mathcal{L}$ $\in $ $\mathbb{N}$, where $\{\mathcal{X}(i)$ $:$ $1$
$\leq $ $i$ $\leq $ $\infty \}$ is a stochastic process. Then $\vartheta ([%
\mathcal{X}_{n}(i)]_{i=1}^{\mathcal{L}_{n}})$ $\overset{d}{\rightarrow }$ $%
\vartheta ([\mathcal{X}(i)]_{i=1}^{\infty })$ for some non-unique sequence
of positive integers $\{\mathcal{L}_{n}\}$, where $\mathcal{L}_{n}$ $%
\rightarrow $ $\infty $ and $\mathcal{L}_{n}$ $=$ $o(n)$.
\end{lemma}

\noindent \textbf{Proof.}\medskip \linebreak \textbf{Step \ 1.\qquad }%
Convergence in finite dimensional distributions, continuity of $\vartheta
(\cdot )$ $\geq $ $0$, and the mapping theorem yield $\vartheta ([\mathcal{X}%
_{n}(i)]_{i=1}^{k})$ $\overset{d}{\rightarrow }$ $\vartheta ([\mathcal{X}%
(i)]_{i=1}^{k})$ for each $k$ $\in $ $\mathbb{N}$. By the definition of
distribution convergence:%
\begin{equation*}
P\left( \vartheta \left( \left[ \mathcal{X}_{n}(i)\right] _{i=1}^{k}\right)
\geq x\right) \rightarrow P\left( \vartheta \left( \left[ \mathcal{X}(i)%
\right] _{i=1}^{k}\right) \geq x\right) \text{ for each }k\in \mathbb{N}%
\text{ and }x\in \lbrack 0,\infty ).
\end{equation*}%
Similarly, by the mapping theorem $P(\exp \{-\vartheta ([\mathcal{X}%
_{n}(i)]_{i=1}^{k})\}$ $\geq $ $x)$ $\rightarrow $ $P(\exp \{-\vartheta ([%
\mathcal{X}(i)]_{i=1}^{k})\}$ $\geq $ $x)$ for each $x$ $\in $ $[0,1].$ Now
define:
\begin{eqnarray*}
&&\mathcal{I}_{k,n}\equiv \int_{0}^{1}P\left( \exp \left\{ -\vartheta \left( %
\left[ \mathcal{X}_{n}(i)\right] _{i=1}^{k}\right) \right\} \geq x\right) dx,
\\
&&\mathcal{I}_{k}\equiv \int_{0}^{1}P\left( \exp \left\{ -\vartheta \left( %
\left[ \mathcal{X}(i)\right] _{i=1}^{k}\right) \right\} \geq x\right) dx%
\text{ and }\mathcal{I}\equiv \int_{0}^{1}P\left( \exp \left\{ -\vartheta
\left( \left[ \mathcal{X}(i)\right] _{i=1}^{\infty }\right) \right\} \geq
x\right) dx.
\end{eqnarray*}%
Notice $\mathcal{I}_{k,n}$, $\mathcal{I}_{k}$\ and $\mathcal{I}$\ are well
defined for any $k$ and $n$ because $\vartheta (\cdot )$ $\geq $ $0$ and $%
\exp \{-\vartheta (\cdot )\}$ $\in $ $[0,1]$.\ Then by Lebesgue's dominated
convergence theorem:%
\begin{equation*}
\lim_{n\rightarrow \infty }\int_{0}^{1}\left\vert P\left( \exp \left\{
-\vartheta \left( \left[ \mathcal{X}_{n}(i)\right] _{i=1}^{k}\right)
\right\} \geq x\right) -P\left( \exp \left\{ -\vartheta \left( \left[
\mathcal{X}(i)\right] _{i=1}^{k}\right) \right\} \geq x\right) \right\vert
dx=0
\end{equation*}%
and%
\begin{equation*}
\lim_{k\rightarrow \infty }\int_{0}^{1}\left\vert P\left( \exp \left\{
-\vartheta \left( \left[ \mathcal{X}(i)\right] _{i=1}^{k}\right) \right\}
\geq x\right) -P\left( \exp \left\{ -\vartheta \left( \left[ \mathcal{X}(i)%
\right] _{i=1}^{\infty }\right) \right\} \geq x\right) \right\vert dx=0.
\end{equation*}%
Scheff\'{e}'s lemma now yields $\lim_{n\rightarrow \infty }\mathcal{I}_{k,n}$
$=$ $\mathcal{I}_{k}$ for each $k$, and $\lim_{k\rightarrow \infty }\mathcal{%
I}_{k}$ $=$ $\mathcal{I}$. Now apply Lemma \ref{lm:array_conv} to $\mathcal{I%
}_{k,n}$ to yield $\lim_{n\rightarrow \infty }\mathcal{I}_{\mathcal{L}_{n},n}
$ $=$ $\mathcal{I}$ for some non-unique, monotonically increasing positive
integer sequence $\{\mathcal{L}_{n}\}$, where $\mathcal{L}_{n}$ $\rightarrow
$ $\infty $. Therefore:%
\begin{equation*}
\lim_{n\rightarrow \infty }\int_{0}^{1}P\left( \exp \left\{ -\vartheta
\left( \left[ \mathcal{X}_{n}(i)\right] _{i=1}^{\mathcal{L}_{n}}\right)
\right\} \geq x\right) dx=\int_{0}^{1}P\left( \exp \left\{ -\vartheta \left( %
\left[ \mathcal{X}(i)\right] _{i=1}^{\infty }\right) \right\} \geq x\right)
dx,
\end{equation*}%
hence identically
\begin{equation}
\lim_{n\rightarrow \infty }E[\exp \{-\vartheta ([\mathcal{X}_{n}(i)]_{i=1}^{%
\mathcal{L}_{n}})\}]=E\left[ \exp \left\{ -\vartheta \left( \left[ \mathcal{X%
}(i)\right] _{i=1}^{\infty }\right) \right\} \right] .  \label{EE_conv}
\end{equation}%
Moreover, in view of Lemma \ref{lm:array_conv}, (\ref{EE_conv}) holds for
any other monotonic sequence of positive integers $\{\mathcal{\tilde{L}}%
_{n}\}_{n=1}^{\infty }$, $\mathcal{\tilde{L}}_{n}$ $\rightarrow $ $\infty $
that satisfies $\mathcal{\tilde{L}}_{n}/\mathcal{L}_{n}$ $\rightarrow $ $0$,
hence $\mathcal{L}_{n}$ $=$ $o(n)$ is always feasible.\medskip \newline
\textbf{Step 2.}\qquad\ Repeat Step 1 to deduce that for each $s$ $\in $ $%
\mathbb{N}$ and some non-unique, positive, monotonically increasing integer
sequences $\{\mathcal{L}_{s}(n)\}$, $\mathcal{L}_{s}(n)$ $=$ $o(n)$ and $%
\mathcal{L}_{s}(n)$ $\rightarrow $ $\infty $:%
\begin{equation}
\lim_{n\rightarrow \infty }E\left[ \exp \left\{ -\vartheta \left( \left[
\mathcal{X}_{n}(i)\right] _{i=1}^{\mathcal{L}_{s}(n)}\right) \right\} ^{s}%
\right] =E\left[ \exp \left\{ -\vartheta \left( \left[ \mathcal{X}(i)\right]
_{i=1}^{\infty }\right) \right\} ^{s}\right] .  \label{EE}
\end{equation}%
Each $\{\mathcal{L}_{s}(n)\}$ is monotonic, and any other sequence $\{%
\mathcal{\tilde{L}}_{n}\}_{n=1}^{\infty }$, $\mathcal{\tilde{L}}_{n}$ $%
\rightarrow $ $\infty $ that satisfies $\mathcal{\tilde{L}}_{n}/\mathcal{L}%
_{s}(n)$ $\rightarrow $ $0$ and $\mathcal{\tilde{L}}_{n}$ $=$ $o(n)$\ is
also valid. Hence, there exists a non-unique monotonic sequence of positive
integers $\{\mathcal{L}_{n}\}$ such that $\mathcal{L}_{n}$ $\rightarrow $ $%
\infty $, $\mathcal{L}_{n}$ $=$ $o(n)$,\ and $\limsup_{n\rightarrow \infty
}\{\mathcal{L}_{n}/\mathcal{L}_{s}(n)\}$ $<$ $1$ for each $s$, that
satisfies (\ref{EE}) for each $s$. Therefore:
\begin{equation}
\lim_{n\rightarrow \infty }E\left[ \exp \left\{ -\vartheta \left( \left[
\mathcal{X}_{n}(i)\right] _{i=1}^{\mathcal{L}_{n}}\right) \right\} ^{s}%
\right] =E\left[ \exp \left\{ -\vartheta \left( \left[ \mathcal{X}(i)\right]
_{i=1}^{\infty }\right) \right\} ^{s}\right] \text{ }\quad \forall s\in
\mathbb{N}.  \label{EsEs}
\end{equation}%
Property (\ref{EsEs}) implies $\exp \{-\vartheta ([\mathcal{X}%
_{n}(i)]_{i=1}^{\mathcal{L}_{n}})\}$ $\overset{d}{\rightarrow }$ $\exp
\{-\vartheta ([\mathcal{X}(i)]_{i=1}^{\infty })\}$
\citep[][Theorem
30.2]{Billingsley1995}. The claim now follows by the mapping theorem. $%
\mathcal{QED}$.\bigskip

Let $h$ $\geq $ $0$. Recall $\rho (h)$ $\equiv $ $E[\epsilon _{t}\epsilon
_{t-h}]/E[\epsilon _{t}^{2}]$ and%
\begin{eqnarray*}
&&G_{t}(\phi )\equiv \left[ \frac{\partial }{\partial \phi ^{\prime }}%
f(x_{t-1},\phi ),\boldsymbol{0}_{k_{\delta }}^{\prime }\right] ^{\prime }\in
\mathbb{R}^{k_{\theta }}\text{ \ and \ }s_{t}(\theta )\equiv \frac{1}{2}%
\frac{\partial }{\partial \theta }\ln \sigma _{t}^{2}(\theta ) \\
&&\mathcal{D}(h)\equiv E\left[ \left( \epsilon _{t}s_{t}+G_{t}/\sigma
_{t}\right) \epsilon _{t-h}\right] +E\left[ \epsilon _{t}\left( \epsilon
_{t-h}s_{t-h}+G_{t-h}/\sigma _{t-h}\right) \right] \in \mathbb{R}^{k_{\theta
}} \\
&&z_{t}(h)\equiv r_{t}(h)-\rho (h)r_{t}(0)\text{ where }r_{t}(h)\equiv \frac{%
\epsilon _{t}\epsilon _{t-h}-E\left[ \epsilon _{t}\epsilon _{t-h}\right] -%
\mathcal{D}(h)^{\prime }\mathcal{A}m_{t}}{E\left[ \epsilon _{t}^{2}\right] },
\end{eqnarray*}%
where $m_{t}$ and $\mathcal{A}$\ appear in plug-in expansion Assumption \ref%
{assum:plug}.c: $\sqrt{n}(\hat{\theta}_{n}$ $-$ $\theta _{0})$ $=$ $\mathcal{%
A}n^{-1/2}\sum_{t=1}^{n}m_{t}(\theta _{0})$ $+$ $O_{p}(n^{-\zeta })$ for
some $\zeta $ $>$ $0$. The following two lemmas are based on standard
arguments and are therefore proved in \citet[Appendix F]{HillMotegi_supp_mat}%
.

\begin{lemma}
\label{lm:expansion}Under Assumptions \ref{assum:dgp} and \ref{assum:plug}:
for some $\zeta $ $>$ $0$ that appears in Assumption \ref{assum:plug}.c, $%
\mathcal{X}_{n}(h)$ $\equiv $ $|\sqrt{n}\{\hat{\rho}_{n}(h)$ $-$ $\rho (h)\}$
$-$ $1/\sqrt{n}\sum\nolimits_{t=1+h}^{n}\{r_{t}(h)$ $-$ $\rho (h)r_{t}(0)\}|$
$=$ $O_{p}(1/n^{\min \{\zeta ,1/2\}})$ for each $h$.
\end{lemma}

\begin{lemma}
\label{lm:conv_fdd}Let Assumptions \ref{assum:dgp} and \ref{assum:plug}
hold, and write $\mathcal{Z}_{n}(h)$ $\equiv $ $1/\sqrt{n}%
\sum_{t=1+h}^{n}z_{t}(h)$. For each $\mathcal{L}$\ $\in \mathbb{N}$ : $\{%
\mathcal{Z}_{n}(h)$ $:$ $1$ $\leq $ $h$ $\leq $ $\mathcal{L\}}$ $\overset{d}{%
\rightarrow }$ $\{\mathcal{Z}(h)$ $:$ $1$ $\leq $ $h$ $\leq $ $\mathcal{L\}}%
, $ where $\{\mathcal{Z}(h)$ $:$ $1$ $\leq $ $h$ $\leq $ $\mathcal{L}\}$ is
a zero mean Gaussian process with variance $\lim_{n\rightarrow \infty
}n^{-1}\sum_{s,t=1}^{n}E[z_{s}(h)z_{t}(h)]$ $\in $ $(0,\infty )$, and
covariance function \linebreak $\lim_{n\rightarrow \infty
}n^{-1}\sum_{s,t=1}^{n}E[z_{s}(h)z_{t}(\tilde{h})]$.
\end{lemma}

\noindent \textbf{Proof of Lemma \ref{lm:corr_expan}.}\qquad Recall
Assumption \ref{assum:dgp}.c states\ $\hat{\omega}_{n}(h)$ $=$ $\omega
(h)+O_{p}(n^{-\kappa })$ for some $\kappa $ $>$ $0$, and under Assumption %
\ref{assum:plug}.c, $\sqrt{n}(\hat{\theta}_{n}$ $-$ $\theta _{0})$ $=$ $%
\mathcal{A}n^{-1/2}\sum_{t=1}^{n}m_{t}(\theta _{0})$ $+$ $O_{p}(n^{-\zeta })$
for some $\zeta $ $>$ $0$.

Property (\ref{NED_var}) applies to $r_{t}(h)$ $-$ $\rho (h)r_{t}(0)$ under
Assumptions \ref{assum:dgp} and \ref{assum:plug}, cf. Theorem 17.8 in \cite%
{Davidson1994}, hence $1/\sqrt{n}\sum_{t=1+h}^{n}\{r_{t}(h)$ $-$ $\rho
(h)r_{t}(0)\}$ $=$ $O_{p}(1)$. By Lemma \ref{lm:expansion}:
\begin{equation*}
\mathcal{X}_{n}(h)\equiv \left\vert \sqrt{n}\left\{ \hat{\rho}_{n}(h)-\rho
(h)\right\} -\frac{1}{\sqrt{n}}\sum_{t=1+h}^{n}\left\{ r_{t}(h)-\rho
(h)r_{t}(0)\right\} \right\vert =O_{p}\left( 1/n^{\min \{\zeta
,1/2\}}\right) .
\end{equation*}%
Therefore:%
\begin{eqnarray}
\left\vert \mathcal{\tilde{X}}_{n}(h)\right\vert  &\equiv &\left\vert \sqrt{n%
}\hat{\omega}_{n}(h)\left\{ \hat{\rho}_{n}(h)-\rho (h)\right\} -\omega (h)%
\frac{1}{\sqrt{n}}\sum_{t=1+h}^{n}\left\{ r_{t}(h)-\rho (h)r_{t}(0)\right\}
\right\vert   \label{rn_ro_h} \\
&\leq &\left\vert \omega (h)\right\vert \times \mathcal{X}_{n}(h)+\left\vert
\hat{\omega}_{n}(h)-\omega (h)\right\vert \times \mathcal{X}_{n}(h)  \notag
\\
&&+\left\vert \hat{\omega}_{n}(h)-\omega (h)\right\vert \times \left\vert
\frac{1}{\sqrt{n}}\sum_{t=1+h}^{n}\left\{ r_{t}(h)-\rho (h)r_{t}(0)\right\}
\right\vert =O_{p}\left( 1/n^{\min \{\zeta ,1/2\}}\right) +O_{p}\left(
1/n^{\kappa }\right)   \notag \\
&=&O_{p}\left( 1/n^{\min \{\zeta ,\kappa ,1/2\}}\right) ,  \notag
\end{eqnarray}%
which implies $(n^{\min \{\zeta ,\kappa ,1/2\}}/\ln (n))|\mathcal{\tilde{X}}%
_{n}(h)|$ $\overset{p}{\rightarrow }$ $0$. Setting $\hat{\rho}_{n}(h)$ $=$ $0
$ for any $h$ $\notin $ $\{0,...,n\}$, the claims now follow by applications
of Lemma \ref{lm:max_p} to $\{|\mathcal{\tilde{X}}_{n}(h)|\}_{h\in \mathbb{N}%
}$. In particular, by (\ref{rn_ro_h}) and Lemma \ref{lm:max_p}, if $(n^{\min
\{\zeta ,\kappa ,1/2\}}/\ln (n))\mathcal{\tilde{X}}_{n}(h)$\ for each $h$ is
uniformly integrable then $\mathcal{L}_{n}$ $=$ $O(n^{\min \{\zeta ,\kappa
,1/2\}}/\ln (n))$. $\mathcal{QED}$.\medskip \newline
\textbf{Proof of Lemma \ref{lm:clt_max}.}\qquad Lemma \ref{lm:conv_fdd}
implies $\{\mathcal{Z}_{n}(h)$ $:$ $1$ $\leq $ $h$ $\leq $ $\mathcal{L}\}$ $%
\overset{d}{\rightarrow }$ $\{\mathcal{Z}(h)$ $:$ $1$ $\leq $ $h$ $\leq $ $%
\mathcal{L}\}$ for each $\mathcal{L}$ $\in $ $\mathbb{N}$, where $\{\mathcal{%
Z}(h)$ $:$ $1$ $\leq $ $h$ $\leq $ $\mathcal{L}\}$ is a zero mean Gaussian
process. Now apply Lemma \ref{lm:max_dist} to prove the claim. $\mathcal{QED}
$.\bigskip

The proof of Theorem \ref{th:p_dep_wild_boot} requires the following uniform
laws and probability bound, and weak convergence for the bootstrapped
p-value. The first result is rudimentary and therefore proved in %
\citet[Appendix F]{HillMotegi_supp_mat}. Recall $m_{t}$ are the Assumption %
\ref{assum:plug}.c$^{\prime }$ estimating equations.

\begin{lemma}
\label{lm:unif_law}Under Assumptions \ref{assum:dgp} and \ref{assum:plug}%
.a,b,c$^{\prime }$,d $\sup_{\theta \in \Theta }||1/n\sum_{t=1}^{n}\omega
_{t}(\partial /\partial \theta )m_{t}(\theta )||$ $\overset{p}{\rightarrow }$
$0$, \linebreak $\sup_{\theta \in \Theta }||1/n\sum_{t=1}^{n}(\partial
/\partial \theta )m_{t}(\theta )$ $-$ $E[(\partial /\partial \theta
)m_{t}(\theta )]||$ $\overset{p}{\rightarrow }$ $0$, and $1/\sqrt{n}%
\sum_{t=1+h}^{n}\omega _{t}m_{t}$ $=$ $O_{p}(1).$
\end{lemma}

Let $\Rightarrow ^{p}$ denote weak convergence in probability on $l_{\infty
} $ (the space of bounded functions) as defined in
\citet[Section
3]{GineZinn1990}. Recall by Lemma \ref{lm:clt_max} that $|\vartheta ([%
\mathcal{Z}_{n}(h)]_{h=1}^{\mathcal{L}_{n}})$ $-$ $\vartheta ([\mathcal{Z}%
(h)]_{h=1}^{\mathcal{L}_{n}})|$ $\overset{p}{\rightarrow }$ $0$ for some
zero mean Gaussian process $\{\mathcal{Z}(h)$ $:$ $h$ $\in $ $\mathbb{N}\}$
with variance $\lim_{n\rightarrow \infty
}n^{-1}\sum_{s,t=1}^{n}E[z_{s}(h)z_{t}(h)]$ $<$ $\infty $. Define the
sample:
\begin{equation*}
\mathfrak{X}_{n}\equiv \left\{ m_{t},x_{t},y_{t}\right\} _{t=1}^{n}.
\end{equation*}

\begin{lemma}
\label{lm:max_cor*}Let Assumptions \ref{assum:dgp} and \ref{assum:plug}.a,b,c%
$^{\prime }$,d hold.\medskip \newline
$a.$ For each $\mathcal{L}$ $\in $ $\mathbb{N}$, $\{\sqrt{n}\hat{\rho}%
_{n}^{(dw)}(h)$ $:$ $1$ $\leq $ $h$ $\leq $ $\mathcal{L}\}$ $\Rightarrow ^{p}
$ $\{\mathcal{\mathring{Z}}(h)$ $:$ $1$ $\leq $ $h$ $\leq $ $\mathcal{L}\},$
where $\{\mathcal{\mathring{Z}}(h)$ $:$ $h$ $\in $ $\mathbb{N}\}$ is an
independent copy of $\{\mathcal{Z}(h)$ $:$ $h$ $\in $ $\mathbb{N}\}$.$%
\medskip $\newline
$b.$ For some monotonic sequence of positive integers $\{\mathcal{L}_{n}\}$,
$\mathcal{L}_{n}$ $\rightarrow $ $\infty $ and $\mathcal{L}_{n}$ $=$ $o(n)$:%
\begin{equation*}
\sup_{c>0}\left\vert P\left( \vartheta \left( \left[ \hat{\omega}_{n}(h)%
\sqrt{n}\hat{\rho}_{n}^{(dw)}(h)\right] _{h=1}^{\mathcal{L}_{n}}\right) \leq
c|\mathfrak{X}_{n}\right) -P\left( \vartheta \left( \left[ \omega (h)%
\mathcal{\mathring{Z}}(h)\right] _{h=1}^{\mathcal{L}_{n}}\right) \leq
c\right) \right\vert \overset{p}{\rightarrow }0.
\end{equation*}%
Finally, $\mathcal{L}_{n}$ $=$ $O(n^{\min \{\zeta ,\kappa ,1/2\}}/\ln (n))$
must be satisfied.
\end{lemma}

\noindent \textbf{Proof.}\qquad \medskip \newline
\textbf{Claim (a).}\qquad Let $\{\varphi _{t}\}_{t=1}^{n}$ be a draw of the
auxiliary variables, and write%
\begin{equation}
\rho _{n}^{\ast }(h)\equiv \frac{1}{E\left[ \epsilon _{t}^{2}\right] }\frac{1%
}{n}\sum_{t=1+h}^{n}\varphi _{t}\left\{ \mathcal{E}_{t,h}-E\left[ \mathcal{E}%
_{1,h}\right] \right\} \text{ where }\mathcal{E}_{t,h}\equiv \epsilon
_{t}\epsilon _{t-h}-\mathcal{D}(h)^{\prime }\mathcal{A}m_{t}.  \label{g*h}
\end{equation}%
Recall $\widehat{\mathcal{E}}_{n,t,h}(\hat{\theta}_{n})$ $\equiv $ $\epsilon
_{t}(\hat{\theta}_{n})\epsilon _{t-h}(\hat{\theta}_{n})$ $-$ $\mathcal{\hat{D%
}}_{n}(h)^{\prime }\widehat{\mathcal{A}}_{n}m_{t}(\hat{\theta}_{n})$, and:%
\begin{equation*}
\hat{\rho}_{n}^{(dw)}(h)\equiv \frac{1}{1/n\sum_{t=1}^{n}\epsilon _{t}^{2}(%
\hat{\theta}_{n})}\frac{1}{n}\sum_{t=1+h}^{n}\varphi _{t}\left\{ \widehat{%
\mathcal{E}}_{n,t,h}(\hat{\theta}_{n})-\frac{1}{n}\sum_{s=1+h}^{n}\widehat{%
\mathcal{E}}_{n,s,h}(\hat{\theta}_{n})\right\} .
\end{equation*}

Let $\{\mathcal{Z}(h)$ $:$ $h$ $\in $ $\mathbb{N}\}$ be the Lemma \ref%
{lm:clt_max} Gaussian process. It suffices to show:%
\begin{eqnarray}
&&\left\{ \sqrt{n}\rho _{n}^{\ast }(h):1\leq h\leq \mathcal{L}\right\}
\Rightarrow ^{p}\left\{ \mathcal{\mathring{Z}}(h):1\leq h\leq \mathcal{L}%
\right\},  \label{weak_r} \\
&&\sqrt{n}\left\vert \hat{\rho}_{n}^{(dw)}(h)-\rho _{n}^{\ast
}(h)\right\vert \overset{p}{\rightarrow }0\text{ for each }h\text{,}
\label{rr}
\end{eqnarray}%
where $\{\mathcal{\mathring{Z}}(h)$ $:$ $h$ $\in $ $\mathbb{N}\}$ is an
independent copy of $\{\mathcal{Z}(h)$ $:$ $h$ $\in $ $\mathbb{N}\}$. We
shorten the proof by letting $\{\xi _{1},\dots ,\xi _{n/b_{n}}\}$ be iid $%
N(0,1)$ random variables. The general case is similar, where $\xi _{i}$ are
iid, $E[\xi _{i}]$ $=$ $0$, $E[\xi _{i}^{2}]$ $=$ $1$ and $E[\xi _{i}^{4}]$ $%
<$ $\infty $, except statements about conditional distribution normality
must be replaced with added steps to show asymptotic convergence in
conditional distribution.\medskip\ \newline
\textbf{Step 1.}\qquad Consider (\ref{weak_r}). Define $\mathbb{L}$ $\equiv $
$\{1,...,\mathcal{L}\}$. It suffices to prove weak convergence on a Polish
space in the sense of \citet{HoffJorg1984,HoffJorg1991}, cf.
\citet[p. 853
and Theorem 3.1.a]{GineZinn1990}. The latter holds\textit{\ if and only if}
there exists a pseudo metric $d$ on $\mathbb{L}$\ such that $(\mathbb{L},d)$
is a totally bounded pseudo metric space; $\{\sqrt{n}\rho _{n}^{\ast }(h)$ $%
: $ $1$ $\leq $ $h$ $\leq $ $\mathcal{L}\}$\ $\overset{d}{\rightarrow }$ $\{%
\mathcal{\mathring{Z}}(h)$ $:$ $1$ $\leq $ $h$ $\leq $ $\mathcal{L}\}$; and
the sequence of distributions governing $\{\sqrt{n}\rho _{n}^{\ast
}(h)\}_{n\geq 1}$ are stochastically equicontinuous on $\mathbb{L}$. $%
\mathbb{L}$\ is compact, so pick the sup-norm $d$. Stochastic equicontinuity
is trivial because $\mathbb{L}$\ is discrete and bounded. It now suffices to
prove convergence in finite dimensional distributions. We follow an argument
given in \citet[proof of Theorem 2]{Hansen1996}.

By construction of $\varphi _{t}$ via $\xi _{t}$:%
\begin{equation*}
\rho _{n}^{\ast }(h)=\frac{1}{E\left[ \epsilon _{t}^{2}\right] }\frac{1}{%
n/b_{n}}\sum_{s=1}^{n/b_{n}}\xi _{s}\frac{1}{b_{n}}%
\sum_{t=(s-1)b_{n}+1+h}^{sb_{n}}\left\{ \mathcal{E}_{t,h}-E\left[ \mathcal{E}%
_{1,h}\right] \right\} =\frac{1}{E\left[ \epsilon _{t}^{2}\right] }\frac{1}{%
n/b_{n}}\sum_{s=1}^{n/b_{n}}\xi _{s}\frac{1}{b_{n}}\mathfrak{E}_{n,h},
\end{equation*}%
say, where $\mathfrak{E}_{n,h}$ $\equiv $ $\sum_{t=(s-1)b_{n}+1+h}^{sb_{n}}\{%
\mathcal{E}_{t,h}$ $-$ $E\left[ \mathcal{E}_{1,h}\right] \}$. Operate
conditionally on $\mathfrak{X}_{n}$ $\equiv $ $\{m_{t},x_{t},y_{t}%
\}_{t=1}^{n}$, and write $E_{\mathfrak{X}_{n}}[\cdot ]$ $\equiv $ $E[\cdot |%
\mathfrak{X}_{n}]$. By joint Gaussianicity and independence of $\xi _{s}$, $%
\{\sqrt{n}\rho _{n}^{\ast }(h)$ $:$ $1$ $\leq $ $h$ $\leq $ $\mathcal{L}\}$
is for each $\mathcal{L}$ $\in $ $\mathbb{N}$ a zero mean Gaussian process
with covariance function $nE_{\mathfrak{X}_{n}}[\rho _{n}^{\ast }(h)\rho
_{n}^{\ast }(\tilde{h})]$ $=$ $1/n\sum_{s=1}^{n/b_{n}}\mathfrak{E}_{n,h}%
\mathfrak{E}_{n,\tilde{h}}/(E[\epsilon _{t}^{2}])^{2}$. Observe:
\begin{eqnarray}
&&\lim_{n\rightarrow \infty }E\left[ nE_{\mathfrak{X}_{n}}\left[ \rho
_{n}^{\ast }(h)\rho _{n}^{\ast }(\tilde{h})\right] \right]  \label{limEnE} \\
&&\text{ \ \ \ \ \ \ \ \ }=\frac{1}{\left[ E\left[ \epsilon _{t}^{2}\right] %
\right] ^{2}}\lim_{n\rightarrow \infty }\frac{1}{n}\sum_{s=1}^{n/b_{n}}%
\sum_{t,u=(s-1)b_{n}+1+h}^{sb_{n}}E\left[ \left\{ \mathcal{E}_{t,h}-E\left[
\mathcal{E}_{1,h}\right] \right\} \left\{ \mathcal{E}_{u,\tilde{h}}-E\left[
\mathcal{E}_{1,\tilde{h}}\right] \right\} \right]  \notag \\
&&\text{ \ \ \ \ \ \ \ \ }=\frac{1}{\left[ E\left[ \epsilon _{t}^{2}\right] %
\right] ^{2}}\sum_{i=0}^{\infty }E\left[ \left\{ \mathcal{E}_{1,h}-E\left[
\mathcal{E}_{1,h}\right] \right\} \left\{ \mathcal{E}_{1+i,\tilde{h}}-E\left[
\mathcal{E}_{1,\tilde{h}}\right] \right\} \right]  \notag \\
&&\text{ \ \ \ \ \ \ \ \ }=\lim_{n\rightarrow \infty }\frac{1}{n}E\left[
\sum_{t=1}^{n}\frac{\left( \mathcal{E}_{t,h}-E\left[ \mathcal{E}_{t,h}\right]
\right) }{E\left[ \epsilon _{t}^{2}\right] }\sum_{t=1}^{n}\frac{\left(
\mathcal{E}_{t,\tilde{h}}-E\left[ \mathcal{E}_{t,\tilde{h}}\right] \right) }{%
E\left[ \epsilon _{t}^{2}\right] }\right] =E\left[ \mathcal{Z}(h)\mathcal{Z}(%
\tilde{h})\right] .  \notag
\end{eqnarray}%
The final equality follows directly from the definition of $\mathcal{Z}(h)$
in Lemma \ref{lm:clt_max}.

Let $\mathfrak{X}$ be the set of samples $\mathfrak{\tilde{X}}_{n}$ such
that $nE_{\mathfrak{\tilde{X}}_{n}}[\rho _{n}^{\ast }(h)\rho _{n}^{\ast }(%
\tilde{h})]$ $\overset{p}{\rightarrow }$ $\lim_{n\rightarrow \infty }E[nE_{%
\mathfrak{\tilde{X}}_{n}}[\rho _{n}^{\ast }(h)\rho _{n}^{\ast }(\tilde{h})]]$
$=$ $E[\mathcal{Z}(h)\mathcal{Z}(\tilde{h})]$. We will prove:
\begin{equation}
P\left( \mathfrak{X}_{n}\in \mathfrak{X}\right) =1.  \label{PXX}
\end{equation}%
In conjunction with (\ref{limEnE}), it then follows that the finite
dimensional distributions of $\{\sqrt{n}\rho _{n}^{\ast }(h)$ $:$ $1$ $\leq $
$h$ $\leq $ $\mathcal{L}\}$ converge to those of $\{\mathcal{\mathring{Z}}%
(h) $ $:$ $1$ $\leq $ $h$ $\leq $ $\mathcal{L}\}$, where $\{\mathcal{%
\mathring{Z}}(h)$ $:$ $1$ $\leq $ $h$ $\leq $ $\mathcal{L}\}$ is a zero mean
Gaussian process with covariance function $E[\mathcal{Z}(h)\mathcal{Z}(%
\tilde{h})]$. Independence of $\xi _{s}$ with respect to the sample $%
\mathfrak{X}_{n}$, Gaussianicity, and the fact that Gaussian processes are
completely determined by their mean and covariance structure, together imply
$\{\mathcal{\mathring{Z}}(h)$ $:$ $1$ $\leq $ $h$ $\leq $ $\mathcal{L}\}$ is
an independent copy of $\{\mathcal{Z}(h)$ $:$ $1$ $\leq $ $h$ $\leq $ $%
\mathcal{L}\}$.

Consider (\ref{PXX}). The following exploits arguments presented in %
\citet[Appendix]{deJong1997}. Let $\{l_{n}\}$ be any sequence of integers $%
l_{n}$ $\in $ $\{1,...,b_{n}\}$ such that $l_{n}$ $\rightarrow $ $\infty $
and $l_{n}$ $=$ $o(b_{n})$. Define:{\small
\begin{equation*}
\mathcal{Y}_{n,s}(h)\equiv \sum_{t=(s-1)b_{n}+l_{n}+1}^{sb_{n}}\left\{
\mathcal{E}_{t,h}-E\left[ \mathcal{E}_{1,h}\right] \right\} \text{, \ }%
\mathcal{U}_{n,s}(h)\equiv \sum_{t=(s-1)b_{n}+1}^{(s-1)b_{n}+l_{n}}\left\{
\mathcal{E}_{t,h}-E\left[ \mathcal{E}_{1,h}\right] \right\} \text{, }%
\mathcal{R}(h)\equiv -\sum_{t=1}^{h}\{\mathcal{E}_{t,h}-E[\mathcal{E}%
_{1,h}]\}.
\end{equation*}%
}By construction $\sum_{t=(s-1)b_{n}+1+h}^{sb_{n}}\{\mathcal{E}_{t,h}$ $-$ $%
E[\mathcal{E}_{1,h}]\}$ $=$ $\mathcal{Y}_{n,s}(h)$ $+$ $\mathcal{U}_{n,s}(h)$
$+$ $\mathcal{R}(h)$, hence%
\begin{eqnarray*}
\frac{1}{n}\sum_{s=1}^{n/b_{n}}\mathfrak{E}_{n,h}\mathfrak{E}_{n,\tilde{h}}
&=&\frac{1}{n}\sum_{s=1}^{n/b_{n}}\mathcal{Y}_{n,s}(h)\mathcal{Y}_{n,s}(%
\tilde{h})+\frac{1}{n}\sum_{s=1}^{n/b_{n}}\mathcal{U}_{n,s}(h)\mathcal{U}%
_{n,s}(\tilde{h})+\frac{1}{b_{n}}\mathcal{R}(h)\mathcal{R}(\tilde{h}) \\
&&+\frac{1}{n}\sum_{s=1}^{n/b_{n}}\mathcal{Y}_{n,s}(h)\mathcal{U}_{n,s}(%
\tilde{h})+\frac{1}{n}\sum_{s=1}^{n/b_{n}}\mathcal{Y}_{n,s}(\tilde{h})%
\mathcal{U}_{n,s}(h)+\frac{1}{n}\sum_{s=1}^{n/b_{n}}\mathcal{Y}_{n,s}(h)%
\mathcal{R}(\tilde{h}) \\
&&+\frac{1}{n}\sum_{s=1}^{n/b_{n}}\mathcal{Y}_{n,s}(\tilde{h})\mathcal{R}(h)+%
\frac{1}{n}\sum_{s=1}^{n/b_{n}}\mathcal{U}_{n,s}(h)\mathcal{R}(\tilde{h})+%
\frac{1}{n}\sum_{s=1}^{n/b_{n}}\mathcal{U}_{n,s}(\tilde{h})\mathcal{R}(h).
\end{eqnarray*}

We will prove all terms are $o_{p}(1)$ save $1/n\sum_{s=1}^{n/b_{n}}\mathcal{%
Y}_{n,s}(h)\mathcal{Y}_{n,s}(\tilde{h})$\ hence:
\begin{equation}
\frac{1}{n}\sum_{s=1}^{n/b_{n}}\mathfrak{E}_{n,h}\mathfrak{E}_{n,\tilde{h}}=%
\frac{1}{n}\sum_{s=1}^{n/b_{n}}\mathcal{Y}_{n,s}(h)\mathcal{Y}_{n,s}(\tilde{h%
})+o_{p}(1).  \label{EEEE_ZU}
\end{equation}%
First, under Assumptions \ref{assum:dgp} and \ref{assum:plug}, $\mathcal{E}%
_{t,h}$ is stationary, ergodic and $L_{2}$-bounded. Therefore $||\mathcal{R}(%
\tilde{h})||_{2}$ $\leq $ $\sum_{t=1}^{\tilde{h}}||\mathcal{E}_{t,h}$ $-$ $E[%
\mathcal{E}_{1,h}]||_{2}$ $\leq $ $K$ for each finite $\tilde{h}$, hence by
the Cauchy-Schwartz inequality $E|b_{n}^{-1}\mathcal{R}(h)\mathcal{R}(\tilde{%
h})|$ $\leq $ $K/b_{n}$ $\rightarrow $ $0$.

Second, the NED and moment properties of $\epsilon _{t}$ and $m_{t}$\ in
Assumptions \ref{assum:dgp} and \ref{assum:plug} imply $\mathcal{E}_{t,h}$ $%
\equiv $ $\epsilon _{t}\epsilon _{t-h}$ $-$ $\mathcal{D}(h)^{\prime }%
\mathcal{A}m_{t}$ is $L_{p}$-bounded, $p$ $>$ $2$, $L_{2}$-NED on an $\alpha
$-mixing base with decay $O(h^{-p/(p-2)})$. Therefore $||1/\sqrt{b_{n}}%
\mathcal{Y}_{n,1}(h)||_{2}$ and $||1/\sqrt{l_{n}}\mathcal{U}_{n,1}(\tilde{h}%
)||_{2}$ are $O(1)$ by (\ref{NED_var}). Multiply and divide $\mathcal{Y}%
_{n,s}(h)$ and $\mathcal{U}_{n,s}(\tilde{h})$\ by $b_{n}$ and $l_{n}$
respectively, and use stationarity, Minkowski and Cauchy-Schwartz
inequalities, and $l_{n}/b_{n}$ $=$ $o(1)$ to yield
\begin{eqnarray*}
&&\left\Vert \frac{1}{n}\sum_{s=1}^{n/b_{n}}\mathcal{Y}_{n,s}(h)\mathcal{U}%
_{n,s}(\tilde{h})\right\Vert _{1}=O\left( \left( \frac{l_{n}}{b_{n}}\right)
^{1/2}\left\Vert \frac{1}{\sqrt{b_{n}}}\mathcal{Y}_{n,1}(h)\right\Vert
_{2}\left\Vert \frac{1}{\sqrt{l_{n}}}\mathcal{U}_{n,1}(\tilde{h})\right\Vert
_{2}\right) =O\left( \left( l_{n}/b_{n}\right) ^{1/2}\right) =o(1), \\
&&\left\Vert \frac{1}{n}\sum_{s=1}^{n/b_{n}}\mathcal{Y}_{n,s}(h)\mathcal{R}%
_{n}(\tilde{h})\right\Vert _{1}=O\left( \left\Vert \frac{1}{\sqrt{b_{n}}}%
\mathcal{Y}_{n,1}(h)\right\Vert _{2}\left\Vert \frac{1}{\sqrt{b_{n}}}%
\mathcal{R}_{n}(\tilde{h})\right\Vert _{2}\right) =o(1), \\
&&\left\Vert \frac{1}{n}\sum_{s=1}^{n/b_{n}}\mathcal{U}_{n,s}(h)\mathcal{U}%
_{n,s}(\tilde{h})\right\Vert _{1}=O\left( \frac{l_{n}}{b_{n}}\left\Vert
\frac{1}{\sqrt{l_{n}}}\mathcal{U}_{n,1}(h)\right\Vert _{2}\left\Vert \frac{1%
}{\sqrt{l_{n}}}\mathcal{U}_{n,1}(\tilde{h})\right\Vert _{2}\right) =o(1), \\
&&\left\Vert \frac{1}{n}\sum_{s=1}^{n/b_{n}}\mathcal{U}_{n,s}(h)\mathcal{R}%
_{n}(\tilde{h})\right\Vert _{1}=O\left( \left( \frac{l_{n}}{b_{n}}\right)
^{1/2}\left\Vert \frac{1}{\sqrt{l_{n}}}\mathcal{U}_{n,1}(h)\right\Vert
_{2}\right) =o(1).
\end{eqnarray*}%
This proves (\ref{EEEE_ZU}).

Next, de Jong's (\citeyear{deJong1997}: Assumption 2) conditions are
satisfied under the given NED property. Hence, by the proof of de Jong's (%
\citeyear{deJong1997}) Theorem 2: $1/n\sum_{s=1}^{n/b_{n}}\mathcal{Y}%
_{n,s}^{2}(h)$ $\overset{p}{\rightarrow }$ $\lim_{n\rightarrow \infty
}n^{-1}E[(\sum_{t=1}^{n}\{\mathcal{E}_{t,h}$ $-$ $E[\mathcal{E}%
_{1,h}]\})^{2}]$. An identical argument can be used to prove that the
product $\mathcal{Y}_{n,s}(h)\mathcal{Y}_{n,s}(\tilde{h})$\ satisfies:
\begin{equation}
\frac{1}{n}\sum_{s=1}^{n/b_{n}}\mathcal{Y}_{n,s}(h)\mathcal{Y}_{n,s}(\tilde{h%
})\overset{p}{\rightarrow }\lim_{n\rightarrow \infty }\frac{1}{n}E\left[
\left( \sum_{t=1}^{n}\left\{ \mathcal{E}_{t,h}-E\left[ \mathcal{E}_{1,h}%
\right] \right\} \right) \left( \sum_{t=1}^{n}\left\{ \mathcal{E}_{t,\tilde{h%
}}-E\left[ \mathcal{E}_{1,\tilde{h}}\right] \right\} \right) \right] .
\label{ZhZh}
\end{equation}%
Property (\ref{PXX}) is proved since combining (\ref{limEnE}), (\ref{EEEE_ZU}%
) and (\ref{ZhZh}) yields%
\begin{equation*}
nE_{\mathfrak{X}_{n}}\left[ \rho _{n}^{\ast }(h)\rho _{n}^{\ast }(\tilde{h})%
\right] \overset{p}{\rightarrow }\lim_{n\rightarrow \infty }E\left[ nE_{%
\mathfrak{X}_{n}}\left[ \rho _{n}^{\ast }(h)\rho _{n}^{\ast }(\tilde{h})%
\right] \right] =E\left[ \mathcal{Z}(h)\mathcal{Z}(\tilde{h})\right] .
\end{equation*}%
\textbf{Step 2.}\qquad Now turn to (\ref{rr}).\medskip

\textbf{Step 2.1}\qquad Recall $\mathcal{E}_{t,h}$ $\equiv $ $\epsilon
_{t}\epsilon _{t-h}$ $-$ $\mathcal{D}(h)^{\prime }\mathcal{A}m_{t}$ and $%
\widehat{\mathcal{E}}_{n,t,h}(\hat{\theta}_{n})$ $\equiv $ $\epsilon _{t}(%
\hat{\theta}_{n})\epsilon _{t-h}(\hat{\theta}_{n})$ $-$ $\mathcal{\hat{D}}%
_{n}(h)^{\prime }\widehat{\mathcal{A}}_{n}m_{t}(\hat{\theta}_{n})$. We will
prove in Step 2.2 that:%
\begin{equation}
\frac{1}{\sqrt{n}}\sum_{t=1+h}^{n}\varphi _{t}\left\{ \widehat{\mathcal{E}}%
_{n,t,h}(\hat{\theta}_{n})-\frac{1}{n}\sum_{s=1+h}^{n}\widehat{\mathcal{E}}%
_{n,s,h}(\hat{\theta}_{n})\right\} =\frac{1}{\sqrt{n}}\sum_{t=1+h}^{n}%
\varphi _{t}\left\{ \mathcal{E}_{t,h}-E\left[ \mathcal{E}_{t,h}\right]
\right\} +o_{p}(1)  \label{omEE omEE}
\end{equation}%
by showing (it is straightforward to show (\ref{omeeDAm})-(\ref{omDAsumm})
imply (\ref{omEE omEE})):
\begin{eqnarray}
&&\frac{1}{\sqrt{n}}\sum_{t=1+h}^{n}\varphi _{t}\epsilon _{t}(\hat{\theta}%
_{n})\epsilon _{t-h}(\hat{\theta}_{n})=\frac{1}{\sqrt{n}}\sum_{t=1+h}^{n}%
\varphi _{t}\epsilon _{t}\epsilon _{t-h}+o_{p}(1),  \label{omeeDAm} \\
&&\mathcal{\hat{D}}_{n}(h)^{\prime }\widehat{\mathcal{A}}_{n}\frac{1}{\sqrt{n%
}}\sum_{t=1+h}^{n}\varphi _{t}m_{t}(\hat{\theta}_{n})=\mathcal{D}(h)^{\prime
}\mathcal{A}\frac{1}{\sqrt{n}}\sum_{t=1+h}^{n}\varphi _{t}m_{t}+o_{p}(1),
\label{DAomm} \\
&&\frac{1}{\sqrt{n}}\sum\nolimits_{t=1+h}^{n}\varphi _{t}\frac{1}{n}%
\sum_{s=1+h}^{n}\epsilon _{s}(\hat{\theta}_{n})\epsilon _{s-h}(\hat{\theta}%
_{n})=\frac{1}{\sqrt{n}}\sum_{t=1+h}^{n}\varphi _{t}E\left[ \epsilon
_{t}\epsilon _{t-h}\right] +o_{p}(1),  \label{omSumee} \\
&&\frac{1}{\sqrt{n}}\sum_{t=1+h}^{n}\varphi _{t}\mathcal{\hat{D}}%
_{n}(h)^{\prime }\widehat{\mathcal{A}}_{n}\frac{1}{n}\sum_{t=1+h}^{n}m_{t}(%
\hat{\theta}_{n})=o_{p}(1).  \label{omDAsumm}
\end{eqnarray}%
By the construction of $\varphi _{t}$, for iid $\xi _{s}$\ distributed $%
N(0,1)$:
\begin{eqnarray*}
E\left[ \left( \frac{1}{\sqrt{n}}\sum_{t=1+h}^{n}\varphi _{t}\left\{
\mathcal{E}_{t,h}-E\left[ \mathcal{E}_{t,h}\right] \right\} \right) ^{2}%
\right] &=&E\left[ \left( \frac{1}{\sqrt{n}}\sum_{s=1}^{n/b_{n}}\xi
_{s}\sum_{t=(s-1)b_{n}+1}^{sb_{n}}\varphi _{t}\left\{ \mathcal{E}_{t,h}-E%
\left[ \mathcal{E}_{t,h}\right] \right\} \right) ^{2}\right] \\
&=&E\left[ \left( \frac{1}{\sqrt{b_{n}}}\sum_{t=1}^{b_{n}}\left\{ \mathcal{E}%
_{t,h}-E\left[ \mathcal{E}_{t,h}\right] \right\} \right) ^{2}\right] .
\end{eqnarray*}%
Under Assumptions \ref{assum:dgp}.b and \ref{assum:plug}.c$^{\prime }$, (\ref%
{NED_var}) applies to $\mathcal{E}_{t,h}$ $-$ $E[\mathcal{E}_{t,h}]$
\citep[Theorems
17.8 and 17.9]{Davidson1994}. Hence $E[(1/\sqrt{b_{n}}\sum_{t=1}^{b_{n}}\{%
\mathcal{E}_{t,h}$ $-$ $E\left[ \mathcal{E}_{t,h}\right] \})^{2}]$ $=$ $O(1)$%
, and therefore:%
\begin{equation}
\frac{1}{\sqrt{n}}\sum_{t=1+h}^{n}\varphi _{t}\left\{ \mathcal{E}_{t,h}-E%
\left[ \mathcal{E}_{t,h}\right] \right\} =O_{p}(1).  \label{omEE}
\end{equation}%
Further, by application of Lemma \ref{lm:expansion}, $\sqrt{n}\{\hat{\gamma}%
_{n}(0)$ $-$ $\gamma (0)\}$ $=$ $n^{-1/2}\sum\nolimits_{t=1}^{n}\{\epsilon
_{t}^{2}$ $-$ $E[\epsilon _{t}^{2}]$ $-$ $\mathcal{D}(0)^{\prime }\mathcal{A}%
m_{t}\}$ $+$ $O_{p}(1/\sqrt{n})$. Coupled with stationarity, ergodicity and
square integrability yields:%
\begin{equation}
\frac{1}{n}\sum_{t=1}^{n}\epsilon _{t}^{2}(\hat{\theta}_{n})=E\left[
\epsilon _{t}^{2}\right] +o_{p}(1).  \label{sume}
\end{equation}%
Combine (\ref{omEE omEE}), (\ref{omEE}) and (\ref{sume})\ to yield (\ref{rr}%
) as required:{\small
\begin{equation*}
\sqrt{n}\hat{\rho}_{n}^{(dw)}(h)=\frac{1}{1/n\sum_{t=1}^{n}\epsilon _{t}^{2}(%
\hat{\theta}_{n})}\frac{1}{\sqrt{n}}\sum_{t=1+h}^{n}\varphi _{t}\left\{
\mathcal{E}_{t,h}-E\left[ \mathcal{E}_{t,h}\right] \right\} +o_{p}(1)=\frac{1%
}{E\left[ \epsilon _{t}^{2}\right] }\frac{1}{\sqrt{n}}\sum_{t=1+h}^{n}%
\varphi _{t}\left\{ \mathcal{E}_{t,h}-E\left[ \mathcal{E}_{t,h}\right]
\right\} +o_{p}(1).
\end{equation*}%
} \qquad

\textbf{Step 2.2}\qquad We now prove (\ref{omeeDAm})-(\ref{omDAsumm}).
Consider (\ref{omeeDAm}). Since $\varphi _{t}$ is zero mean Gaussian and
independent of the sample, the proof of Lemma \ref{lm:corr_expan} carries
over verbatim to show:
\begin{eqnarray}
\frac{1}{\sqrt{n}}\sum_{t=1+h}^{n}\varphi _{t}\epsilon _{t}(\hat{\theta}%
_{n})\epsilon _{t-h}(\hat{\theta}_{n}) &=&\frac{1}{\sqrt{n}}%
\sum_{t=1+h}^{n}\varphi _{t}\epsilon _{t}\epsilon _{t-h}-\sqrt{n}\left( \hat{%
\theta}_{n}-\theta _{0}\right) ^{\prime }\frac{1}{n}\sum_{t=1+h}^{n}\varphi
_{t}\left( \epsilon _{t}s_{t}+G_{t}/\sigma _{t}\right) \epsilon _{t-h}
\notag \\
&&-\sqrt{n}\left( \hat{\theta}_{n}-\theta _{0}\right) ^{\prime }\frac{1}{n}%
\sum_{t=1+h}^{n}\varphi _{t}\epsilon _{t}\left( \epsilon _{t-h}s_{t-h}+\frac{%
G_{t-h}}{\sigma _{t-h}}\right) +o_{p}(1).  \label{omee_ee_r}
\end{eqnarray}%
By the stated moment bounds and the construction of $\varphi _{t}$ we have:
\begin{eqnarray*}
\frac{1}{n}\sum_{t=1+h}^{n}\varphi _{t}\left( \epsilon
_{t}s_{t}+G_{t}/\sigma _{t}\right) \epsilon _{t-h} &=&\frac{1}{n}%
\sum_{t=1}^{n}\varphi _{t}\left( \epsilon _{t}s_{t}+G_{t}/\sigma _{t}\right)
\epsilon _{t-h}+o_{p}(1) \\
&=&\frac{1}{n}\sum_{s=1}^{n/b_{n}}\xi
_{s}\sum_{t=(s-1)b_{n}+1}^{sb_{n}}\left( \epsilon _{t}s_{t}+G_{t}/\sigma
_{t}\right) \epsilon _{t-h}+o_{p}(1).
\end{eqnarray*}%
Stationarity, independence of $\xi _{s}$, and $E[(\epsilon _{t}s_{t}$ $+$ $%
G_{t}/\sigma _{t})^{2}\epsilon _{t-h}^{2}]$ $<$ $\infty $ under Assumptions %
\ref{assum:dgp}.b and \ref{assum:plug}.a,b yield:
\begin{eqnarray*}
E\left[ \left( \frac{1}{n}\sum_{s=1}^{n/b_{n}}\xi _{s}\left\{
\sum_{t=(s-1)b_{n}+1+h}^{sb_{n}}\left( \epsilon _{t}s_{t}+\frac{G_{t}}{%
\sigma _{t}}\right) \epsilon _{t-h}\right\} \right) ^{2}\right] &=&\frac{%
b_{n}}{n}E\left[ \left\{ \frac{1}{b_{n}}\sum_{t=1}^{b_{n}}\left( \epsilon
_{t}s_{t}+\frac{G_{t}}{\sigma _{t}}\right) \epsilon _{t-h}\right\} ^{2}%
\right] \\
&\leq &\frac{b_{n}}{n}\left( \left\Vert \left( \epsilon _{t}s_{t}+\frac{G_{t}%
}{\sigma _{t}}\right) \epsilon _{t-h}\right\Vert _{2}\right) ^{2}=o(1).
\end{eqnarray*}%
Hence $1/n\sum_{t=1+h}^{n}\varphi _{t}(\epsilon _{t}s_{t}+G_{t}/\sigma
_{t})\epsilon _{t-h}$ $\overset{p}{\rightarrow }$ $0$. Combining that with $%
\sqrt{n}(\hat{\theta}_{n}-\theta _{0})$ $=$ $O_{p}(1)$ and (\ref{omee_ee_r})
yields (\ref{omeeDAm}).

Next, (\ref{DAomm}). By Lemma \ref{lm:unif_law}:%
\begin{equation}
\sup_{\theta \in \Theta }\left\Vert \frac{1}{n}\sum_{t=1}^{n}\varphi _{t}%
\frac{\partial }{\partial \theta }m_{t}(\theta )\right\Vert \overset{p}{%
\rightarrow }0\text{ and }\frac{1}{\sqrt{n}}\sum_{t=1+h}^{n}\varphi
_{t}m_{t}=O_{p}(1).  \label{domulln}
\end{equation}%
Now write:%
\begin{eqnarray*}
\mathcal{\hat{D}}_{n}(h)^{\prime }\widehat{\mathcal{A}}_{n}\frac{1}{\sqrt{n}}%
\sum_{t=1+h}^{n}\varphi _{t}m_{t}(\hat{\theta}_{n}) &=&\mathcal{D}%
(h)^{\prime }\mathcal{A}\frac{1}{\sqrt{n}}\sum_{t=1+h}^{n}\varphi _{t}m_{t}+%
\mathcal{D}(h)^{\prime }\mathcal{A}\frac{1}{\sqrt{n}}\sum_{t=1+h}^{n}\varphi
_{t}\left\{ m_{t}(\hat{\theta}_{n})-m_{t}\right\} \\
&&+\left\{ \mathcal{\hat{D}}_{n}(h)^{\prime }\widehat{\mathcal{A}}_{n}-%
\mathcal{D}(h)^{\prime }\mathcal{A}\right\} \frac{1}{\sqrt{n}}%
\sum_{t=1+h}^{n}\varphi _{t}m_{t} \\
&&+\left\{ \mathcal{\hat{D}}_{n}(h)^{\prime }\widehat{\mathcal{A}}_{n}-%
\mathcal{D}(h)^{\prime }\mathcal{A}\right\} \frac{1}{\sqrt{n}}%
\sum_{t=1+h}^{n}\varphi _{t}\left\{ m_{t}(\hat{\theta}_{n})-m_{t}\right\} .
\end{eqnarray*}%
Note that $\mathcal{\hat{D}}_{n}(h)$ $\overset{p}{\rightarrow }$ $\mathcal{D}%
(h)$ by arguments in the proof of Lemma \ref{lm:corr_expan}, and by
supposition $\widehat{\mathcal{A}}_{n}$ $\overset{p}{\rightarrow }$ $%
\mathcal{A}$. Moreover, by a mean value theorem argument, Assumption \ref%
{assum:plug}.c$^{\prime }$, and (\ref{domulln}):%
\begin{equation*}
\left\Vert \frac{1}{\sqrt{n}}\sum_{t=1+h}^{n}\varphi _{t}\left\{ m_{t}(\hat{%
\theta}_{n})-m_{t}\right\} \right\Vert \leq \sqrt{n}\left\Vert \hat{\theta}%
_{n}-\theta _{0}\right\Vert \times \sup_{\theta \in \Theta }\left\Vert \frac{%
1}{n}\sum_{t=1+h}^{n}\varphi _{t}\frac{\partial }{\partial \theta }%
m_{t}(\theta )\right\Vert \overset{p}{\rightarrow }0.
\end{equation*}%
The latter convergence in probability, combined with (\ref{domulln}),
suffice to prove (\ref{DAomm}).

Proceeding to (\ref{omSumee}), first note that%
\begin{equation}
\frac{1}{\sqrt{n}}\sum_{t=1}^{n}\varphi _{t}=b_{n}\frac{1}{\sqrt{n}}%
\sum_{s=1}^{n/b_{n}}\xi _{s}=\sqrt{b_{n}}\left( n/b_{n}\right)
^{-1/2}\sum_{s=1}^{n/b_{n}}\xi _{s}=O_{p}\left( \sqrt{b_{n}}\right) .
\label{sum_om}
\end{equation}%
Second, by equation (F.5) in the proof of Lemma \ref{lm:expansion} in \cite%
{HillMotegi_supp_mat}:%
\begin{equation}
\left\vert \sqrt{n}\hat{\gamma}_{n}(h)-n^{-1/2}\sum_{t=1+h}^{n}\epsilon
_{t}\epsilon _{t-h}+\sqrt{n}\left( \hat{\theta}_{n}-\theta _{0}\right)
^{\prime }\mathcal{D}(h)\right\vert \overset{p}{\rightarrow }0.
\label{root(n)g}
\end{equation}%
Use (\ref{root(n)g}), and $\hat{\theta}_{n}$ $=$ $\theta _{0}$ $+$ $O_{p}(1/%
\sqrt{n})$ to deduce $1/n\sum_{s=1+h}^{n}\epsilon _{s}(\hat{\theta}%
_{n})\epsilon _{s-h}(\hat{\theta}_{n})$ $=$ $1/n\sum_{t=1+h}^{n}\epsilon
_{t}\epsilon _{t-h}$ $+$ $O_{p}(1/\sqrt{n})$. Therefore%
\begin{equation*}
\frac{1}{\sqrt{n}}\sum_{t=1+h}^{n}\varphi _{t}\frac{1}{n}\sum_{s=1+h}^{n}%
\epsilon _{s}(\hat{\theta}_{n})\epsilon _{s-h}(\hat{\theta}_{n})=\frac{1}{%
\sqrt{n}}\sum_{t=1+h}^{n}\varphi _{t}\frac{1}{n}\sum_{t=1+h}^{n}\epsilon
_{t}\epsilon _{t-h}+O_{p}\left( 1/\sqrt{n/b_{n}}\right) .
\end{equation*}

It remains to show%
\begin{equation}
\frac{1}{\sqrt{n}}\sum_{t=1+h}^{n}\varphi _{t}\frac{1}{n}\sum_{t=1+h}^{n}%
\epsilon _{t}\epsilon _{t-h}=\frac{1}{\sqrt{n}}\sum_{t=1+h}^{n}\varphi _{t}E%
\left[ \epsilon _{t}\epsilon _{t-h}\right] +o_{p}(1).  \label{eomEe}
\end{equation}%
Under Assumptions \ref{assum:dgp}.b, $\epsilon _{t}\epsilon _{t-h}$ $-$ $%
E[\epsilon _{t}\epsilon _{t-h}]$\ satisfies (\ref{NED_var}), hence $E[(1/%
\sqrt{n}\sum_{t=1}^{n}\{\epsilon _{t}\epsilon _{t-h}$ $-$ $E\left[ \epsilon
_{t}\epsilon _{t-h}\right] \})^{2}]$ $=$ $O(1)$. Further $1/\sqrt{n}%
\sum_{t=1+h}^{n}\varphi _{t}$ $=$ $O_{p}(\sqrt{b_{n}})$ from (\ref{sum_om}).
Hence%
\begin{equation*}
\frac{1}{\sqrt{n}}\sum_{t=1+h}^{n}\varphi _{t}\frac{1}{n}\sum_{t=1+h}^{n}%
\left\{ \epsilon _{t}\epsilon _{t-h}-E\left[ \epsilon _{t}\epsilon _{t-h}%
\right] \right\} =\frac{1}{\sqrt{n}}\sum_{t=1+h}^{n}\varphi _{t}\times
O_{p}(1/\sqrt{n})=O_{p}\left( 1/\sqrt{n/b_{n}}\right) .
\end{equation*}%
Since $b_{n}/n$ $\rightarrow $ $0$, (\ref{eomEe}) follows directly.

Finally, for (\ref{omDAsumm}), since $1/\sqrt{n}\sum_{t=1+h}^{n}\varphi _{t}$
$=$ $O_{p}(\sqrt{b_{n}})$ and $\mathcal{\hat{D}}_{n}(h)^{\prime }\widehat{%
\mathcal{A}}_{n}$ $\overset{p}{\rightarrow }$ $\mathcal{D}(h)^{\prime }%
\mathcal{A}$ we need only show $1/n\sum_{t=1}^{n}m_{t}(\hat{\theta}_{n})$ $=$
$o_{p}(1/\sqrt{b_{n}})$. A first order expansion and the mean value theorem
yield:%
\begin{equation*}
\left\Vert \frac{1}{n}\sum_{t=1+h}^{n}m_{t}(\hat{\theta}_{n})-\frac{1}{n}%
\sum_{t=1+h}^{n}m_{t}\right\Vert \leq \sup_{\theta \in \Theta }\left\Vert
\frac{1}{n}\sum_{t=1+h}^{n}\frac{\partial }{\partial \theta }m_{t}(\theta
_{n}^{\ast })\right\Vert \left\Vert \hat{\theta}_{n}-\theta _{0}\right\Vert .
\end{equation*}%
By Lemma \ref{lm:unif_law}: $\sup_{\theta \in \Theta
}||1/n\sum_{t=1}^{n}(\partial /\partial \theta )m_{t}(\theta )$ $-$ $%
E[(\partial /\partial \theta )m_{t}(\theta )]||$ $\overset{p}{\rightarrow }$
$0$, and $\sup_{\theta \in \Theta }||E[(\partial /\partial \theta
)m_{t}(\theta )]||$ $<$ $\infty $ and $\hat{\theta}_{n}$ $-$ $\theta _{0}$ $=
$ $O_{p}(1/\sqrt{n})$\ under Assumption \ref{assum:plug}.c$^{\prime }$.
Moreover, by Assumption \ref{assum:plug}.c$^{\prime }$, $m_{t}$ $=$ $%
[m_{i,t}]_{i=1}^{k_{m}}$\ satisfies (\ref{NED_var}), hence $E[(1/\sqrt{n}%
\sum_{t=1}^{n}m_{i,t}^{2}]$ $=$ $O(1)$. This yields $1/n%
\sum_{t=1+h}^{n}m_{t}(\hat{\theta}_{n})$ $=$ $1/n\sum_{t=1+h}^{n}m_{t}$ $+$ $%
O_{p}(1/\sqrt{n})$ $=$ $O_{p}(1/\sqrt{n})$. Since $b_{n}$ $=$ $o(n)$ the
proof is complete.\medskip \newline
\textbf{Claim (b).}\qquad Weak convergence in probability Claim (a), the
mapping theorem and Slutsky's theorem yield for each $\mathcal{L}$ $\in $ $%
\mathbb{N}$:%
\begin{equation}
\vartheta \left( \left[ \sqrt{n}\hat{\omega}_{n}(h)\hat{\rho}_{n}^{(dw)}(h)%
\right] _{h=1}^{\mathcal{L}}\right) \Rightarrow ^{p}\vartheta \left( \left[
\omega (h)\mathcal{\mathring{Z}}(h)\right] _{h=1}^{\mathcal{L}}\right) .
\label{maxp_maxR_weak_p}
\end{equation}%
Therefore \citep[see, e.g.,][eq. (3.4)]{GineZinn1990}:%
\begin{equation*}
\mathcal{X}_{n}(\mathcal{L})\equiv \sup_{c>0}\left\vert P\left( \vartheta
\left( \left[ \sqrt{n}\hat{\omega}_{n}(h)\hat{\rho}_{n}^{(dw)}(h)\right]
_{h=1}^{\mathcal{L}}\right) \leq c|\mathfrak{X}_{n}\right) -P\left(
\vartheta \left( \left[ \omega (h)\mathcal{\mathring{Z}}(h)\right] _{h=1}^{%
\mathcal{L}}\right) \leq c\right) \right\vert \overset{p}{\rightarrow }0.
\end{equation*}%
Apply Lemma \ref{lm:max_p}.a to $\mathcal{X}_{n}(\mathcal{L})$\ to yield for
some monotonic sequence of positive integers $\{\mathcal{L}_{n}\}_{n\geq 1}$%
, $\mathcal{L}_{n}$ $\rightarrow $ $\infty $ and $\mathcal{L}_{n}$ $=$ $o(n)$%
:
\begin{equation*}
\mathcal{X}_{n}(\mathcal{L}_{n})\leq \max_{1\leq \mathcal{L}\leq \mathcal{L}%
_{n}}\mathcal{X}_{n}(\mathcal{L})\overset{p}{\rightarrow }0.
\end{equation*}%
Therefore $\mathcal{X}_{n}(\mathcal{L}_{n})$ $\overset{p}{\rightarrow }$ $0$%
. Finally, since $\mathcal{X}_{n}(\mathcal{L})$ is bounded, it is uniformly
integrable. Hence $\mathcal{L}_{n}$ $=$ $O(n^{\min \{\zeta ,\kappa
,1/2\}}/\ln (n))$ must be satisfied for $\max_{1\leq \mathcal{L}\leq
\mathcal{L}_{n}}\mathcal{X}_{n}(\mathcal{L})$ $\overset{p}{\rightarrow }$ $0$
and therefore for $\mathcal{X}_{n}(\mathcal{L}_{n})$ $\overset{p}{%
\rightarrow }$ $0$, cf. Lemma \ref{lm:max_p}.a. $\mathcal{QED}$.\bigskip
\newline
\textbf{Proof of Theorem \ref{th:p_dep_wild_boot}.}\qquad Assume the weights
$\hat{\omega}_{n}(h)$ $=$ $1$ to conserve notation, without loss of
generality. Operate conditionally on $\mathfrak{X}_{n}$ $\equiv $ $%
\{m_{t},x_{t},y_{t}\}_{t=1}^{n}$, and recall $\hat{p}_{n,M}^{(dw)}$ $\equiv $
$1/M\sum_{i=1}^{M}I(\mathcal{\hat{T}}_{n,i}^{(dw)}$ $\geq $ $\mathcal{\hat{T}%
}_{n})$. First, by the Glivenko-Cantelli theorem:%
\begin{equation}
\hat{p}_{n,M}^{(dw)}\overset{p}{\rightarrow }P\left( \vartheta \left( \left[
\sqrt{n}\hat{\rho}_{n}^{(dw)}(h)\right] _{h=1}^{\mathcal{L}_{n}}\right) \geq
\vartheta \left( \left[ \sqrt{n}\hat{\rho}_{n}(h)\right] _{h=1}^{\mathcal{L}%
_{n}}\right) |\mathfrak{X}_{n}\right) \text{ as }M\rightarrow \infty .
\label{p_dw_p}
\end{equation}%
Second, by Lemma \ref{lm:max_cor*}:
\begin{equation}
\sup_{c>0}\left\vert P\left( \vartheta \left( \left[ \sqrt{n}\hat{\rho}%
_{n}^{(dw)}(h)\right] _{h=1}^{\mathcal{L}_{n}}\right) \leq c|\mathfrak{X}%
_{n}\right) -P\left( \vartheta \left( \left[ \mathcal{\mathring{Z}}(h)\right]
_{h=1}^{\mathcal{L}_{n}}\right) \leq c\right) \right\vert \overset{p}{%
\rightarrow }0,  \label{p_dw_R1}
\end{equation}%
where $\{\mathcal{Z}(h)$ $:$ $h$ $\in $ $\mathbb{N}\}$ is a zero mean
Gaussian process with variance $E[\mathcal{Z}(h)^{2}]$ $<$ $\infty $, $\{%
\mathcal{\mathring{Z}}(h)$ $:$ $h$ $\in $ $\mathbb{N}\}$ is an independent
copy of $\{\mathcal{Z}(h)$ $:$ $h$ $\in $ $\mathbb{N}\}$, and $\{\mathcal{L}%
_{n}\}$ is a monotonic sequence of positive integers, $\mathcal{L}_{n}$ $%
\rightarrow $ $\infty $ and $\mathcal{L}_{n}$ $=$ $o(n)$. In particular, $%
\mathcal{L}_{n}$ $=$ $O(n^{\min \{\zeta ,\kappa ,1/2\}}/\ln (n))$ must be
satisfied.

Impose $H_{0}$ $:$ $\rho (h)$ $=$ $0$ $\forall h$ $\in $ $\mathbb{N}$.
Define $\bar{F}_{n}^{(0)}(c)$ $\equiv $ $P(\vartheta ([\mathcal{\mathring{Z}}%
(h)]_{h=1}^{\mathcal{L}_{n}})$ $>$ $c)$. Note that (\ref{p_dw_R1}) implies:%
\begin{equation*}
P\left( \vartheta \left( \left[ \sqrt{n}\hat{\rho}_{n}^{(dw)}(h)\right]
_{h=1}^{\mathcal{L}_{n}}\right) \geq \vartheta \left( \left[ \sqrt{n}\hat{%
\rho}_{n}(h)\right] _{h=1}^{\mathcal{L}_{n}}\right) |\mathfrak{X}_{n}\right)
-P\left( \vartheta \left( \left[ \mathcal{\mathring{Z}}(h)\right] _{h=1}^{%
\mathcal{L}_{n}}\right) \geq \vartheta \left( \left[ \sqrt{n}\hat{\rho}%
_{n}(h)\right] _{h=1}^{\mathcal{L}_{n}}\right) \right) \overset{p}{%
\rightarrow }0.
\end{equation*}%
Since $[\mathcal{\mathring{Z}}(h)]_{h=1}^{\mathcal{L}_{n}}$ is independent
of the sample $\mathfrak{X}_{n}$, we therefore have:
\begin{equation}
P\left( \vartheta \left( \left[ \sqrt{n}\hat{\rho}_{n}^{(dw)}(h)\right]
_{h=1}^{\mathcal{L}_{n}}\right) \geq \vartheta \left( \left[ \sqrt{n}\hat{%
\rho}_{n}(h)\right] _{h=1}^{\mathcal{L}_{n}}\right) |\mathfrak{X}_{n}\right)
-\bar{F}_{n}^{(0)}\left( \vartheta \left( \left[ \sqrt{n}\hat{\rho}_{n}(h)%
\right] _{h=1}^{\mathcal{L}_{n}}\right) \right) \overset{p}{\rightarrow }0.
\label{PF}
\end{equation}%
$\bar{F}_{n}^{(0)}$ is continuous by Gaussianicity. Theorem \ref%
{th:max_corr_expan} and Slutsky's theorem therefore yield:
\begin{equation}
\left\vert \bar{F}_{n}^{(0)}\left( \vartheta \left( \left[ \sqrt{n}\hat{\rho}%
_{n}(h)\right] _{h=1}^{\mathcal{L}_{n}}\right) \right) -\bar{F}%
_{n}^{(0)}\left( \vartheta \left( \left[ \mathcal{Z}(h)\right] _{h=1}^{%
\mathcal{L}_{n}}\right) \right) \right\vert \overset{p}{\rightarrow }0.
\label{FF}
\end{equation}%
Together, (\ref{p_dw_p}), (\ref{PF}) and (\ref{FF}) yield for any sequence
of positive integers $\{M_{n}\}$, $M_{n}$ $\rightarrow $ $\infty $:%
\begin{eqnarray}
\hat{p}_{n,M_{n}}^{(dw)} &=&P\left( \vartheta \left( \left[ \sqrt{n}\hat{\rho%
}_{n}^{(dw)}(h)\right] _{h=1}^{\mathcal{L}_{n}}\right) \geq \vartheta \left( %
\left[ \sqrt{n}\hat{\rho}_{n}(h)\right] _{h=1}^{\mathcal{L}_{n}}\right) |%
\mathfrak{X}_{n}\right) +o_{p}(1)  \label{p_dw_F} \\
&=&\bar{F}_{n}^{(0)}\left( \vartheta \left( \left[ \mathcal{Z}(h)\right]
_{h=1}^{\mathcal{L}_{n}}\right) \right) +o_{p}(1).  \notag
\end{eqnarray}%
Since$\{\mathcal{\mathring{Z}}(h)$ $:$ $h$ $\in $ $\mathbb{N}\}$ is an
independent copy of $\{\mathcal{Z}(h)$ $:$ $h$ $\in $ $\mathbb{N}\}$, $\bar{F%
}_{n}^{(0)}(\vartheta ([\mathcal{Z}(h)]_{h=1}^{\mathcal{L}_{n}}))$ is
distributed uniform on $[0,1]$. Now use (\ref{p_dw_F}) to conclude $P(\hat{p}%
_{n,M_{n}}^{(dw)}$ $<$ $\alpha )$ $=$ $P(\bar{F}_{n}^{(0)}\left( (\vartheta %
\left[ \mathcal{Z}(h)\right] _{h=1}^{\mathcal{L}_{n}})\right) $ $<$ $\alpha
) $ $+$ $o(1)$ $=$ $\alpha $ $+$ $o(1)$ $\rightarrow $ $\alpha .$

Impose $H_{1}$ $:$ $\rho (h)$ $\neq $ $0$ for some $h$ $\in $ $\mathbb{N}$.
Recall $\vartheta $ satisfies the triangle inequality, and divergence $%
\vartheta (a)$ $\rightarrow $ $\infty $ as $||a||$ $\rightarrow $ $\infty $.
Theorem \ref{th:max_corr_expan} therefore yields: $\vartheta ([\sqrt{n}\hat{%
\rho}_{n}(h)]_{h=1}^{\mathcal{L}_{n}})$ $\leq $ $\vartheta ([\sqrt{n}\{\hat{%
\rho}_{n}(h)$ $-$ $\rho (h)\}]_{h=1}^{\mathcal{L}_{n}})$ $+$ $\vartheta ([%
\sqrt{n}\rho (h)]_{h=1}^{\mathcal{L}_{n}})$ $=$ $\vartheta ([\mathcal{Z}%
(h)]_{h=1}^{\mathcal{L}_{n}})+\vartheta ([\sqrt{n}\rho (h)]_{h=1}^{\mathcal{L%
}_{n}})$ $+$ $o_{p}(1)$, and $\vartheta ([\sqrt{n}\rho (h)]_{h=1}^{\mathcal{L%
}_{n}})$ $\leq $ $\vartheta ([\sqrt{n}\{\hat{\rho}_{n}(h)$ $-$ $\rho
(h)\}]_{h=1}^{\mathcal{L}_{n}})$ $+$ $\vartheta ([\sqrt{n}\hat{\rho}%
_{n}(h)]_{h=1}^{\mathcal{L}_{n}})$ $=$ $\vartheta ([\mathcal{Z}(h)]_{h=1}^{%
\mathcal{L}_{n}})$ $+$ $\vartheta ([\sqrt{n}\hat{\rho}_{n}(h)]_{h=1}^{%
\mathcal{L}_{n}})$ $+$ $o_{p}(1)$ $\overset{p}{\rightarrow }$ $\infty $.
Hence:%
\begin{equation}
\infty \overset{p}{\leftarrow }\vartheta \left( \left[ \mathcal{Z}(h)\right]
_{h=1}^{\mathcal{L}_{n}}\right) +\vartheta \left( \left[ \sqrt{n}\rho (h)%
\right] _{h=1}^{\mathcal{L}_{n}}\right) +o_{p}(1)\geq \vartheta \left( \left[
\sqrt{n}\rho (h)\right] _{h=1}^{\mathcal{L}_{n}}\right) -\vartheta \left( %
\left[ \mathcal{Z}(h)\right] _{h=1}^{\mathcal{L}_{n}}\right) +o_{p}(1)%
\overset{p}{\rightarrow }\infty .  \label{phi(rho)}
\end{equation}%
Combine (\ref{p_dw_p}), (\ref{p_dw_R1}) and (\ref{phi(rho)}) to deduce $P(%
\hat{p}_{n,M_{n}}^{(dw)}$ $<$ $\alpha )$ $\rightarrow $ $1$ for any $\alpha $
$\in $ $(0,1)$\ because:{\small
\begin{eqnarray*}
\hat{p}_{n,M_{n}}^{(dw)} &=&P\left( \vartheta \left( \left[ \sqrt{n}\hat{\rho%
}_{n}^{(dw)}(h)\right] _{h=1}^{\mathcal{L}_{n}}\right) \geq \vartheta \left( %
\left[ \sqrt{n}\hat{\rho}_{n}(h)\right] _{h=1}^{\mathcal{L}_{n}}\right) |%
\mathfrak{X}_{n}\right) +o_{p}(1) \\
&=&P\left( \vartheta \left( \left[ \mathcal{\mathring{Z}}(h)\right] _{h=1}^{%
\mathcal{L}_{n}}\right) \geq \vartheta \left( \left[ \sqrt{n}\hat{\rho}%
_{n}(h)\right] _{h=1}^{\mathcal{L}_{n}}\right) \right) +o_{p}(1)=\bar{F}%
_{n}^{(0)}\left( \vartheta \left( \left[ \sqrt{n}\hat{\rho}_{n}(h)\right]
_{h=1}^{\mathcal{L}_{n}}\right) \right) +o_{p}(1)\overset{p}{\rightarrow }0.%
\text{ }\mathcal{QED}.
\end{eqnarray*}%
}\textbf{Proof of Theorem \ref{th:lag_select}}.\qquad Let $q$ be any fixed
positive constant. Recall that the penalty $\mathcal{P}_{n}(\mathcal{L})$ $=$
$\sqrt{\mathcal{L}\ln n}$ if $\mathcal{\hat{T}}_{n}(\mathcal{L})$ $\leq $ $%
\sqrt{q\ln n}$, else $\mathcal{P}_{n}(\mathcal{L})$ $=$ $\sqrt{2\mathcal{L}}$%
.\medskip \newline
\textbf{Claim (a).}\qquad Let $H_{0}$\ be true. It suffices to prove the
following. First, for any $\{\mathcal{L}_{n}\}$, $\mathcal{L}_{n}$ $%
\rightarrow $ $(0,\infty ]$ and $\mathcal{L}_{n}/\mathcal{\bar{L}}_{n}$ $%
\rightarrow $ $[0,1]$, the penalty term satisfies:%
\begin{equation}
P\left( \mathcal{P}_{n}(\mathcal{L}_{n})=\sqrt{\mathcal{L}_{n}\ln (n)}%
\right) \rightarrow 1.  \label{PP_lnn}
\end{equation}%
Hence $\mathcal{\hat{T}}_{n}^{\mathcal{P}}(\mathcal{L})$ $\equiv $ $\mathcal{%
\hat{T}}_{n}(\mathcal{L})$ $-$ $\sqrt{\mathcal{L}\ln n}$ asymptotically with
probability approaching one. Second, for such $\{\mathcal{L}_{n}\}$ the
following holds:%
\begin{eqnarray}
&&P\left( \mathcal{\hat{T}}_{n}(\mathcal{L}_{n})-\mathcal{\hat{T}}%
_{n}(l)\geq \left( \sqrt{\mathcal{L}_{n}}-\sqrt{l}\right) \sqrt{\ln (n)}%
\right) \rightarrow 1\text{ if }l\geq \mathcal{L}_{n}  \label{PTT} \\
&&P\left( \mathcal{\hat{T}}_{n}(\mathcal{L}_{n})-\mathcal{\hat{T}}%
_{n}(l)\geq \left( \sqrt{\mathcal{L}_{n}}-\sqrt{l}\right) \sqrt{\ln (n)}%
\right) \rightarrow 0\text{ for fixed }l=1,...,\mathcal{L}_{n}-1.  \notag
\end{eqnarray}%
Together (\ref{PP_lnn}) and (\ref{PTT}) prove the claim $P(\mathcal{L}%
_{n}^{\ast }$ $=$ $1)$ $\rightarrow $ $1$ since\ the following holds \textit{%
for every} $l$ $=$ $1,...,\mathcal{\bar{L}}_{n}$ \textit{if and only if} $%
\mathcal{L}_{n}$ $\rightarrow $ $1$:
\begin{equation}
\lim_{n\rightarrow \infty }P\left( \mathcal{\hat{T}}_{n}^{\mathcal{P}}(%
\mathcal{L}_{n})\geq \mathcal{\hat{T}}_{n}^{\mathcal{P}}(l)\right)
=\lim_{n\rightarrow \infty }P\left( \mathcal{\hat{T}}_{n}(\mathcal{L}_{n})-%
\mathcal{\hat{T}}_{n}(l)\geq \left( \sqrt{\mathcal{L}_{n}}-\sqrt{l}\right)
\sqrt{\ln (n)}\right) =1,  \label{limPlimP}
\end{equation}%
while $\mathcal{L}_{n}^{\ast }$ is the least of sequences that satisfy (\ref%
{limPlimP}) \textit{for every} $l$ $=$ $1,...,\mathcal{\bar{L}}_{n}$.

Now consider (\ref{PP_lnn}). By construction of $\mathcal{P}_{n}(\mathcal{L}%
_{n})$ it suffices to prove $P(\mathcal{\hat{T}}_{n}(\mathcal{L}_{n})$ $>$ $%
\sqrt{q\ln n})$ $\rightarrow $ $0$. Under $H_{0}$, $\sqrt{n}\hat{\rho}%
_{n}(h) $ $=$ $O_{p}(1)$ by (\ref{rn_ro_h}), hence $\sqrt{n}\hat{\rho}%
_{n}(h)/\sqrt{q\ln n}$ $\overset{p}{\rightarrow }$ $0$ for any fixed $q$ $%
\in $ $(0,\infty )$. Therefore, by Lemma \ref{lm:max_p} for some $\{\mathcal{%
\bar{L}}_{n}\}$ that satisfies $\mathcal{\bar{L}}_{n}$ $\rightarrow $ $%
\infty $ and $\mathcal{\bar{L}}_{n}$ $=$ $o(n)$:
\begin{equation}
\frac{\mathcal{\hat{T}}_{n}(\mathcal{\bar{L}}_{n})}{\sqrt{q\ln n}}=\frac{%
\sqrt{n}\max_{1\leq h\leq \mathcal{\bar{L}}_{n}}\left\vert \hat{\rho}%
_{n}(h)\right\vert }{\sqrt{q\ln n}}\overset{p}{\rightarrow }0.  \label{T/lnn}
\end{equation}%
By the same lemma, if $(n^{\min \{\zeta ,\kappa ,1/2\}}/\ln (n))\mathcal{%
\tilde{X}}_{n}(h)$ for all $h$\ is uniformly integrable, where $\mathcal{%
\tilde{X}}_{n}(h)$ is defined in (\ref{expans_rate}), then $\mathcal{\bar{L}}%
_{n}$ $=$ $O(n^{\min \{\zeta ,\kappa ,1/2\}}/\ln (n))$ must hold. By
monotonicity of $\mathcal{\hat{T}}_{n}(\cdot )$ $\geq $ $0$, (\ref{T/lnn})
holds for any $\{\mathcal{L}_{n}\}$, $\mathcal{L}_{n}$ $\rightarrow $ $%
(0,\infty ]$ and $\mathcal{L}_{n}/\mathcal{\bar{L}}_{n}$ $\rightarrow $ $%
[0,1]$. Thus $\mathcal{\hat{T}}_{n}(\mathcal{L}_{n})/\sqrt{q\ln n}$ $\overset%
{p}{\rightarrow }$ $0$ for all such $\{\mathcal{L}_{n}\}$.

Now consider (\ref{PTT}). Suppose $l$ $>$ $\mathcal{L}_{n}$. By (\ref{T/lnn}%
), $\mathcal{\hat{T}}_{n}(\mathcal{\bar{L}}_{n})/\sqrt{\ln n}$ $=$ $o_{p}(1)$
and therefore $\mathcal{\hat{T}}_{n}(\mathcal{L}_{n})$ $-$ $\mathcal{\hat{T}}%
_{n}(l)$ $=$ $o_{p}(\sqrt{\ln (n)})$ for any $\{\mathcal{L}_{n}\}$, $%
\mathcal{L}_{n}$ $\rightarrow $ $(0,\infty ]$ and $\mathcal{L}_{n}/\mathcal{%
\bar{L}}_{n}$ $\rightarrow $ $[0,1]$, and any $1$ $\leq $ $l$ $\leq $ $%
\mathcal{\bar{L}}_{n}$. Now use (\ref{PP_lnn}), monotonicity of $\mathcal{%
\hat{T}}_{n}(\cdot )$, and $\inf_{n\geq 1}\{\sqrt{l}-\sqrt{\mathcal{L}_{n}}%
\} $ $>$ $0$, to yield that as $n$ $\rightarrow $ $\infty $:
\begin{eqnarray*}
P\left( \mathcal{\hat{T}}_{n}(\mathcal{L}_{n})-\mathcal{\hat{T}}_{n}(l)\geq
\left( \sqrt{\mathcal{L}_{n}}-\sqrt{l}\right) \sqrt{\ln (n)}\right)
&=&P\left( \frac{\mathcal{\hat{T}}_{n}(\mathcal{L}_{n})-\mathcal{\hat{T}}%
_{n}(l)}{\sqrt{\ln (n)}}\geq \sqrt{\mathcal{L}_{n}}-\sqrt{l}\right) \\
&& \\
&=&P\left( \sqrt{l}-\sqrt{\mathcal{L}_{n}}\geq \frac{\mathcal{\hat{T}}%
_{n}(l)-\mathcal{\hat{T}}_{n}(\mathcal{L}_{n})}{\sqrt{\ln (n)}}\right)
\rightarrow 1.
\end{eqnarray*}%
Similarly, if $l$ $=$ $\mathcal{L}_{n}$ then $\sqrt{l}$ $-$ $\sqrt{\mathcal{L%
}_{n}}$ $=$ $0$ and $\mathcal{\hat{T}}_{n}(l)$ $-$ $\mathcal{\hat{T}}_{n}(%
\mathcal{L}_{n})$ $=$ $0$ hence the above limit holds.

Conversely, suppose $l\in \{1,...,\mathcal{L}_{n}$ $-$ $1\}$\ and $\mathcal{L%
}_{n}$ $>$ $1$. Then from $\mathcal{\hat{T}}_{n}(\mathcal{L}_{n})$ $=$ $%
o_{p}(\sqrt{q\ln n})$\ and $1$ $-$ $\sqrt{l/\mathcal{L}_{n}}>$ $0$ it
follows:
\begin{equation*}
P\left( \mathcal{\hat{T}}_{n}(\mathcal{L}_{n})-\mathcal{\hat{T}}_{n}(l)\geq
\left( \sqrt{\mathcal{L}_{n}}-\sqrt{l}\right) \sqrt{\ln (n)}\right) =P\left(
\frac{\mathcal{\hat{T}}_{n}(\mathcal{L}_{n})-\mathcal{\hat{T}}_{n}(l)}{\sqrt{%
\mathcal{L}_{n}}\sqrt{\ln (n)}}\geq \left( 1-\sqrt{\frac{l}{\mathcal{L}_{n}}}%
\right) \right) \rightarrow 0.
\end{equation*}%
Claim (\ref{PTT}) follows directly.\medskip \newline
\textbf{Claim (b).}\qquad Let $H_{1}$ hold. Let $ap1$ denote \textit{%
asymptotically with probability approaching one}. Define $h_{n}^{\ast }$ $%
\equiv $ $\min \{h_{n}$ $:$ $h_{n}$ $=$ $\arg \max_{1\leq h\leq \mathcal{%
\bar{L}}_{n}}|\hat{\rho}_{n}(h)|\}$, the smallest lag at which the largest
sample correlation in magnitude over lags $1$ $\leq $ $h$ $\leq $ $\mathcal{%
\bar{L}}_{n}$ occurs.

Define $\mathbb{N}_{1}$ $\equiv $ $\{h$ $\in $ $\mathbb{N}$ $:$ $E[\epsilon
_{t}\epsilon _{t-h}]$ $\neq $ $0\}$ and $\mathbb{\text{\b{N}}}_{1}$ $\equiv $
$\min \{\mathbb{N}_{1}\}$, the smallest lag at which the autocorrelation is
not zero. We prove in Step 1 that for any integer sequence $\{\mathcal{L}%
_{n}\}$ such that $\mathcal{L}_{n}$ $\rightarrow $ $[\mathbb{\text{\b{N}}}%
_{1},\infty ]$ and $\mathcal{L}_{n}/\mathcal{\bar{L}}_{n}$ $\rightarrow $ $%
[0,1]$:%
\begin{equation}
P\left( \mathcal{P}_{n}(\mathcal{L}_{n})=\sqrt{2\mathcal{L}_{n}}\right)
\rightarrow 1.  \label{P2L}
\end{equation}%
We then prove in Step 2 that \textit{if and only if} $\mathcal{L}%
_{n}/h_{n}^{\ast }$ $\overset{p}{\rightarrow }$ $[1,\infty ]$:
\begin{equation}
P\left( \mathcal{\hat{T}}_{n}(\mathcal{L}_{n})\geq \mathcal{\hat{T}}%
_{n}(l)+2(\sqrt{\mathcal{L}_{n}}-\sqrt{l})\right) \rightarrow 1\text{
\textit{for each} }1\leq l\leq \mathcal{\bar{L}}_{n}.  \label{TTLl}
\end{equation}%
Moreover, $h_{n}^{\ast }$ $\overset{p}{\rightarrow }$ $h^{\ast }$ $\equiv $ $%
\min \{h$ $:$ $h$ $=$ $\arg \max_{1\leq h\leq \infty }|\rho (h)|\}$ is an
easy consequence of $\mathcal{\bar{L}}_{n}$ $\rightarrow $ $\infty $,
consistency of the sample correlation under the stated assumptions, and
Slutsky's theorem. Notice $h^{\ast }$ $\in $ $[\mathbb{\text{\b{N}}}%
_{1},\infty )$ by construction of $\mathbb{\text{\b{N}}}_{1}$.

The proof of the claim then proceeds as follows. Take any integer sequence $%
\{\mathcal{L}_{n}\}$, $\mathcal{L}_{n}/h_{n}^{\ast }$ $\overset{p}{%
\rightarrow }$ $[1,\infty ]$ and $\mathcal{L}_{n}/\mathcal{\bar{L}}_{n}$ $%
\rightarrow $ $[0,1]$. Then (\ref{P2L}) holds because $h^{\ast }$ $\in $ $[%
\mathbb{\text{\b{N}}}_{1},\infty )$, hence $\mathcal{\hat{T}}_{n}^{\mathcal{P%
}}(\mathcal{L}_{n})$ $\equiv $ $\mathcal{\hat{T}}_{n}(\mathcal{L}_{.})$ $-$ $%
\sqrt{2\mathcal{L}_{n}}$ $ap1$. Since such a sequence implies (\ref{TTLl}),
we have $\mathcal{\hat{T}}_{n}^{\mathcal{P}}(\mathcal{L}_{n})$ $\geq $ $%
\mathcal{\hat{T}}_{n}^{\mathcal{P}}(l)$ $ap1$\ for each $l$ $=$ $1,...,%
\mathcal{\bar{L}}_{n}$. Conversely, if (\ref{TTLl}) holds then $\mathcal{L}%
_{n}/h_{n}^{\ast }$ $\overset{p}{\rightarrow }$ $[1,\infty ]$. This yields (%
\ref{P2L}) because $h^{\ast }$ $\in $ $[\mathbb{\text{\b{N}}}_{1},\infty )$.
Therefore $\mathcal{\hat{T}}_{n}^{\mathcal{P}}(\mathcal{L}_{n})$ $\geq $ $%
\mathcal{\hat{T}}_{n}^{\mathcal{P}}(l)$ $ap1$\ for each $l$ $=$ $1,...,%
\mathcal{\bar{L}}_{n}$ \textit{if and only if} $\mathcal{L}_{n}/h_{n}^{\ast
} $ $\overset{p}{\rightarrow }$ $[1,\infty ]$. Since the optimal $\{\mathcal{%
L}_{n}^{\ast }\}$ is the least of such sequences, the selection $\mathcal{L}%
_{n}^{\ast }$ satisfies $\mathcal{L}_{n}/h_{n}^{\ast }$ $\overset{p}{%
\rightarrow }$ $1$. Together $\mathcal{L}_{n}/h_{n}^{\ast }$ $\overset{p}{%
\rightarrow }$ $1$ and $h_{n}^{\ast }$ $\overset{p}{\rightarrow }$ $h^{\ast
} $ prove the claim.\medskip

\textbf{Step 1:}\qquad Consider (\ref{P2L}). Use (\ref{rn_ro_h}) to deduce $%
\hat{\rho}_{n}(h)$ $-$ $\rho (h)\overset{p}{\rightarrow }0$ for each $h$.
Lemma \ref{lm:max_p} therefore yields for some integer sequence $\{\mathcal{%
\bar{L}}_{n}\}$, $\mathcal{\bar{L}}_{n}$ $\rightarrow $ $\infty $:%
\begin{equation*}
\left\vert \max_{1\leq h\leq \mathcal{\bar{L}}_{n}}\left\vert \hat{\rho}%
_{n}(h)\right\vert -\max_{1\leq h\leq \mathcal{\bar{L}}_{n}}\left\vert \rho
(h)\right\vert \right\vert \leq \left\vert \max_{1\leq h\leq \mathcal{\bar{L}%
}_{n}}\left\vert \hat{\rho}_{n}(h)-\rho (h)\right\vert \right\vert \overset{p%
}{\rightarrow }0,
\end{equation*}%
where $\lim_{n\rightarrow \infty }\max_{1\leq h\leq \mathcal{\bar{L}}%
_{n}}\left\vert \rho (h)\right\vert $ $\in $ $(0,\infty )$. By monotonicity,
for any $\{\mathcal{L}_{n}\}$, $\mathcal{L}_{n}$ $\rightarrow $ $(0,\infty ]$
and $\mathcal{L}_{n}/\mathcal{\bar{L}}_{n}$ $\rightarrow $ $[0,1]$, and
sufficiently large $n$:%
\begin{equation*}
\left\vert \max_{1\leq h\leq \mathcal{L}_{n}}\left\vert \hat{\rho}%
_{n}(h)\right\vert -\max_{1\leq h\leq \mathcal{L}_{n}}\left\vert \rho
(h)\right\vert \right\vert \leq \left\vert \max_{1\leq h\leq \mathcal{L}%
_{n}}\left\vert \hat{\rho}_{n}(h)-\rho (h)\right\vert \right\vert \leq
\left\vert \max_{1\leq h\leq \mathcal{\bar{L}}_{n}}\left\vert \hat{\rho}%
_{n}(h)-\rho (h)\right\vert \right\vert \overset{p}{\rightarrow }0.
\end{equation*}%
Therefore for any $\{\mathcal{L}_{n}\}$, $\mathcal{L}_{n}$ $\rightarrow $ $[%
\mathbb{\text{\b{N}}}_{1},\infty ]$ and $\mathcal{L}_{n}/\mathcal{\bar{L}}%
_{n}$ $\rightarrow $ $[0,1]$:
\begin{equation*}
\frac{\mathcal{\hat{T}}_{n}(\mathcal{L}_{n})}{\sqrt{q\ln n}}=\frac{\sqrt{n}%
\max_{1\leq h\leq \mathcal{L}_{n}}\left\vert \hat{\rho}_{n}(h)\right\vert }{%
\sqrt{q\ln n}}\overset{p}{\rightarrow }\infty .
\end{equation*}%
This proves (\ref{P2L}) by construction (\ref{Pn}) of the penalty term $%
\mathcal{P}_{n}(\mathcal{L}_{n})$.\medskip

\textbf{Step 2:}\qquad Next we prove (\ref{TTLl}). First, note that by
Theorem \ref{th:max_corr_expan} $\mathcal{\hat{T}}_{n}(\mathcal{L}_{n})/%
\sqrt{n}$ $\overset{p}{\rightarrow }$ $(0,1)$ for any $\{\mathcal{L}_{n}\}$,
$\mathcal{L}_{n}$ $\rightarrow $ $[\mathbb{\text{\b{N}}}_{1},\infty ]$ and $%
\mathcal{L}_{n}/\mathcal{\bar{L}}_{n}$ $\rightarrow $ $[0,1]$. Hence $%
\mathcal{\hat{T}}_{n}(\mathcal{L}_{n})/\sqrt{n/\ln (n)}$ $\overset{p}{%
\rightarrow }$ $\infty $ for any $\mathcal{L}_{n}$ $\rightarrow $ $[\mathbb{%
\text{\b{N}}}_{1},\infty ]$, where $\mathcal{L}_{n}$ $=$ $o(n/\ln (n))$ by
assumption. Monotonicity ensures $\mathcal{\hat{T}}_{n}(\mathcal{L}_{n})$ $%
\geq $ $\mathcal{\hat{T}}_{n}(l)$ for each $l$ $\leq $ $\mathcal{L}_{n}$,
hence $\mathcal{\hat{T}}_{n}(l)/\mathcal{\hat{T}}_{n}(\mathcal{L}_{n})$ $=$ $%
[\mathcal{\hat{T}}_{n}(l)/\sqrt{n}]/[\mathcal{\hat{T}}_{n}(\mathcal{L}_{n})/%
\sqrt{n}]$ $\overset{p}{\rightarrow }$ $[0,1]$ for such $l$. Indeed, if both
$(l,\mathcal{L}_{n})\geq h_{n}^{\ast }$ $\equiv $ $\min \{h_{n}$ $:$ $h_{n}$
$=$ $\arg \max_{1\leq h\leq \mathcal{\bar{L}}_{n}}|\hat{\rho}_{n}(h)|\}$
then by construction $\mathcal{\hat{T}}_{n}(l)/\mathcal{\hat{T}}_{n}(%
\mathcal{L}_{n})$ $=$ $1$.

Now suppose $1$ $\leq $ $l$ and $l/\mathcal{L}_{n}$ $\rightarrow $ $[0,1)$,
and $\mathcal{L}_{n}/h_{n}^{\ast }$ $\overset{p}{\rightarrow }$ $[0,1)$,
hence $1$ $\leq $ $l$ $<$ $\mathcal{L}_{n}$ $<$ $h_{n}^{\ast }$ as $n$ $%
\rightarrow $ $\infty $ $ap1$. Then $\mathcal{\hat{T}}_{n}(l)/\mathcal{\hat{T%
}}_{n}(\mathcal{L}_{n})$ $\overset{p}{\rightarrow }$ $[0,1)$ by monotonicity
and the construction of $h_{n}^{\ast }$. Now use $\mathcal{L}_{n}$ $\leq $ $%
\mathcal{\bar{L}}_{n}$ $=$ $o(n/\ln (n))$ by assumption, and $\mathcal{\hat{T%
}}_{n}(\mathcal{L}_{n})/\sqrt{n/\ln (n)}$ $\overset{p}{\rightarrow }$ $%
\infty $ to yield:
\begin{eqnarray}
P\left( \mathcal{\hat{T}}_{n}(\mathcal{L}_{n})\geq \mathcal{\hat{T}}%
_{n}(l)+2\left( \sqrt{\mathcal{L}_{n}}-\sqrt{l}\right) \right)  &=&P\left(
\mathcal{\hat{T}}_{n}(\mathcal{L}_{n})\left( 1-\frac{\mathcal{\hat{T}}_{n}(l)%
}{\mathcal{\hat{T}}_{n}(\mathcal{L}_{n})}\right) \geq 2\sqrt{\mathcal{L}_{n}}%
\left( 1-\sqrt{\frac{l}{\mathcal{L}_{n}}}\right) \right) \text{ \ \ \ \ \ \
\ \ \ }  \label{PPP1} \\
&\geq &P\left( \frac{\mathcal{\hat{T}}_{n}(\mathcal{L}_{n})}{\sqrt{n/\ln (n)}%
}\left( 1-\frac{\mathcal{\hat{T}}_{n}(l)}{\mathcal{\hat{T}}_{n}(\mathcal{L}%
_{n})}\right) \geq 2\sqrt{\frac{\mathcal{L}_{n}}{n/\ln (n)}}\right)
\rightarrow 1.  \notag
\end{eqnarray}

Next, consider $1$ $\leq $ $l$ and $l/h_{n}^{\ast }$ $\overset{p}{%
\rightarrow }$ $[0,1)$, and $\mathcal{L}_{n}/h_{n}^{\ast }$ $\overset{p}{%
\rightarrow }$ $[1,\infty ]$, hence $1$ $\leq $ $l$ $\leq $ $h_{n}^{\ast }$ $%
-$ $1$ $ap1$ and $\mathcal{L}_{n}$ $\geq $ $h_{n}^{\ast }$ $ap1$. Then $P(%
\mathcal{\hat{T}}_{n}(l)$ $=$ $\mathcal{\hat{T}}_{n}(\mathcal{L}_{n})$ $%
\rightarrow $ $0$ since by construction $h_{n}^{\ast }$ is the smallest lag
at which the maximum correlation occurs. Monotonicity therefore yields $%
\mathcal{\hat{T}}_{n}(l)/\mathcal{\hat{T}}_{n}(\mathcal{L}_{n})$ $\overset{p}%
{\rightarrow }$ $[0,1)$, and again we deduce (\ref{PPP1}).

Now let $(l,\mathcal{L}_{n})$ $\geq $ $h_{n}^{\ast }$ $ap1$. Then by
construction $\mathcal{\hat{T}}_{n}(\mathcal{L}_{n})$ $=$ $\mathcal{\hat{T}}%
_{n}(l)$ $ap1$. Trivially if $l$ $<$ $\mathcal{L}_{n}$ $(l$ $\geq $ $%
\mathcal{L}_{n})$ then $\sqrt{\mathcal{L}_{n}}$ $-$ $\sqrt{l}$ $>$ $0$ ($%
\sqrt{\mathcal{L}_{n}}$ $-$ $\sqrt{l}$ $\leq $ $0$). Hence $P(\mathcal{\hat{T%
}}_{n}(\mathcal{L}_{n})$ $\geq $ $\mathcal{\hat{T}}_{n}(l)$ $+$ $2[\sqrt{%
\mathcal{L}_{n}}$ $-$ $\sqrt{l}])$ $\rightarrow $ $1$ \textit{if and only if}
$l$ $\geq $ $\mathcal{L}_{n}$.

Next, let $\mathcal{L}_{n}$ $<$ $h_{n}^{\ast }$ $\leq $ $l$ $ap1$ such that $%
\mathcal{\hat{T}}_{n}(l)$ $=$ $\mathcal{\hat{T}}_{n}(h_{n}^{\ast })$ $ap1$.
Use $\mathcal{L}_{n}/l$ $\rightarrow $ $[0,1)$, $l$ $=$ $o(n/\ln (n))$, $%
\mathcal{\hat{T}}_{n}(h_{n}^{\ast })/\sqrt{n/\ln (n)}$ $\overset{p}{%
\rightarrow }$ $\infty $, and $\mathcal{\hat{T}}_{n}(\mathcal{L}_{n})/%
\mathcal{\hat{T}}_{n}(h_{n}^{\ast })$ $\overset{p}{\rightarrow }$ $[0,1)$ to
yield:
\begin{equation*}
P\left( \mathcal{\hat{T}}_{n}(\mathcal{L}_{n})\geq \mathcal{\hat{T}}%
_{n}(l)+2\left( \sqrt{\mathcal{L}_{n}}-\sqrt{l}\right) \right) =P\left(
2\left( 1-\sqrt{\frac{\mathcal{L}_{n}}{l}}\right) \sqrt{\frac{l}{n/\ln (n)}}%
\geq \frac{\mathcal{\hat{T}}_{n}(h_{n}^{\ast })}{\sqrt{n/\ln (n)}}\left( 1-%
\frac{\mathcal{\hat{T}}_{n}(\mathcal{L}_{n})}{\mathcal{\hat{T}}%
_{n}(h_{n}^{\ast })}\right) \right) \rightarrow 0.
\end{equation*}

Finally, generally $\mathcal{\hat{T}}_{n}(l)$ $=$ $\mathcal{\hat{T}}_{n}(%
\mathcal{L}_{n})$ $a.s.$ for some $\{l,\mathcal{L}_{n}\}$ and all but a
finite number of $n$ is possible. For example when $l$ $=$ $\mathcal{L}_{n}$%
. In this case $P(\mathcal{\hat{T}}_{n}(\mathcal{L}_{n})$ $\geq $ $\mathcal{%
\hat{T}}_{n}(l)$ $+$ $2(\sqrt{\mathcal{L}_{n}}$ $-$ $\sqrt{l}))$ $=$ $P(0$ $%
\geq $ $2(\sqrt{\mathcal{L}_{n}}$ $-$ $\sqrt{l}))$ $\rightarrow $ $1$
\textit{if and only if} $l$ $\geq $ $\mathcal{L}_{n}$.

Combining the above results, we deduce $P(\mathcal{\hat{T}}_{n}(\mathcal{L}%
_{n})$ $\geq $ $\mathcal{\hat{T}}_{n}(l)$ $+$ $2[\sqrt{\mathcal{L}_{n}}$ $-$
$\sqrt{l}])$ $\rightarrow $ $1$ \textit{for every} $1$ $\leq $ $l$ $\leq $ $%
\mathcal{\bar{L}}_{n}$\ \textit{if and only if} $\mathcal{L}_{n}$ $\geq $ $%
h_{n}^{\ast }$, proving (\ref{TTLl}). $\mathcal{QED}$.

\setstretch{.8}
\bibliographystyle{econometrica}
\bibliography{ref_max_corr}

\clearpage

%%%%%%%%%%%%%%%%%%%%%%%%%
%%%%%%%%%%%%%%%%%%%%%%%%%

\begin{table}[th]
\caption{Median of Automatically Selected Lags $\mathcal{L}_{n}^{*}$ }
\label{table:median_automatic_lag}
\begin{center}
{\fontsize{10.5pt}{19pt} \selectfont
\begin{tabular}{c|c|c|c|c}
\hline
$e_{t}$ & IID & GARCH(1,1) & MA(2) & AR(1) \\ \hline
$n$ & $\{ 100, 250, 500, 1000 \}$ & $\{ 100, 250, 500, 1000 \}$ & $\{ 100,
250, 500, 1000 \}$ & $\{ 100, 250, 500, 1000 \}$ \\ \hline
\#1 & $\{ 1, 1, 1, 1 \}$ & $\{ 1, 1, 1, 1 \}$ & $\{ 1, 1, 1, 1 \}$ & $\{ 1,
1, 1, 1 \}$ \\
& $H_{0}$, $h^{*} = 1$ & $H_{0}$, $h^{*} = 1$ & $H_{1}$, $h^{*} = 1$ & $%
H_{1} $, $h^{*} = 1$ \\ \hline
\#2 & $\{ 1, 1, 1, 1 \}$ & $\{ 1, 2, 2, 2 \}$ & $\{ 1, 1, 1, 1 \}$ & $\{ 1,
1, 1, 1 \}$ \\
& $H_{0}$, $h^{*} = 1$ & $H_{1}$, $\hat{h}^{*} = 4$ & $H_{1}$, $\hat{h}^{*}
= 1$ & $H_{1}$, $\hat{h}^{*} = 1$ \\ \hline
\#3 & $\{ 1, 1, 1, 1 \}$ & $\{ 1, 1, 1, 1 \}$ & $\{ 1, 1, 1, 1 \}$ & $\{ 1,
1, 1, 2 \}$ \\
& $H_{0}$, $h^{*} = 1$ & $H_{0}$, $h^{*} = 1$ & $H_{1}$, $h^{*} = 1$ & $%
H_{1} $, $h^{*} = 1$ \\ \hline
\#4 & $\{ 2, 2, 2, 2 \}$ & $\{ 2, 2, 2, 2 \}$ & $\{ 1, 1, 2, 1 \}$ & $\{ 1,
1, 1, 1 \}$ \\
& $H_{1}$, $\hat{h}^{*} = 1$ & $H_{1}$, $\hat{h}^{*} = 1$ & $H_{1}$, $\hat{h}%
^{*} = 1$ & $H_{1}$, $\hat{h}^{*} = 1$ \\ \hline
\#5 & $\{ 1, 1, 1, 1 \}$ & $\{ 1, 1, 2, 2 \}$ & $\{ 1, 1, 1, 1 \}$ & $\{ 1,
1, 1, 1 \}$ \\
& $H_{0}$, $h^{*} = 1$ & $H_{1}$, $\hat{h}^{*} = 4$ & $H_{1}$, $\hat{h}^{*}
= 1$ & $H_{1}$, $\hat{h}^{*} = 1$ \\ \hline
\#6 & $\{ 1, 1, 1, 1 \}$ & $\{ 1, 1, 1, 1 \}$ & $\{ 1, 1, 1, 1 \}$ & $\{ 1,
1, 1, 1 \}$ \\
& $H_{0}$, $h^{*} = 1$ & $H_{0}$, $h^{*} = 1$ & $H_{1}$, $h^{*} = 1$ & $%
H_{1} $, $h^{*} = 1$ \\ \hline
\#7 & $\{ 1, 1, 6, 6 \}$ & - & - & - \\
& $H_{1}$, $h^{*} = 6$ & - & - & - \\ \hline
\#8 & $\{ 1, 1, 12, 12 \}$ & - & - & - \\
& $H_{1}$, $h^{*} = 12$ & - & - & - \\ \hline
\#9 & $\{ 1, 1, 1, 24 \}$ & - & - & - \\
& $H_{1}$, $h^{*} = 24$ & - & - & - \\ \hline
\end{tabular}
}
\end{center}
\par
{\fontsize{9.5pt}{13pt} \selectfont
\#1: simple $y_{t}=e_{t}$ with a mean filter. \#2: bilinear process with a
mean filter. \#3: AR(2) process with an AR(2) filter. \#4: AR(2) process
with an AR(1) filter. \#5: GARCH(1,1) process without a filter. \#6:
GARCH(1,1) process with a GARCH filter. \#7: Remote MA(6) process with a
mean filter. \#8: Remote MA(12) process with a mean filter. \#9: Remote
MA(24) process with a mean filter. The error term $e_{t}$ is IID,
GARCH(1,1), MA(2), or AR(1) in Scenarios \#1--\#6, while it is IID in
Scenarios \#7--\#9. This table reports the median of automatic lags for
actual test statistics, $\mathcal{L}_{n}^{\ast }$, across $J=1000$ Monte
Carlo samples. The largest possible lag length is $\bar{\mathcal{L}}_{n} =
[10 \sqrt{n} / (\ln n)]$. The tuning parameter that affects the penalty term
$\mathcal{P}_{n}(\mathcal{L})$ is $q=3$. $H_{0}$ implies the test variable $%
\{\epsilon _{t}\}$ is white noise, while $H_{1}$ implies $\{\epsilon_{t}\}$
is serially correlated. The smallest lag at which the largest correlation
occurs, $h^{\ast }$, is reported if it can be computed analytically.
Otherwise, we report a simulation-based $\hat{h}^{\ast}$ as follows. $J=50000
$ Monte Carlo samples of size $n=50000$ are generated, and sample
autocorrelations of $\{\epsilon _{t}\}$ at $h=1,\dots ,20$ are computed. Let
$\hat{h}_{j}^{\ast }$ be the smallest lag at which the largest correlation
occurs for the $j^{th}$ sample, then the reported $\hat{h}^{\ast }$ is the
median of $\{\hat{h}_{1}^{\ast },\dots ,\hat{h}_{J}^{\ast}\}$. }
\end{table}

\clearpage

\begin{table}[th]
\caption{Rejection Frequencies of Max-Correlation Test with Automatic Lag $%
\mathcal{L}_{n}^{*}$ (Scenarios \#1--\#6) }
\label{table:main_paper_rf_automatic_scenario123456}
\begin{center}
{\fontsize{9.5pt}{15.5pt} \selectfont
\begin{tabular}{c|c|c|c|c|c|c}
\multicolumn{7}{c}{IID Error: $e_{t} = \nu_{t}$} \\ \hline
& \#1. Simple & \#2. Bilin & \#3. AR2/AR2 & \#4. AR2/AR1 & \#5. GARCH/wo &
\#6. GARCH/w \\ \hline
$n$ & 1\%, 5\%, 10\% & 1\%, 5\%, 10\% & 1\%, 5\%, 10\% & 1\%, 5\%, 10\% &
1\%, 5\%, 10\% & 1\%, 5\%, 10\% \\ \hline
$100$ & .017, .068, .128 & .008, .047, .090 & .002, .061, .129 & .034, .197,
.327 & .006, .031, .068 & .026, .091, .140 \\
$250$ & .011, .045, .087 & .012, .042, .087 & .005, .048, .093 & .178, .479,
.634 & .005, .029, .063 & .019, .064, .125 \\
$500$ & .007, .047, .096 & .010, .036, .077 & .004, .045, .094 & .462, .803,
.894 & .005, .033, .083 & .016, .053, .096 \\
$1000$ & .012, .050, .109 & .008, .056, .103 & .006, .057, .097 & .929,
.996, .998 & .008, .038, .077 & .011, .047, .101 \\ \hline
\multicolumn{7}{c}{} \\
\multicolumn{7}{c}{GARCH(1,1) Error: $e_{t} = \nu_{t} w_{t}$ with $w_{t}^{2}
= 1 + 0.2 e_{t-1}^{2} + 0.5 w_{t-1}^{2}$} \\ \hline
& \#1. Simple & \#2. Bilin & \#3. AR2/AR2 & \#4. AR2/AR1 & \#5. GARCH/wo &
\#6. GARCH/w \\ \hline
$n$ & 1\%, 5\%, 10\% & 1\%, 5\%, 10\% & 1\%, 5\%, 10\% & 1\%, 5\%, 10\% &
1\%, 5\%, 10\% & 1\%, 5\%, 10\% \\ \hline
$100$ & .005, .026, .075 & .007, .021, .040 & .004, .054, .109 & .027, .150,
.248 & .001, .003, .012 & .026, .090, .162 \\
$250$ & .004, .031, .069 & .008, .023, .038 & .007, .040, .084 & .116, .312,
.451 & .004, .010, .015 & .011, .063, .107 \\
$500$ & .001, .031, .063 & .017, .029, .042 & .008, .032, .078 & .283, .588,
.733 & .003, .005, .006 & .012, .046, .089 \\
$1000$ & .006, .032, .071 & .014, .026, .031 & .008, .033, .085 & .741,
.925, .961 & .002, .002, .002 & .006, .051, .102 \\ \hline
\multicolumn{7}{c}{} \\
\multicolumn{7}{c}{MA(2) Error: $e_{t} = \nu_{t} + 0.5 \nu_{t-1} + 0.25
\nu_{t-2}$} \\ \hline
& \#1. Simple & \#2. Bilin & \#3. AR2/AR2 & \#4. AR2/AR1 & \#5. GARCH/wo &
\#6. GARCH/w \\ \hline
$n$ & 1\%, 5\%, 10\% & 1\%, 5\%, 10\% & 1\%, 5\%, 10\% & 1\%, 5\%, 10\% &
1\%, 5\%, 10\% & 1\%, 5\%, 10\% \\ \hline
$100$ & .693, .901, .951 & .582, .769, .825 & .012, .068, .135 & .242, .601,
.762 & .461, .707, .788 & .908, .966, .979 \\
$250$ & .993, .998, 1.00 & .841, .935, .962 & .006, .060, .114 & .677, .927,
.982 & .707, .834, .868 & .992, .993, .993 \\
$500$ & 1.00, 1.00, 1.00 & .932, .965, .976 & .019, .086, .151 & .972, .999,
.999 & .798, .874, .908 & 1.00, 1.00, 1.00 \\
$1000$ & 1.00, 1.00, 1.00 & .968, .982, .986 & .063, .184, .257 & 1.00,
1.00, 1.00 & .900, .934, .952 & 1.00, 1.00, 1.00 \\ \hline
\multicolumn{7}{c}{} \\
\multicolumn{7}{c}{AR(1) Error: $e_{t} = 0.7 e_{t-1} + \nu_{t}$} \\ \hline
& \#1. Simple & \#2. Bilin & \#3. AR2/AR2 & \#4. AR2/AR1 & \#5. GARCH/wo &
\#6. GARCH/w \\ \hline
$n$ & 1\%, 5\%, 10\% & 1\%, 5\%, 10\% & 1\%, 5\%, 10\% & 1\%, 5\%, 10\% &
1\%, 5\%, 10\% & 1\%, 5\%, 10\% \\ \hline
$100$ & .531, .752, .847 & .477, .637, .685 & .021, .128, .227 & .263, .636,
.788 & .179, .345, .432 & .987, .991, .991 \\
$250$ & .903, .979, .990 & .676, .774, .812 & .041, .217, .355 & .714, .954,
.984 & .156, .271, .334 & 1.00, 1.00, 1.00 \\
$500$ & .998, 1.00, 1.00 & .715, .819, .860 & .181, .512, .636 & .991, 1.00,
1.00 & .111, .176, .230 & 1.00, 1.00, 1.00 \\
$1000$ & 1.00, 1.00, 1.00 & .723, .823, .864 & .599, .847, .922 & 1.00,
1.00, 1.00 & .066, .126, .158 & 1.00, 1.00, 1.00 \\ \hline
\end{tabular}
}
\end{center}
\par
{\fontsize{9.5pt}{13pt} \selectfont
\#1: Simple $y_{t} = e_{t}$ with a mean filter. \#2: Bilinear $y_{t} = 0.5
e_{t-1} y_{t-2} + e_{t}$ with a mean filter. \#3: AR(2) $y_{t} = 0.3 y_{t-1}
- 0.15 y_{t-2} + e_{t}$ with an AR(2) filter. \#4: AR(2) $y_{t} = 0.3
y_{t-1} - 0.15 y_{t-2} + e_{t}$ with an AR(1) filter. \#5: GARCH(1,1) $y_{t}
= \sigma_{t} e_{t}$, $\sigma_{t}^{2} = 1 + 0.2 y_{t-1}^{2} + 0.5
\sigma_{t-1}^{2}$ without (wo) a filter. \#6: GARCH(1,1) $y_{t} = \sigma_{t}
e_{t}$, $\sigma_{t}^{2} = 1 + 0.2 y_{t-1}^{2} + 0.5 \sigma_{t-1}^{2}$ with
(w) a GARCH filter. For each scenario, $\nu_{t} \overset{i.i.d.}{\sim}
N(0,1) $. The largest possible lag length is $\bar{\mathcal{L}}_{n} = [10
\sqrt{n} / (\ln n)]$, and the tuning parameter that affects the penalty term
$\mathcal{P}_{n} (\mathcal{L})$ is $q = 3$. This table reports rejection
frequencies with respect to nominal size $\alpha \in \{ 0.01, 0.05, 0.10 \}$
across $J = 1000$ Monte Carlo samples, where the number of bootstrap samples
is $M = 500$. }
\end{table}

\clearpage

\begin{table}[th]
\caption{Rejection Frequencies of Cram\'{e}r-von Mises Test $CvM^{dw}$
(Scenarios \#1--\#6) }
\label{table:main_paper_rf_CvM_scenario123456}
\begin{center}
{\fontsize{9.5pt}{15.5pt} \selectfont
\begin{tabular}{c|c|c|c|c|c|c}
\multicolumn{7}{c}{IID Error: $e_{t} = \nu_{t}$} \\ \hline
& \#1. Simple & \#2. Bilin & \#3. AR2/AR2 & \#4. AR2/AR1 & \#5. GARCH/wo &
\#6. GARCH/w \\ \hline
$n$ & 1\%, 5\%, 10\% & 1\%, 5\%, 10\% & 1\%, 5\%, 10\% & 1\%, 5\%, 10\% &
1\%, 5\%, 10\% & 1\%, 5\%, 10\% \\ \hline
100 & .023, .081, .138 & .018, .076, .149 & .020, .086, .167 & .133, .338,
.483 & .021, .077, .141 & .034, .087, .144 \\
250 & .016, .072, .144 & .030, .085, .154 & .011, .065, .127 & .370, .615,
.735 & .011, .058, .118 & .019, .065, .112 \\
500 & .010, .051, .102 & .014, .072, .124 & .012, .059, .132 & .710, .882,
.939 & .009, .053, .103 & .016, .072, .141 \\
1000 & .008, .060, .108 & .016, .063, .106 & .010, .049, .102 & .974, .991,
.993 & .015, .058, .107 & .013, .057, .103 \\ \hline
\multicolumn{7}{c}{} \\
\multicolumn{7}{c}{GARCH(1,1) Error: $e_{t} = \nu_{t} w_{t}$ with $w_{t}^{2}
= 1 + 0.2 e_{t-1}^{2} + 0.5 w_{t-1}^{2}$} \\ \hline
& \#1. Simple & \#2. Bilin & \#3. AR2/AR2 & \#4. AR2/AR1 & \#5. GARCH/wo &
\#6. GARCH/w \\ \hline
$n$ & 1\%, 5\%, 10\% & 1\%, 5\%, 10\% & 1\%, 5\%, 10\% & 1\%, 5\%, 10\% &
1\%, 5\%, 10\% & 1\%, 5\%, 10\% \\ \hline
100 & .017, .081, .149 & .002, .030, .070 & .026, .086, .168 & .118, .287,
.430 & .006, .049, .103 & .036, .100, .168 \\
250 & .013, .059, .108 & .029, .048, .083 & .012, .058, .127 & .242, .501,
.648 & .009, .037, .080 & .020, .075, .132 \\
500 & .015, .066, .115 & .026, .038, .075 & .011, .051, .104 & .550, .802,
.881 & .013, .052, .111 & .026, .072, .143 \\
1000 & .010, .060, .116 & .004, .014, .028 & .008, .056, .105 & .880, .973,
.993 & .006, .032, .065 & .049, .065, .073 \\ \hline
\multicolumn{7}{c}{} \\
\multicolumn{7}{c}{MA(2) Error: $e_{t} = \nu_{t} + 0.5 \nu_{t-1} + 0.25
\nu_{t-2}$} \\ \hline
& \#1. Simple & \#2. Bilin & \#3. AR2/AR2 & \#4. AR2/AR1 & \#5. GARCH/wo &
\#6. GARCH/w \\ \hline
$n$ & 1\%, 5\%, 10\% & 1\%, 5\%, 10\% & 1\%, 5\%, 10\% & 1\%, 5\%, 10\% &
1\%, 5\%, 10\% & 1\%, 5\%, 10\% \\ \hline
100 & .898, .984, .995 & .450, .743, .866 & .029, .113, .182 & .570, .805,
.898 & .681, .908, .969 & .878, .927, .940 \\
250 & .999, 1.00, 1.00 & .769, .924, .968 & .019, .086, .189 & .951, .996,
.999 & .903, .979, .994 & .983, .989, .991 \\
500 & 1.00, 1.00, 1.00 & .884, .966, .990 & .032, .144, .250 & 1.00, 1.00,
1.00 & .959, .994, .995 & .995, .998, .998 \\
1000 & 1.00, 1.00, 1.00 & .974, .994, .997 & .068, .295, .471 & 1.00, 1.00,
1.00 & .986, .997, 1.00 & .998, .998, .998 \\ \hline
\multicolumn{7}{c}{} \\
\multicolumn{7}{c}{AR(1) Error: $e_{t} = 0.7 e_{t-1} + \nu_{t}$} \\ \hline
& \#1. Simple & \#2. Bilin & \#3. AR2/AR2 & \#4. AR2/AR1 & \#5. GARCH/wo &
\#6. GARCH/w \\ \hline
$n$ & 1\%, 5\%, 10\% & 1\%, 5\%, 10\% & 1\%, 5\%, 10\% & 1\%, 5\%, 10\% &
1\%, 5\%, 10\% & 1\%, 5\%, 10\% \\ \hline
100 & .925, .996, 1.00 & .282, .567, .741 & .064, .193, .299 & .472, .741,
.849 & .564, .818, .923 & .958, .970, .973 \\
250 & .999, 1.00, 1.00 & .341, .572, .718 & .136, .341, .465 & .935, .991,
.999 & .680, .849, .912 & .984, .987, .988 \\
500 & 1.00, 1.00, 1.00 & .393, .630, .781 & .325, .592, .700 & .999, 1.00,
1.00 & .700, .852, .918 & .999, 1.00, 1.00 \\
1000 & 1.00, 1.00, 1.00 & .474, .697, .810 & .688, .876, .923 & 1.00, 1.00,
1.00 & .750, .877, .929 & .998, .999, .999 \\ \hline
\end{tabular}
}
\end{center}
\par
{\fontsize{9.5pt}{13pt} \selectfont
\#1: Simple $y_{t} = e_{t}$ with a mean filter. \#2: Bilinear $y_{t} = 0.5
e_{t-1} y_{t-2} + e_{t}$ with a mean filter. \#3: AR(2) $y_{t} = 0.3 y_{t-1}
- 0.15 y_{t-2} + e_{t}$ with an AR(2) filter. \#4: AR(2) $y_{t} = 0.3
y_{t-1} - 0.15 y_{t-2} + e_{t}$ with an AR(1) filter. \#5: GARCH(1,1) $y_{t}
= \sigma_{t} e_{t}$, $\sigma_{t}^{2} = 1 + 0.2 y_{t-1}^{2} + 0.5
\sigma_{t-1}^{2}$ without (wo) a filter. \#6: GARCH(1,1) $y_{t} = \sigma_{t}
e_{t}$, $\sigma_{t}^{2} = 1 + 0.2 y_{t-1}^{2} + 0.5 \sigma_{t-1}^{2}$ with
(w) a GARCH filter. For each scenario, $\nu_{t} \overset{i.i.d.}{\sim}
N(0,1) $. The dependent wild bootstrap with $M = 500$ samples is used to
compute an approximate p-value of the Cram\'{e}r-von Mises test. All $%
\mathcal{L}_{n} = n - 1$ lags are used. This table reports rejection
frequencies with respect to nominal size $\alpha \in \{ 0.01, 0.05, 0.10 \}$
across $J = 1000$ Monte Carlo samples. }
\end{table}

\clearpage

\begin{table}[th]
\caption{Rejection Frequencies of $\hat{\mathcal{T}}^{dw} (\mathcal{L}%
_{n}^{*})$ and $CvM^{dw}$ (Scenarios \#7--\#9) }
\label{table:main_paper_rf_scenario789}
\begin{center}
{\fontsize{11pt}{18pt} \selectfont
\begin{tabular}{c|c||c|c|c||c|c|c||c|c|c}
\multicolumn{11}{c}{Max-correlation test with automatically selected lag $%
\hat{\mathcal{T}}^{dw} (\mathcal{L}_{n}^{*})$} \\ \hline
&  & \multicolumn{3}{c||}{\#7. MA(6)} & \multicolumn{3}{c||}{\#8. MA(12)} &
\multicolumn{3}{c}{\#9. MA(24)} \\ \hline
$n$ & $\bar{\mathcal{L}}_{n}$ & 1\% & 5\% & 10\% & 1\% & 5\% & 10\% & 1\% &
5\% & 10\% \\ \hline
100 & 21 & .016 & .084 & .139 & .013 & .067 & .117 & .017 & .065 & .118 \\
250 & 28 & .155 & .289 & .352 & .024 & .134 & .244 & .012 & .042 & .088 \\
500 & 35 & .710 & .812 & .826 & .371 & .673 & .770 & .024 & .097 & .192 \\
1000 & 45 & .999 & 1.00 & 1.00 & .983 & .997 & .997 & .578 & .833 & .918 \\
\hline
\multicolumn{11}{c}{} \\
\multicolumn{11}{c}{Cram\'{e}r-von Mises test $CvM^{dw}$} \\ \hline
&  & \multicolumn{3}{c||}{\#7. MA(6)} & \multicolumn{3}{c||}{\#8. MA(12)} &
\multicolumn{3}{c}{\#9. MA(24)} \\ \hline
$n$ & $\mathcal{L}_{n}$ & 1\% & 5\% & 10\% & 1\% & 5\% & 10\% & 1\% & 5\% &
10\% \\ \hline
100 & 99 & .040 & .098 & .171 & .034 & .110 & .179 & .029 & .098 & .186 \\
250 & 249 & .026 & .080 & .142 & .025 & .087 & .155 & .022 & .088 & .143 \\
500 & 499 & .014 & .087 & .175 & .026 & .092 & .161 & .024 & .071 & .133 \\
1000 & 999 & .038 & .160 & .320 & .017 & .083 & .166 & .028 & .079 & .144 \\
\hline
\end{tabular}
}
\end{center}
\par
{\fontsize{9.5pt}{13pt} \selectfont
%Scenario \#7: Remote MA(6) $y_{t} = \nu_{t} + 0.25 \nu_{t-6}$ with a mean filter $\epsilon_{t} = y_{t} - E[y_{t}]$.
%Scenario \#8: Remote MA(12) $y_{t} = \nu_{t} + 0.25 \nu_{t-12}$ with a mean filter $\epsilon_{t} = y_{t} - E[y_{t}]$.
%Scenario \#9: Remote MA(24) $y_{t} = \nu_{t} + 0.25 \nu_{t-24}$ with a mean filter $\epsilon_{t} = y_{t} - E[y_{t}]$.
Scenario \#7: Remote MA(6) $y_{t} = e_{t} + 0.25 e_{t-6}$ with a mean
filter. Scenario \#8: Remote MA(12) $y_{t} = e_{t} + 0.25 e_{t-12}$ with a
mean filter. Scenario \#9: Remote MA(24) $y_{t} = e_{t} + 0.25 e_{t-24}$
with a mean filter. For each scenario, $e_{t} \overset{i.i.d.}{\sim} N(0,1)$%
. For each test, the dependent wild bootstrap is used to compute an
approximate p-value. For the max-correlation test, the largest possible lag
length is $\bar{\mathcal{L}}_{n} = [10 \sqrt{n} / (\ln n)]$, and the tuning
parameter that affects the penalty term $\mathcal{P}_{n} (\mathcal{L})$ is $%
q = 3$. For the Cram\'{e}r-von Mises test, all $\mathcal{L}_{n} = n - 1$
lags are used. We report rejection frequencies with respect to nominal size $%
\alpha \in \{ 0.01, 0.05, 0.10 \}$ across $J = 1000$ Monte Carlo samples. }
\end{table}

\clearpage

\begin{figure}[h!]
\caption{Empirical Size and Size-Adjusted Power of $\hat{\mathcal{T}}^{dw} (%
\mathcal{L}_{n}^{*})$ with $\protect\alpha = 0.05$ }
\label{fig:size_power_maxcorr_automatic_lag}
\begin{center}
\captionsetup[subfigure]{labelformat=empty}
\begin{minipage}[b]{72mm}
      \centering
      \includegraphics[width = 72mm, height = 55mm]{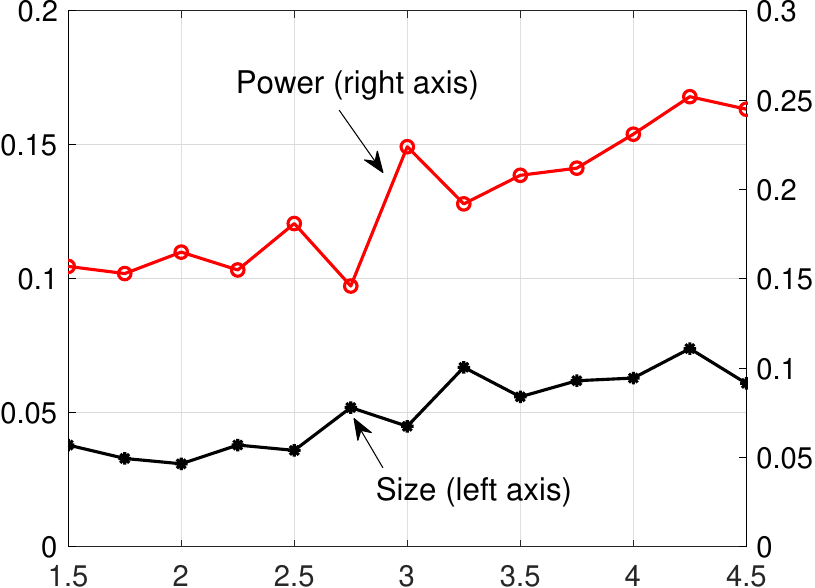}
      \vspace{0.0cm}
      \subcaption{Case 1, $n = 100$}
      \end{minipage}
\quad \quad \quad
\begin{minipage}[b]{72mm}
      \centering
      \includegraphics[width = 72mm, height = 55mm]{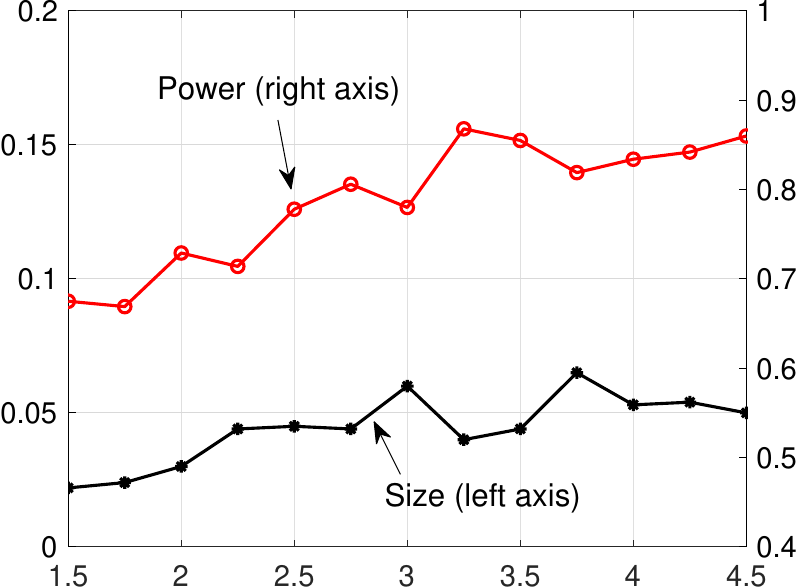}
      \vspace{0.0cm}
      \subcaption{Case 1, $n = 500$}
      \end{minipage} \\[0pt]
\par
\vspace{0.3cm}
\par
\begin{minipage}[b]{72mm}
      \centering
      \includegraphics[width = 72mm, height = 55mm]{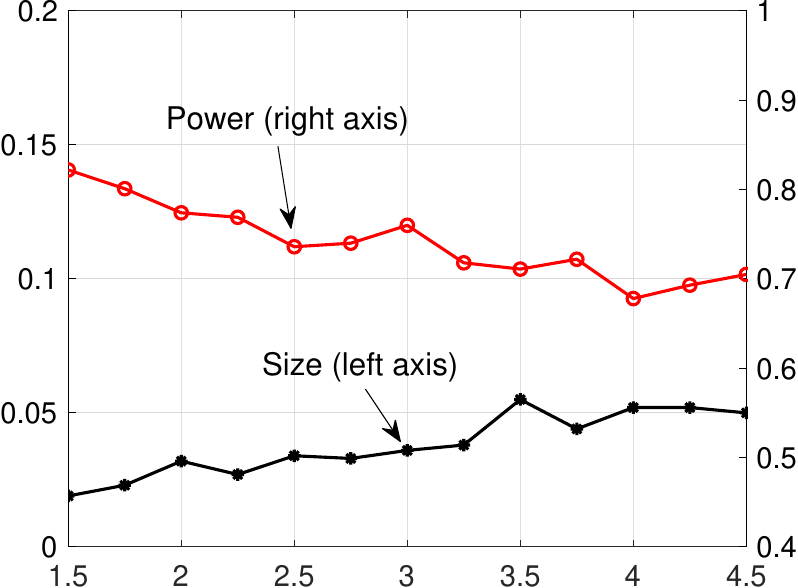}
      \vspace{0.0cm}
      \subcaption{Case 2, $n = 100$}
      \end{minipage}
\quad \quad \quad
\begin{minipage}[b]{72mm}
      \centering
      \includegraphics[width = 72mm, height = 55mm]{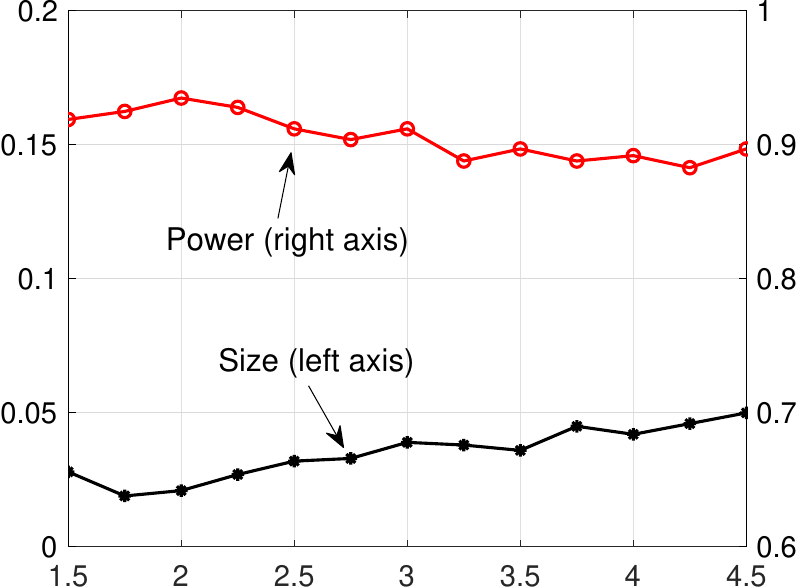}
      \vspace{0.0cm}
      \subcaption{Case 2, $n = 500$}
      \end{minipage} \\[0pt]
\end{center}
\par
{\fontsize{9.5pt}{13pt} \selectfont
We plot empirical size and size-adjusted power of the bootstrapped
max-correlation test with automatic lag selection given nominal size 5\%. In
Case 1, the empirical size and empirical quantiles for size adjustment are
computed under Scenario \#1 (iid $y_{t}$ and mean filter) with i.i.d. error;
then the size-adjusted power is computed under Scenario \#4 (AR(2) $y_{t}$
and AR(1) filter) with i.i.d. error. In Case 2, the empirical size and
empirical quantiles for size adjustment are computed under Scenario \#5
(GARCH $y_{t}$ and no filter) with i.i.d. error; then the size-adjusted
power is computed under Scenario \#5 with MA(2) error. The tuning parameter
that affects the penalty term $\mathcal{P}_{n} (\mathcal{L}) $ is $q \in \{
1.50, 1.75, \dots, 4.50 \}$. The largest possible lag length is $\bar{%
\mathcal{L}}_{n} = [10 \sqrt{n} / (\ln n)]$, which implies that $\bar{%
\mathcal{L}}_{100} = 21$ and $\bar{\mathcal{L}}_{500} = 35$. $J = 1000$
Monte Carlo samples and $M = 500$ bootstrap samples are generated. }
\end{figure}

\clearpage

\end{document}